\documentclass[prb,twocolumn,showpacs,superscriptaddress,amssymb]{revtex4}
\usepackage[colorlinks=true,linkcolor=magenta,citecolor=magenta,urlcolor=magenta,linktocpage=true]{hyperref}
\usepackage[utf8]{inputenc}
\usepackage[english]{babel}
\usepackage[T1]{fontenc}
\usepackage{amsmath}
\usepackage{cleveref}
\usepackage{svg}
\usepackage{tabularx}
\usepackage{array, makecell}

\usepackage{listings}
\usepackage{xcolor}
\definecolor{codegreen}{rgb}{0,0.6,0}
\definecolor{codegray}{rgb}{0.5,0.5,0.5}
\definecolor{codepurple}{rgb}{0.58,0,0.82}
\definecolor{backcolour}{rgb}{0.95,0.95,0.92}
\lstdefinestyle{mystyle}{
    backgroundcolor=\color{backcolour},   
    commentstyle=\color{codegreen},
    keywordstyle=\color{magenta},
    numberstyle=\tiny\color{codegray},
    stringstyle=\color{codepurple},
    basicstyle=\ttfamily\footnotesize,
    breakatwhitespace=false,         
    breaklines=true,                 
    captionpos=b,                    
    keepspaces=true,                 
    numbers=none,                    
    numbersep=5pt,                  
    showspaces=false,                
    showstringspaces=false,
    showtabs=false,                  
    tabsize=2
}
\usepackage{psfrag,graphicx}
\usepackage{dcolumn}
\usepackage{bm}
\usepackage{amsfonts,amssymb,amsmath}       
\usepackage{slashed}
\usepackage[font=small,labelfont=bf]{caption}
\usepackage[utf8]{inputenc}
\usepackage{graphicx}
\usepackage[caption=false]{subfig}
\usepackage{cancel}

\usepackage{xcolor}

\usepackage{amsmath}
\setcitestyle{numbers,square}

\numberwithin{equation}{section}
\newcommand\numberthis{\addtocounter{equation}{1}\tag{\theequation}}

\newcommand{\figref}[1]{\mbox{Fig.~\ref{#1}}}
\newcommand{\be}{\begin{equation}}
\newcommand{\ee}{\end{equation}}
\newcommand{\bq}{\begin{eqnarray}}
\newcommand{\eq}{\end{eqnarray}}

\newcommand{\ket}[1]{\left |#1 \right\rangle}
\newcommand{\bra}[1]{\left \langle #1 \right |}
\newcommand{\ketbra}[2]{\left|#1\right\rangle\left\langle#2\right|}

\newcommand{\hhat}[1]{\hat{\hat{#1}}}

\begin{document}
\title{The Fermionic influence superoperator: a canonical derivation for the development of methods to simulate the influence of a Fermionic environment on a quantum system with arbitrary parity symmetry}
\author{Mauro Cirio}
\email{cirio.mauro@gmail.com}
\affiliation{Graduate School of China Academy of Engineering Physics, Haidian District, Beijing, 100193, China}
\author{Po-Chen Kuo}
\affiliation{Department of Physics, National Cheng Kung University, 701 Tainan, Taiwan}
\affiliation{Center for Quantum Frontiers of Research \& Technology, NCKU, 70101 Tainan, Taiwan}
\author{Yueh-Nan Chen}
\affiliation{Department of Physics, National Cheng Kung University, 701 Tainan, Taiwan}
\affiliation{Center for Quantum Frontiers of Research \& Technology, NCKU, 70101 Tainan, Taiwan}
\author{Franco Nori}
\affiliation{Theoretical Quantum Physics Laboratory, RIKEN Cluster for Pioneering Research, Wako-shi, Saitama 351-0198, Japan}
\affiliation{RIKEN Center for Quantum Computing (RQC), Wakoshi, Saitama 351-0198, Japan}
\affiliation{Physics Department, The University of Michigan, Ann Arbor, Michigan 48109-1040, USA.}
\author{Neill Lambert}
\email{nwlambert@gmail.com}
\affiliation{Theoretical Quantum Physics Laboratory, RIKEN Cluster for Pioneering Research, Wako-shi, Saitama 351-0198, Japan}

\date{\today}

\begin{abstract}
We present a canonical derivation of an influence superoperator which generates the reduced dynamics of a Fermionic quantum system linearly coupled to a Fermionic environment initially at thermal equilibrium. We use this formalism to derive a generalized-Lindblad master equation (in the Markovian limit) and a generalized version of the hierarchical equations of motion valid in arbitrary parity-symmetry conditions, important for the correct evaluation of system correlation functions and spectra.
\end{abstract}
\maketitle
\tableofcontents
\vspace{5mm}

The reduced dynamics of a quantum system linearly coupled to Bosonic and Fermionic baths at thermal equilibrium can be fully specified by correlation functions characterizing the environments. Using path integral techniques, Feynman and Vernon \cite{Vernon} used these correlations to define influence functionals able to generate the effective dynamics of the system after tracing out the  degrees of freedom of a Bosonic environment \cite{Caldeira_Leggett_1,Caldeira_Leggett_2}. 
The capability  of the path integral formalism to intrinsically encode the Fermionic anticommutation relations 
using Grassman variables enabled the extension of the the original derivation to the Fermionic case \cite{Caldeira_0,Chen_YC,Bonig,Yan_5,Schinabeck}. As an alternative to these path-integral approaches, the influence of the environment on the system can also be derived through stochastic \cite{Hsieh_1,Hsieh_2,Shao,Chernyak1,Chernyak2} and algebraic \cite{Yan_1,Yan_2,Yan_3,Yan_4}  techniques or by mapping the bath into physical \cite{Garg, Martinazzo,PhysRevA.90.032114,Strasberg_2016,Woods,Melina,Chin_2010,PhysRevLett.105.050404,PhysRevLett.123.090402,PhysRevB.101.155134} or unphysical \cite{PhysRevA.55.2290,PhysRevLett.120.030402,Lemmer_2018,Lambert,PhysRevA.101.052108,PhysRevX.10.031040} degrees of freedom. In particular, for Bosonic environments, it is also possible to derive \emph{``influence superoperators''} using a \emph{canonical, i.e., purely operator-based}, formalism \cite{Petruccione, Ma_2012, Aurell,JianMa}. 

In order to generalize these canonical methods to the Fermionic case, it is necessary to model anticommutation rules throughout the time evolution, akin to the strategies involving Grassman variables in path integrals. 
To achieve this, we use a parity-based formalism to present \emph{a purely canonical derivation of an influence superoperator which describes the effects of Fermionic environments initially at thermal equilibrium linearly coupled to a quantum system.} The resulting expression allows the computation of the system's dynamics even when the initial state is in a superposition of an even-odd number of Fermions. This is, in principle, generally prevented by parity and charge superselection rules \cite{Wick_1,Wick_2}. However, by refraining  from making  this (usually) physical assumption, we allow the formalism to be  used in more general contexts, such as the computation of correlation functions \cite{KondoPRL} (where fictitious states evolve in time).

To demonstrate the utility of this formal result, we use it  to: $(i)$ derive a generalized Gorini–Kossakowski–Sudarshan–Lindblad  master equation \cite{Lindblad, Gorini} (valid in the Markovian regime) and to $(ii)$ derive a generalized version (without parity-symmetry restrictions) of  another formally exact method: Hierarchical Equations of Motion (HEOM) \cite{Tanimura_1,Tanimura_2,Tanimura_3,Ishizaki_1,Ishizaki_2,Yan_5,Hartle2013,Schinabeck,Gauger_2020}.  As mentioned earlier, relaxing parity-symmetry restrictions is important for the correct evaluation of system correlation functions and spectra, as demonstrated in the application of the HEOM method to single-impurity Anderson models and Kondo physics \cite{KondoPRL}.\\

This article is organized as follows. The main article focuses on the logic of the derivation, highlighting the key conceptual steps. At the same time, each section is associated to a supplementary one presenting technical details which are necessary to justify the proof but not essential to its overall understanding.

The results are described in two main sections. Section \ref{sec:InfluenceSuperoperator} presents the canonical derivation of the influence superoperator which we split into four parts: In subsection \ref{sec:Parity}, we introduce a parity-based formalism and analyze Fermionic partial traces. In subsection \ref{sec:Dyson}, we use this setup to trace out the Fermionic bath and to further expand the reduced dynamics in terms of a Dyson series. In subsection \ref{sec:Wick}, we explicitly highlight the dependence of each $n$-point correlation function appearing in the Dyson series with respect to the $2$-point correlations, i.e., we invoke a version of the Wick's theorem for Fermionic superoperators. Finally, in subsection \ref{sec:Influence}, the resulting expression is formally re-summed into a compact expression written in terms of an influence superoperator, which is the main result of this article. 

In section \ref{sec:Applications}, we use this result to derive a Lindblad master equation in the Markovian regime (subsection \ref{sec:Markovian}) and to derive the HEOM (subsection \ref{sec:HEOM}). In section \ref{sec:correlations}, we discuss the importance of  arbitrary parity-symmetry for computing correlation functions.
\begin{figure}
    \centering
    \includegraphics[width=1\columnwidth]{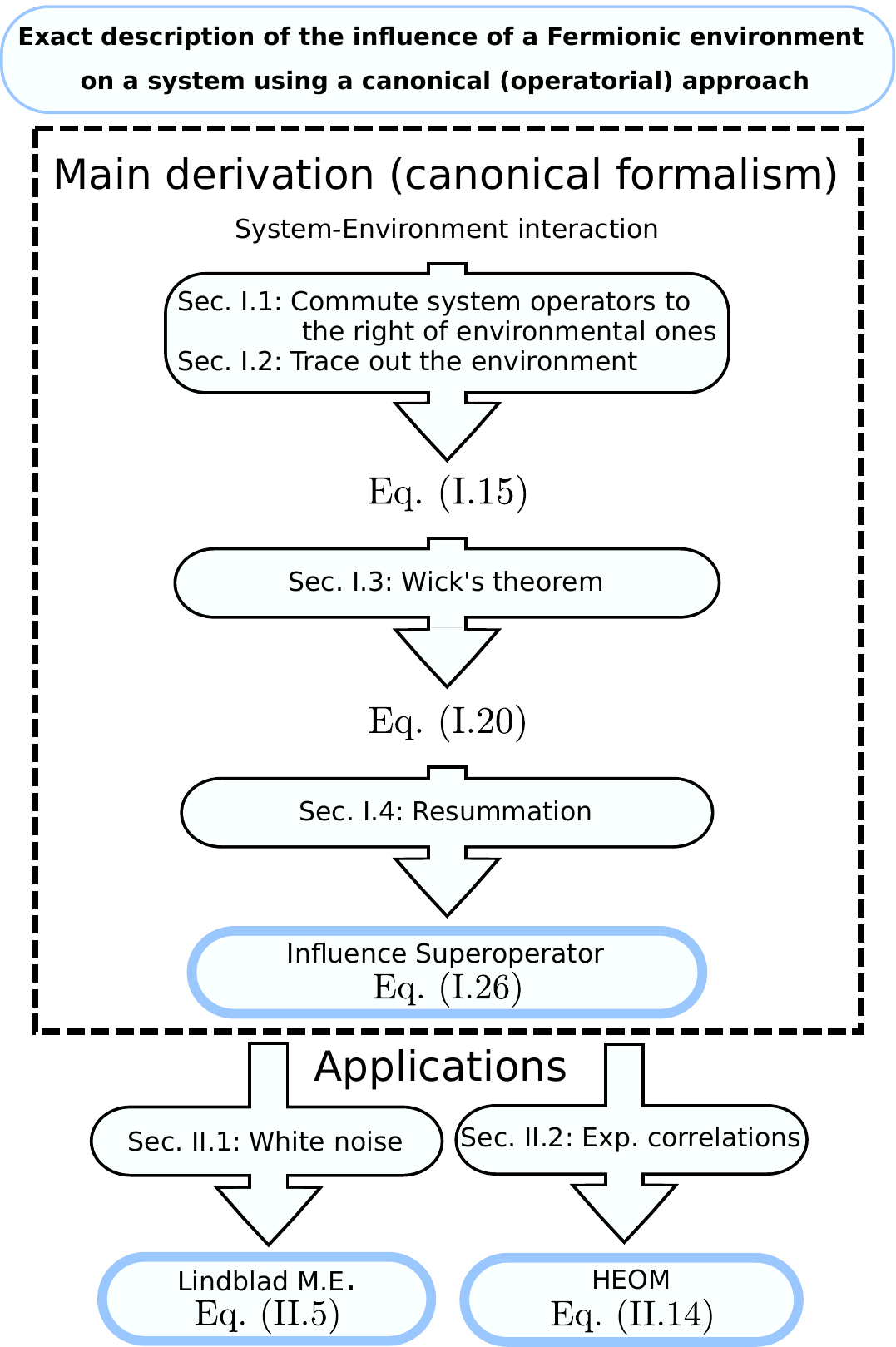}
    \caption{Diagram highlighting the milestones for the derivation and the most important equations in this work.}
    \label{fig:my_label}
\end{figure}
\section{Fermionic influence superoperator}
\label{sec:InfluenceSuperoperator}
We start by introducing the physical setting which we are going to analyze.
We consider an open quantum system \cite{Gardiner,Petruccione,Lidar,PhysRevA.98.063815,Xiong2015,PhysRevA.100.042112,PhysRevA.100.062131,PhysRevA.101.062112,PhysRevLett.116.120402,PhysRevB.78.235311,PhysRevLett.109.170402} described by the Hamiltonian ($\hbar=1$ throughout the article)
\begin{equation}
\label{eq:H_full}
H= H_S + H_E +H_I\;\;,
\end{equation}
where $H_S$ is the system Hamiltonian (which we assume to have even Fermionic parity) and $H_E = \sum_k \omega_k c^\dagger_k c_k$ is the Hamiltonian of the environment in which the $k$th Fermion has energy $\omega_k$ and it is associated with the destruction operator $c_k$. Here, the even/odd parity projections of the operator $O_\mathcal{S}$ in the domain $\mathcal{S}=S/E$, are defined as
\begin{equation}
 O_{\mathcal{S}}^\text{e/o} =\hhat{P}^{\text{e/o}}_{\mathcal{S}}[ O_{\mathcal{S}}]\;\;,
\end{equation}
where $\hhat{P}^{\text{e/o}}_{\mathcal{S}}[\cdot]$ is the projector onto the even/odd subspaces. Throughout this article, we use the double hat notation to label superoperators. Explicitly,
\begin{equation}
\hhat{P}_{\mathcal{S}}^{\text{e/o}}[\cdot]=P_{\mathcal{S}}^{\text{e}}[\cdot]P_{\mathcal{S}}^{\text{e/o}}+P^{\text{o/e}}_{\mathcal{S}}[\cdot]P^{\text{o}}_{\mathcal{S}}\;\;,
\end{equation}
where 
\begin{equation}
    P^{\text{e/o}}_{\mathcal{S}}=(1\pm P_{\mathcal{S}})/2\;\;,
\end{equation}
with
\begin{equation}
    P_{\mathcal{S}}=\prod_{k\in\mathcal{S}} \exp[i\pi f^\dagger_k f_k]\;\;.
\end{equation}
Here, $f_k$ destroys a Fermion in the domain $\mathcal{S}$ (for example, when $\mathcal{S}=E$, $f_k\rightarrow c_k$). We further assume the interaction Hamiltonian to be
\begin{equation}
\label{eq:H_I}
H_I =\displaystyle \sum_k g_k(s c^\dagger_k - s^\dagger c_k )\;\;,
\end{equation}
where $s$ is an (odd-parity) Fermionic operator for the system and $g_k$ quantifies the interaction strength between the system and the $k$th Fermionic mode. 

We define $\rho(t)$ to be the density matrix of the full system+environment, i.e., the solution of the Shr\"odinger equation with Hamiltonian in Eq.~(\ref{eq:H_full}) and subject to the initial condition
\begin{equation}
\label{eq:product}
\rho(0)=\rho^\text{eq}_E\;\rho_S(0)\;\;,
\end{equation} 
where $\rho^\text{eq}_E$ characterizes the environment in thermal equilibrium. While this implies the state of the environment to have even parity, we are not going to assume any parity symmetry for the system's state $\rho_S(0)$.

\emph{The main quantity of interest of this article is the reduced density matrix} $\rho_S(t)$ which is the one containing the same information as $\rho(t)$ as far as expectation values of system operators are concerned, i.e., which fulfills
\begin{equation}
\label{eq:density_matrix_implicit}
\text{Tr}_{ES}[A_S \rho(t)]\equiv\text{Tr}_S[A_S\rho_S(t)]\;\;,
\end{equation}
for all operators $A_S$ with support on the system.

Before attempting to find a formal solution for $\rho_S$, it is important to observe that the Fermionic anticommutation rules require a careful analysis of the concepts of partial trace and tensor product  \cite{Friis_1,Friis_2,Schwarz}). For example, the operators $s$ and $c_k$ in Eq.~(\ref{eq:H_I}) cannot be interpreted as acting independently (as they would in a tensor product) on the system and the environment Hilbert spaces due to the fundamental fact that independent Fermions anticommute rather than commute between each other. In parallel, when the full density matrix $\rho(t)$ has both even and odd parity contributions, the  usual definition $\rho_S=\text{Tr}_E\rho(t)$ cannot be deduced from Eq.~(\ref{eq:density_matrix_implicit}) because of the properties of the  partial trace.\\

Following \cite{Schwarz}, these Fermionic properties can be modeled by a formalism which keeps track of the parity of operators throughout the time-evolution and which we introduce in the next section.

\subsection{A parity-friendly formalism}
\label{sec:Parity}
As a direct consequence of the Fermionic anticommutation rules, two Fermionic operators are, in general, not independent  even when they have support on different physical spaces (here  the environment and the system). At the same time, Fermionic systems come equipped with a $\mathbb{Z}_2$ graded structure, i.e., a decomposition of the Hilbert space into an even- and odd-parity sector. Following Schwarz and collaborators \cite{Schwarz}, it is possible to take advantage of this structure in order to define a parity-formalism in which the system's operators can be effectively treated as independent from the environmental ones, while still accounting for all Fermionic effects. 

To do this, we define (see \cite{Schwarz}) ``hat'' system operators $\hat{O}_S$ as being the same as $O_S$ but commuted to the right of all environmental operators, i.e., $\hat{O}_S O_E=O_E \hat{O}_S$ for all environmental operators $O_E$.
This definition is non-trivial only when ${O}_S$ has odd parity, in which case, the relation with $\hat{O}_S$ depends on the number of environmental Fermions present on the right of $O_S$. Explicitly, we can write, see Eq.~(\ref{eq:hat_app}),
\begin{equation}
\label{eq:simple_but_powerful}
O_S = \hat{O}_S^\text{e}+P_E \hat{O}_S^\text{o}\;\;,
\end{equation}  
where 
\begin{equation}
    P_E=\prod_{k\in E} \exp{[i\pi c^\dagger_k c_k]}\;\;.
\end{equation}
The simplicity of Eq.~(\ref{eq:simple_but_powerful}) should not hide its ability to introduce a Bosonic-like structure in the formalism as ``hat'' system operators commute with environmental ones by construction. 

This definition also allows us to write the following identity for partial traces, see Appendix \ref{sec:Parity_2}, Eq.~(\ref{eq:key_relation_app}), 
\bq
\label{eq:effective_0}
\text{Tr}_{ES} [A_S O_E \hat{O}_S]&=&\text{Tr}_S\left\{A_S\text{Tr}_E[O_E] \hat{O}^\text{e}_S\right.\nonumber\\
&+&\left.\text{Tr}_E [P_E O_E] \hat{O}^\text{o}_S\right\},
\eq
where the operators $O_E$, $A_S$, and $O_S$ have support on the environment and system. By taking $A_S$ to be arbitrary, this equation can be used to implicitly define properties of Fermionic partial traces (at least when ``hat'' operators are present) thereby overcoming the issues originating from the fact that, in general $\text{Tr}[O_EO_S]\neq \text{Tr}_E[O_E]O_S$ in Fermionic systems, see Eq.~(\ref{eq:tr_ineq}).
\subsubsection{Strategy to solve for the reduced system dynamics}
Taken together, Eq.~(\ref{eq:simple_but_powerful}) and Eq.~(\ref{eq:effective_0}) give us a practical strategy to find a formal solution for $\rho_S(t)$. The first step, is to use Eq.~(\ref{eq:simple_but_powerful}) to write the initial condition in Eq.~(\ref{eq:product}) as
\begin{equation}
\label{eq:initial_condition}
\rho(0)=\rho^\text{eq}_E\rho_S(0) = \rho^\text{eq}_E\hat{\rho}_S^\text{e}(0)+\rho^\text{eq}_E P_E \hat{\rho}_S^\text{o}(0)\;\;,
\end{equation} 
and the  interaction Hamiltonian in Eq.~(\ref{eq:H_I}) as 
\begin{equation}
\label{eq:decomposition_Hamiltonian}
H_I=\displaystyle P_E B^\dagger \hat{s} -P_E B  \hat{s}^\dagger\;\;,
\end{equation}
where $B\equiv\sum_k g_k c_k$. The second step, analyzed in the next section, is to formally solve the Shr$\ddot{o}$dinger equation for $\rho(t)$  to find a decomposition of the full density matrix in the form
\begin{equation}
\label{eq:decomposition_0}
\rho(t)= \sum_i{\rho}^i_E\hat{{\rho}}^i_S\;\;.
\end{equation} 
Using the substitution $O_E\hat{O}_S\rightarrow\rho(t)$ in the left hand-side of Eq.~(\ref{eq:effective_0}), direct comparison between the right-hand side of Eq.~(\ref{eq:density_matrix_implicit}) and Eq.~(\ref{eq:effective_0}) gives the following explicit definition of reduced density matrix as
\begin{equation}
\label{eq:effective_1}
{\rho}_S(t)=\sum_i\text{Tr}_E[{\rho}^i_E] {\hat{{\rho}}}^{i,\text{e}}_S+\text{Tr}_E [P_E {\rho}^i_E] {\hat{{\rho}}}^{i,\text{o}}_S\;\;,
\end{equation}
see derivation of Eq.~(\ref{eq:effective_1_app}) in Appendix \ref{sec:Parity_2} for more details.\\

As promised, in the next section we are going to find the explicit expression for the terms in Eq.~(\ref{eq:decomposition_0}) which, used in the equation above, will return the expansion of $\rho_S(t)$ in terms of a ``reduced'' Dyson series.
\subsection{Reduced Dyson series}
\label{sec:Dyson}
In the interaction frame, the full density matrix $\rho(t)$ can be written as the Dyson series
\begin{equation}
\label{eq:series_0}
{\rho}(t)=\displaystyle\sum_{n=0}^\infty\frac{(-i)^n}{n!}\hhat{T}^\text{b}\int_0^t \left[\prod_{i=1}^n d t_i\hhat{H}^\times_{I}(t_i)\right]\rho(0)\;\;,
\end{equation}
where, using Eq.~(\ref{eq:decomposition_Hamiltonian}), $H_I(t)=P_E B^\dagger(t) \hat{s}(t)-P_EB(t)\hat{s}^\dagger(t)$ (in which we used the invariance of $P_E$ under the free dynamics of the bath) in terms of $\hat{s}(t)=U_S^\dagger(t) \hat{s}U_S(t)$ and $B(t)=\sum_k g_k c_k e^{-i\omega_k t}$ with $U(t)=\exp[-i H_S t]$. Here $\hhat{H}^X_I(t)=[H,\cdot]$, where we recall that the double hat notation is  used to label superoperators. Here, the time ordering $\hhat{T}^b$ is the same one used for Bosonic variables. This is due to the fact that the Hamiltonian is even in the fields, see  \cite{Greiner} pag.~217 and pag.~132. We further stress that the time-ordering is defined as acting at the level of superoperators, see Eq.~(\ref{eq:Tb_app}).

The main ingredients of Eq.~(\ref{eq:series_0}) are contributions of the form $\hhat{H}^\times_{I}(t_n)\cdots \hhat{H}^\times_{I}(t_1)\rho(0)$ which, using Eq.~(\ref{eq:decomposition_Hamiltonian}) for the Hamiltonian and Eq.~(\ref{eq:initial_condition}) for the initial state, can be written as a sum over terms with the following structure
\begin{equation}
\label{eq:temp_main_11}
\hhat{T}_E\hhat{\rho}_E^{ \prime n}[\rho_E^\text{eq}]\hhat{T}_S\hhat{\rho}_S^{\prime n}[\hat{\rho}_S^\text{e}]+\hhat{T}_E\hhat{\rho}_E^{\prime \prime  n}[\rho_E^\text{eq}P_E]\hhat{T}_S\hhat{\rho}_S^{\prime \prime n}[\hat{\rho}_S^\text{o}],
\end{equation}
where we omitted the zero-time specification in $\rho^\text{e/o}_S(0)$ and where the explicit expressions are presented in Appendix \ref{app:Dyson}, see Eq.~(\ref{eq:def_from_main}). Here, we highlight that $\hhat{\rho}_E^{\prime n}$ and $\hhat{\rho}_E^{\prime\prime n}$ involve environmental superoperators and that $\hhat{\rho}_S^{\prime n}$ and $\hhat{\rho}_S^{\prime\prime n}$ are defined as the product of $n$ superoperators each evaluated at a different point in time. 

Using the decomposition $\hhat{T}^b=\hhat{T}_E\hhat{T}_S$, these products are time-ordered in terms of both the environment and the system superoperators. Interestingly,  since $H_I$ is even in the fields, $\hhat{T}_E$  and $\hhat{T}_S$ can be chosen as Fermionic, i.e., producing an extra-minus sign each time they apply a swap. This choice is made in order to keep the symmetries explicitly consistent with the application of the Fermionic Wick's theorem as we will see in the next section. 

By using Eq.~(\ref{eq:simple_but_powerful}) into the expressions for $\hhat{\rho}^{\prime n}_S$ and $\hhat{\rho}^{\prime\prime n}_S$ present in Eq.~(\ref{eq:temp_main_11}), the full density matrix $\rho(t)$ is written in terms of the decomposition presented in Eq.~(\ref{eq:decomposition_0}), see Eq.~(\ref{eq:identifications_1}) for further details. In turn, this  justifies the use of Eq.~(\ref{eq:effective_1}) for the reduced density matrix, ultimately allowing  to finally trace out the environmental degrees of freedom to get, see Eq.~(\ref{eq:Dyson_app}),
\begin{equation}
\label{eq:reduced_density_0}
\begin{array}{lll}
\rho_S&=&\displaystyle\sum_{n=0}^\infty\frac{(-1)^n}{n!}\int_0^t \left(\prod_{i=1}^n d t_i\right)\sum_{q_n,\lambda_n\cdots q_1,\lambda_1}\\
&&\displaystyle C^{\lambda_n\cdots \lambda_1}_{q_n\cdots q_1}\hhat{T}_S\left[\hhat{S}_{q_n}^{\bar{\lambda}_n}\cdots\hhat{S}_{q_1}^{\bar{\lambda}_1}\right]{\rho}_S(0)\;\;,
\end{array}
\end{equation}
where we observed that there is no need to keep the ``hat'' notation for system operators once the environment has been traced out, and where 
\begin{equation}
\label{eq:correlation}
C^{\lambda_n\cdots \lambda_1}_{q_n\cdots q_1}=\text{Tr}_E\left[\hhat{T}_E\hhat{B}^{ \lambda_n}_{q_n}\cdots\hhat{B}^{ \lambda_1}_{q_1}\right][\rho^\text{eq}_E]\;\;,
\end{equation}
 with $\bar{\lambda}=-\lambda$.
For notational convenience, we also hide the time-dependence of the superoperators $\hhat{B}$ and $\hhat{S}$ defined as
\begin{equation}
\label{eq:PEB_main}
\begin{array}{lll}
\hhat{B}^{\lambda}_{q}[\cdot]=\delta_{q,1}B^\lambda[\cdot]+\delta_{q,-1}\hhat{P}_E[\cdot B^\lambda]\\
\hhat{S}^{\lambda}_{q}[\cdot]=\delta_{q,1}s^\lambda[\cdot]-\delta_{q,-1}\hhat{P}_S[\cdot s^\lambda]\;\;,
\end{array}
\end{equation}
where, for clarity, we omitted the time dependence (see Eq.~(\ref{eq:PEB_app}) for a more explicit version). This notation uses the upper indexes to denote the presence ($\lambda=1$) or absence ($\lambda=-1$) of a Hermitian conjugation, and lower indexes to characterize the left ($q=+1$) or right ($q=-1$) action. Here, $P_S=\prod_{k\in S} \exp{[i\pi c^\dagger_k c_k]}$ is the parity operator for the system. Remarkably, the disjoint action on the odd and even initial conditions originally present in the terms described in Eq.~(\ref{eq:temp_main_11}) has now been completely encoded into the correlation $C_{q_n\cdots q_1}^{\lambda_n\cdots \lambda_1}$ and the superoperators $\hhat{S}$ which act  on the full $\rho_S(0)$ directly.\\

The environment considered here is described by a quadratic Hamiltonian and it is initially at thermal equilibrium. These characteristics specify the Gaussian nature of the bath, i.e., the possibility to reduce the $n-$point correlation functions appearing in Eq.~(\ref{eq:reduced_density_0}) in terms of 2-point ones. We analyze this in more detail in the next section.
\subsection{Wick's theorem}
\label{sec:Wick}
At first sight, it is not obvious how to prove a Wick's theorem for the correlations defined in Eq.~(\ref{eq:correlation}). The issue is that the usual derivation (see for example, \cite{Rammer}, pag.~243) fails because superoperators do not obey clear-cut commutation or anticommutation relations. For example, superoperators which create different Fermionic particles on different sides of their argument trivially commute, while they anticommute when acting on the same side. To deal with this, we use the elegant techniques developed by Saptsov \emph{et al.} in \cite{Saptsov}. There, see also the analysis done at the end of Appendix (\ref{sec:Wick:super:elegant}), it is shown that a form of Wick's theorem holds when the correlations are written in terms of linear combinations of the fields $\hhat{B}^\lambda_q$ we used in the previous section, see Eq.~(\ref{eq:PEB_main}). For this reason, following \cite{Saptsov}, see also Appendix \ref{app:Wick}, it is then possible to apply  Wick's theorem to write
\begin{equation}
\label{eq:Wicks_theorem}
\begin{array}{lll}
{C}^{\lambda_n\cdots \lambda_1}_{q_n\cdots q_1}&=&\displaystyle\sum_{c\in \bar{C}_n}(-1)^{\#c}\prod_{i,j\in c}{C}^{\lambda_{i},\lambda_{j}}_{q_{i},q_{j}}\;\;,
\end{array}
\end{equation}
in terms of two-point correlation functions which, using, Eq.~(\ref{eq:correlation}), read
\begin{equation}
\label{eq:two_point_correlations}
{C}^{\lambda_{i},\lambda_{j}}_{q_{i},q_{j}}=\text{Tr}_E\left[\hhat{T}_E\hhat{B}^{ \lambda_{i}}_{q_{i}}(t_{i})\hhat{B}^{ \lambda_{j}}_{q_{j}}(t_{j})\rho^\text{eq}_E\right]\;\;.
\end{equation}
 Here, each full-contraction $c\in \bar{C}_n$ is one of the possible sets of ordered pairs (or just, contractions) $(i,j)$, $i<j$ over the set $\bar{\mathbb{N}}_n=\{n,\cdots,1\}$. We further denote by $\#c$ the parity of the full-contraction $c$, i.e., the parity of the permutation needed to order the set $\bar{\mathbb{N}}_n$, such that all pairs in $c$ are adjacent.

\subsection{Influence superoperator}
\label{sec:Influence}
We now have all the tools needed to derive the formal expression of an influence superoperator which generates the reduced dynamics of the system.

In fact, using Wick's theorem, Eq.~(\ref{eq:Wicks_theorem}), in the expression for the reduced density matrix in Eq.~(\ref{eq:reduced_density_0}) we explicitly obtain the following expression for the reduced density matrix
\begin{equation}
\label{eq:temp_reduced_main}
\begin{array}{lll}
\rho_S&=&\displaystyle\sum_{n=0}^\infty\frac{(-1)^n}{n!}\int_0^t \left(\prod_{i=1}^n d t_i\right)\sum_{q,\lambda}\sum_{c\in \bar{C}_n}(-1)^{\#c}\\
&&\displaystyle\left(\prod_{(i,j)\in c}C^{\lambda_i,\lambda_j}_{q_i,q_j}\right)\hhat{T}_S\left[\hhat{S}_{q_n}^{\bar{\lambda}_n}\cdots\hhat{S}_{q_1}^{\bar{\lambda}_1}\right]{\rho}_S(0),
\end{array}
\end{equation}
where $\sum_{q,\lambda}\equiv\sum_{q_1,\lambda_1\cdots q_n,\lambda_n}$.
Our goal is now to formally re-sum this expression. To make progress, we  recall the meaning of the factor $(-1)^{\# c}$. This sign depends on the parity of the permutation needed to bring the set $\bar{\mathbb{N}}_n=\{n,\cdots,1\}$ into one in which all the pairs $(i,j)\in c$ are adjacent. Quite conveniently, this is exactly the same sign acquired when re-ordering the sequence of operators $\hhat{S}_{q_n}^{\bar{\lambda}_n}\cdots\hhat{S}_{q_1}^{\bar{\lambda}_1}$, such that all the pairs $\hhat{S}_{q_i}^{\bar{\lambda}_i}\hhat{S}_{q_j}^{\bar{\lambda}_j}$ with $(i,j)\in c$ are adjacent. The origin of this latter extra-minus sign lies in the Fermionic nature of the time ordering $\hhat{T}_S$, justifying the choice made in section \ref{sec:Dyson}. This means that we can write, see Appendix \ref{sec:func_sup_app},
\begin{equation}
\label{eq:F_first}
\begin{array}{lll}
\rho_S(t)
&=&\displaystyle \sum_{n=0}^\infty\frac{(-1)^{2n}}{(2n)!}\sum_{c\in C_{2n}}\prod_{(i,j)\in c}2\hhat{T}_S \hhat{\mathcal{F}}(t){\rho}_S(0)\;\;,
\end{array}
\end{equation}
where we also used the fact that correlations are non-zero only for even $n$. The previous expression is written in terms of the influence superoperator
\begin{equation}
\label{eq:F}
\displaystyle\hhat{\mathcal{F}}(t)=\int_0^t dt_{2}\int_0^{t_2} dt_{1} \hhat{W}(t_2,t_1)\;\;,
\end{equation}
in which we enforced partial time-ordering by constraining the integration bounds which gives rise to the factor $2$ in Eq.~(\ref{eq:F_first}). We also defined, see Appendix \ref{sec:W_app},
\begin{equation}
\label{eq:KernelW}
\begin{array}{lll}
\hhat{W}(t_2,t_1)&=&\displaystyle\sum_{q_1,q_2,\lambda_1,\lambda_2} C^{\lambda_2,\lambda_1}_{q_2,q_1}\hhat{S}_{q_2}^{\prime\bar{\lambda}_2}(t_2)\hhat{S}_{q_1}^{\prime\bar{\lambda}_1}(t_1)\\
&=&\displaystyle\sum_{\sigma=\pm}\hhat{A}^\sigma(t_2) \hhat{B}^\sigma(t_2,t_1)\;\;.
\end{array}
\end{equation}
Here, the superoperators $\hhat{A}^\sigma$ and $\hhat{B}^\sigma$ are defined as 
\begin{equation}
\label{eq:A_and_B}
\begin{array}{lll}
\hhat{A}^\sigma(t)[\cdot]&=&\displaystyle{s}^{\bar{\sigma}}(t)[\cdot]-\hhat{P}_S[\cdot{s}^{\bar{\sigma}}(t)]\\
\hhat{B}^{\sigma}(t_2,t_1)[\cdot]&=&-C^{\sigma}{s}^\sigma(t_1)[\cdot]-\bar{C}^{\bar{\sigma}}\hhat{P}_S[\cdot{s}^\sigma(t_1)],
\end{array}
\end{equation}
with $\bar{\sigma}=-\sigma$ and where $C^\sigma\equiv C^\sigma(t_2,t_1)$ with
\begin{equation}
\label{eq:correlations_main}
\begin{array}{lll}
C^{\sigma=1}(t_2,t_1)&=&\text{Tr}_E\left[B^\dagger(t_2)B(t_1)\rho_E(0)\right]\\
C^{\sigma=-1}(t_2,t_1)&=&\text{Tr}_E\left[B(t_2)B^\dagger(t_1)\rho_E(0)\right]\;\;.
\end{array}
\end{equation}
We now observe that in Eq.~(\ref{eq:F_first}) there is no actual dependence on the contraction $c$ (in $\hhat{\mathcal{F}}(t) $, all indexes are contracted and all times are integrated over). In this way the product over the pairs $(i,j)\in c$ effectively simply amounts in taking the $n$th power of the influence superoperator. For the same reason, the sum over $c\in C_{2n}$ effectively amounts in just counting the number of contractions in a list of $2n$ elements, which is $(2n-1)!!$. With this in mind, we can write
\begin{equation}
\label{eq:main_result}
\begin{array}{lll}
\displaystyle\rho_S(t)&=&\displaystyle\sum_{n=0}^\infty\frac{(-1)^{2n}(2n-1)!!}{(2n)!} 2^n\hhat{T}_S\hhat{\mathcal{F}}(t)^n{\rho}_S(0)\\
&=&\displaystyle\hhat{T}_Se^{\hhat{\mathcal{F}}(t)}{\rho}_S(0)\;\;,
\end{array}
\end{equation}
where we used the identity $(2n-1)!!/(2n)!=1/(2^n n!)$, see Appendix \ref{app:factorial}.
\emph{The formal expression in Eq.~(\ref{eq:main_result}) is the main result of this article} and, for this reason, we highlight its explicit form as 
\begin{equation}
\label{eq:F_widetext}
\displaystyle\rho_S(t)=
\displaystyle\sum_{p=\pm}\hhat{T}_S\exp\left\{\int_0^t dt_{2}\int_0^{t_2} dt_{1} \hhat{W}_p(t_2,t_1)[\cdot]\right\}\rho^p_S(0)\;\;,
\end{equation}
with $\rho^\pm_S(0)=\rho^{\text{e}/\text{o}}_S(0)$, and where (see Appendix \ref{app:W_explicit})
\begin{equation}
\label{eq:W_widetext}
    \begin{array}{lll}
    \hhat{W}_\pm(t_2,t_1)[\cdot]&=&\displaystyle -\sum_{\sigma=\pm}C^{\sigma}(t_2,t_1)[s^{\bar{\sigma}}(t_2),s^\sigma(t_1)\cdot]_{\mp}\\
    &&\displaystyle -\sum_{\sigma=\pm}C^{\sigma}(t_2,t_1)[s^{\bar{\sigma}}(t_1),s^\sigma(t_2)\cdot]_{\mp},
    \end{array}
\end{equation}
in which $[\cdot,\cdot]_\pm$ denotes a commutator/anticommutator.
This equation describes \emph{the reduced dynamics of the system in terms of the exponential of an influence superoperator and it can be applied to both even and odd parity sectors}. We note that, even when restricting to the physical even-parity sector, it is in general not possible to use parity symmetry to further simplify the final expression. To better analyze this point, we can consider the application of the operator $\hhat{W}$ to an even state. In this case, the $\hhat{P}_S$ appearing in the definition for $\hhat{B}^\sigma$ ($\hhat{A}^\sigma$) can be effectively replaced by $-1$ ($+1$). However, this is not the case when $\hhat{W}$ appears in Eq.~(\ref{eq:main_result}), i.e., in the expression for the reduced dynamics. In fact, in this case, the time-ordering might end up introducing further superoperators in between any of the $\hhat{A}^\sigma$ and $\hhat{B}^\sigma$, thereby making the alleged simplifications simply not correct (unless the superoperators in $\hhat{W}$ are evaluated at the same point in time, as in the Markovian regime).\\

Given the generality of Eq.~(\ref{eq:main_result}), it is opportune to show that we can recover well-known results in some specific limits. In the next section, we show that Eq.~(\ref{eq:main_result}) \emph{leads to a generalized-Lindblad Master equation in the Markovian regime and that it is a sufficient condition to derive a generalized version of the Hierarchical Equations of Motion.} The mentioned generalization consists in the possibility to apply the formalism to initial states with arbitrary parity symmetry and it recovers the usual Lindblad and HEOM form when restricted to the even-parity sector.

\section{Applications}
\label{sec:Applications}
Despite its innocent appearance, the exponentiation of the Fermionic influence superoperator in Eq.~(\ref{eq:main_result}) is not easy to solve. One reason is the  presence of the time-ordering operator $\hhat{T}_S$ which prevents the direct computation of the  integral in the expression for $\hhat{\mathcal{F}}(t) $. In turn this makes Eq.~(\ref{eq:main_result}) a formal expression ultimately referring back to the reduced Dyson series. 

In this section we analyze two different ways in which this problem can be approached. One, is to operate in a Markovian regime in which the action of the time-ordering is trivial, allowing to derive a master equation in Lindblad form. In more general regimes, it is instead possible to iteratively postpone the evaluation of the time-ordering leading to the Hierarchical Equations of Motion.
\subsection{Markovian regime}
\label{sec:Markovian}
The formal expression in Eq.~(\ref{eq:main_result}) describes \emph{all the effects of the environment on the system.} Among them, is an effective memory emerging when the correlation functions in Eq.~(\ref{eq:correlations_main}) are non-trivial for $t_2\neq t_1$. Because of these memory effects, the time-ordering applied to the terms $\hhat{\mathcal{F}}^n(t)$ in the reduced Dyson series might not leave the two superoperators appearing in each $\hhat{\mathcal{F}}(t)$ adjacent to each other thereby preventing the direct computation of the integrals in the influence superoperator.
The opposite regime is when $C^\sigma(t_2,t_1)\propto\delta(t_2-t_1)$, i.e., when we can neglect these memory effects. To better describe this Markovian case, we first introduce the spectral density
\begin{equation}
J(\omega)=\pi\sum_k g_k^2\delta(\omega-\omega_k)\;\;,
\end{equation}
which characterizes the strength of the system-environment interaction in the continuum limit, and in terms of which the correlations take the form (see Eq.~(\ref{eq:temp_corr_app}) in Appendix \ref{app:corr})
\begin{equation}
\label{eq:temp_corr}
\begin{array}{lll}
C^{\sigma}(t_2,t_1)=\displaystyle\int_{-\infty}^\infty \frac{d\omega}{\pi} J(\omega) e^{i\sigma\omega(t_2-t_1)}n^\sigma(\omega)\;\;,
\end{array}
\end{equation}
where $n^\sigma(\omega)= {[1-\sigma+2\sigma n^\text{eq}(\omega)]}/{2}$ in terms of the equilibrium Fermi-Dirac distribution $n^\text{eq}(\omega)$, see Eq.~(\ref{eq:FermiDirac}). From this expression, we see that a sufficient condition to define a Markovian regime is to have both $J(\omega)$ and $n^\text{eq}(\omega)$ constant in frequency, i.e., 
\begin{equation}
\label{eq:Markov_1}
\begin{array}{lll}
J(\omega)&=&\Gamma\\
n^\text{eq}(\omega)&=&n_0\;\;,
\end{array}
\end{equation}
where $\Gamma$ is a constant decay rate and $0\leq n_0\leq 1$. The assumption of a constant spectral density is usually named  first Markov approximation (\cite{Gardiner}, pag.~160). 

On the other hand, the assumption of a constant Fermi-Dirac distribution implies, using Eq.~(\ref{eq:FermiDirac}) that an environmental Fermion with energy $\omega$ must be in an initial state with an energy-dependent temperature $\beta(\omega) = {\log(1/n_0-1)}/{\omega}$. This explicitly shows how such a condition is not compatible with a true thermal equilibrium (except in special limiting cases such as for quantum transport at infinite bias, see \cite{PhysRevB.53.1050,PhysRevB.53.15932}). For this reason, the Markov regime defined here is an idealization in which the  environmental degrees of freedom act as an effective quantum white noise (see \cite{Gardiner} pag.~164). 

Using Eq.~(\ref{eq:Markov_1}) into Eq.~(\ref{eq:temp_corr}), we can write
\begin{equation}
\label{eq:corr_delta}
C^\sigma(t_2,t_1)= \Gamma (1-\sigma +2\sigma n_0)\delta(t_2-t_1)\;\;,
\end{equation}
where we used the exponential representation of the Dirac delta which can be found, for example, in \cite{Gardiner}, Eq.~(5.3.11). As mentioned, this expression shows the absence of memory effects, hence explicitly representing the Markovian regime. 

The delta-correlated environment considered in this section allows drastic simplifications in Eq.~(\ref{eq:main_result}). This is mainly due to the fact that all superoperators in Eq.~(\ref{eq:KernelW}) are evaluated at the same point in time, leading to a simpler time-ordering action. As it can be explicitly seen in Appendix \ref{app:Markovian},  using Eq.~(\ref{eq:corr_delta}) into Eq.~(\ref{eq:main_result}) leads to the following master equation in a generalized Lindblad form
\begin{equation}
\label{eq:Lindblad}
\begin{array}{lll}
\dot{\rho}_S
&=&-i[H_S,\rho_S]\\
&&+\displaystyle\Gamma\sum_{r=\pm} \{(1-n_0)D^r[\rho^r_S]+ n_0 D^r_{s^\dagger}[\rho^r_S]\}\;\;,
\end{array}
\end{equation}
where we omitted the time dependence for clarity. Here, with an abuse of notation, the density matrix refers to the Shr\"odinger picture and we defined $\rho^{r}=\delta_{r,1}\rho^\text{e}_S+\delta_{r,-1}\rho^\text{o}_S$ and $D^r[\cdot]=2 r O[\cdot]O^\dagger-O^\dagger O[\cdot]-[\cdot]O^\dagger O$, for a generic  operator $O$. \emph{The generalization of this equation with respect to the more commonly used Lindblad equation lies in the presence of an extra minus sign in the jump-term present in the dissipator in the odd-parity sector, consistent with  Eq.~(13) in \cite{Schwarz}.}
\subsection{Hierarchical equations of motion}
\label{sec:HEOM}
In this section we apply Eq.~(\ref{eq:F}) to derive a generalized version of the HEOM which can be applied to initial states with arbitrary parity symmetry. When applied to density matrices with even parity, this recovers the HEOM in their usual form.

The HEOM~\cite{Tanimura_1,Tanimura_2,Tanimura_3,Yan_5,Hartle2013,Schinabeck} are iterative equations which are based on the following ansatz for the correlation functions~\cite{Chen2015,Fruchtman}
\begin{equation}
\label{eq:ansats_correlations}
C^\sigma(t_2,t_1)=\sum_m a^\sigma_m e^{-b^\sigma_m (t_2-t_1)}\;\;.
\end{equation}
where $a^\sigma_m,b^\sigma_m\in \mathbb{C}$.
In the continuum limit, the previous expression has no loss of generality making the HEOM a formally exact method.

Nevertheless, for practical applications, the number of non-trivial exponents in this expression needs to be truncated. This  leads to an approximation to the full functional form of the correlation which can be non-negligible for heavily structured environments or at zero temperature due to the number of Matsubara frequencies approaching a continuum, see \cite{Strumpfer,Tang,Moix,Lambert} (although other, possibly more optimized, decompositions are possible \cite{Ding}).

By using the ansatz above, the influence superoperator in Eq.~(\ref{eq:F}) takes the form, see Appendix \ref{sec:expression_influence_HEOM_app},
\begin{equation}
\label{eq:F_with_exp}
\begin{array}{lll}
\hhat{\mathcal{F}}(t) &=& \displaystyle\int_0^t d t_2\int_0^{t_2} dt_1\sum_{j}  \hhat{A}^j(t_2) e^{-b_j(t_2-t_1)}\hhat{\mathcal{B}}_j(t_1),
\end{array}
\end{equation}
in terms of the multi-index $j=(m,\sigma)$ and where $\hhat{A}^j\equiv \hhat{A}^\sigma$ as defined in Eq.~(\ref{eq:A_and_B}), $b_j\equiv b^\sigma_m$, and
\begin{equation}
\label{eq:AB}
\hhat{\mathcal{B}}_j(t)[\cdot]\equiv\hhat{\mathcal{B}}^{\sigma}_m(t)[\cdot]=\displaystyle -\left(a_m^{\sigma}{s}^\sigma(t)[\cdot]+\bar{a}_m^{\bar{\sigma}}P_S[[\cdot]{s}^\sigma(t)]\right),
\end{equation}
with $\bar{\sigma}=-\sigma$. Using this expression into Eq.~(\ref{eq:F}) and taking a time derivative \cite{Ma_2012}, see Eq.~(\ref{eq:der_2_app}), we arrive at the following self-referential  equation of motion
\begin{equation}
\label{eq:motion}
\displaystyle\dot{\rho}_S(t)=\displaystyle\sum_j \hhat{A}^j(t) \hhat{T}_S  \hhat{\Theta}_j (t)\rho_S(t)\;\;,
\end{equation}
with $\hhat{\Theta}_j(t)=\displaystyle\int_0^{t} d\tau e^{-b_j(t-\tau)}\hhat{\mathcal{B}}_j(\tau)$, which satisfies the key property
\begin{equation}
\label{eq:HEOM_ansats}
\displaystyle\frac{d}{dt}{\hhat{\Theta}}_j(t)=\displaystyle -b_j \hhat{\Theta}_j(t)+ \hhat{\mathcal{B}}_j(t)\;\;.
\end{equation}
The self-referential nature of Eq.~(\ref{eq:motion}) can be formally lifted by writing
\begin{equation}
\label{eq:HEOM_0}
\displaystyle\dot{\rho}_S(t)=\alpha^{-1}\displaystyle\sum_{j} \hhat{A}^{j}(t) \rho_{j}(t)\;\;,
\end{equation}
in term of the auxiliary density matrix 
\begin{equation}
\label{eq:HEOM_00}
\rho_{j}(t)=\alpha \hhat{T}_S  \hhat{\Theta}_{j} (t)\rho_S(t)\;\;,
\end{equation}
where we introduced the parameter $\alpha\in\mathbb{C}$ upon which the system's dynamics does not depend. 
In fact, the auxiliary density matrices in the HEOM are unphysical degrees of freedom which can be re-scaled (see also \cite{Qiang_Shi}).

An interesting feature of  Eq.~(\ref{eq:HEOM_0})  is that it involves the time-ordering through the definition of the auxiliary density matrix $\rho_j$, leading to the possibility of \emph{iteratively postponing} its challenging evaluation. In fact, we can define the $n$th auxiliary density matrix as
\begin{equation}
\label{eq:rho_n}
\rho^{(n)}_{j_n\cdots j_1}(t)=\alpha^n\hhat{T}_S \hhat{\Theta}_{j_n}(t)\cdots \hhat{\Theta}_{j_1}(t)\rho_S(t)\;\;,
\end{equation}
so that, in this notation, $\rho_S(t)=\rho^{(0)}(t)$.
Its derivative can be computed by using Eq.~(\ref{eq:HEOM_ansats}) and Eq.~(\ref{eq:HEOM_0}) and leads to the following generalized version of the HEOM
\begin{equation}
\label{eq:generalized_HEOM}
\begin{array}{lll}
\displaystyle\dot{\rho}^{(n)}_{j_n\cdots j_1}\!\!\!\!&=&\!\!\!\!\displaystyle-\sum_{k=1}^nb_{j_k}\rho^{(n)}_{j_n\cdots j_1}+\alpha^{-1}\sum_{j_{n+1}}\hhat{A}^{\sigma_{n+1}}\rho^{(n+1)}_{j_{n+1}\cdots j_1}\\
&&+\alpha\displaystyle\sum_{k=1}^n (-1)^{n-k}\hhat{\mathcal{B}}_{j_k}\rho^{(n-1)}_{j_n\cdots j_{k+1}j_{k-1}\cdots j_1}\;\;,
\end{array}
\end{equation}
\emph{which is valid for both even- and odd-parity initial conditions}, see Eq.~(\ref{eq:rho_n_app}). 
However, if we now assume $\rho^{(0)}(t)$ to have even parity, then the parity superoperators inside the definitions in Eq.~(\ref{eq:AB}) translate into signs dependent on the iteration index $n$.  By moving to the Shr$\ddot{o}$dinger picture and making the choice $\alpha=i$, this leads to
\begin{equation}
\begin{array}{lll}
\displaystyle\dot{\rho}^{(n)}_{j_n\cdots j_1}
&=&\displaystyle\left(\hhat{\mathcal{L}}-\sum_{k=1}^n b_{j_k}\right)\rho^{(n)}_{j_n\cdots j_1}\\
&&-\displaystyle i\sum_{j_{n+1}}\hhat{\mathcal{A}}_n^{\sigma_{n+1}}\rho^{(n+1)}_{j_{n+1}\cdots j_1}\displaystyle \\
&&-i\displaystyle\sum_{k=1}^n (-1)^{n-k}\hhat{\mathcal{C}}_n^{j_k}\rho^{(n-1)}_{j_n\cdots j_{k+1}j_{k-1}\cdots j_1}\;\;,
\end{array}
\end{equation}
see Eq.~(\ref{eq:rho_n_simplified_app}). Here, $\hhat{\mathcal{L}}=-i[H_S,\cdot]$, and
\begin{equation}
\begin{array}{lll}
\hhat{\mathcal{A}}_n^\sigma[\cdot]&=&\displaystyle{s}^{\bar{\sigma}}[\cdot]+(-1)^n[\cdot]{s}^{\bar{\sigma}}\\
\hhat{\mathcal{C}}_n^j[\cdot]&=&\displaystyle a_n^{\sigma}{s}^\sigma[\cdot]-(-1)^n\bar{a}_n^{\bar{\sigma}}[\cdot]{s}^\sigma\;\;.
\end{array}
\end{equation}
The equation above is one of the standard forms for the \emph{hierarchical equations of motion}, see, for example, Eq.~(38) in \cite{Lambert_Bofin}.
\subsection{Computing system correlation functions}
\label{sec:correlations}
The influence superoperator defined in the previous section allows one to generate the reduced system dynamics without restrictions on the parity of the initial state. This feature can be convenient when computing correlation functions of the kind
\begin{equation}
C_{XY}(t)=\text{Tr}_{SE}[X_S(t) Y_S\rho(0)]\;\;,
\end{equation}
where $X_S$ and $Y_S$ are system operators. Here, the time dependence is intended in the full system+environment space, i.e., $X_S(t)=U^\dagger(t)X_SU(t)$, where $U=\exp[-iHt]$ in terms of the Hamiltonian in Eq.~(\ref{eq:H_full}). Consequently, we can write
\begin{equation}
\label{eq:corr:special}
C_{XY}(t)=\text{Tr}_{SE}[X_S U(t) Y_S\rho(0)U^\dagger(t)]\;\;.
\end{equation}
Supposing Eq.~(\ref{eq:product}) and using Eq.~(\ref{eq:density_matrix_implicit}) and Eq.~(\ref{eq:main_result}), we can compute this quantity as
\begin{equation}
\label{eq:corr_system}
C_{XY}(t)=\text{Tr}_{S}[X_S \rho^{\prime}_S(t)]=\text{Tr}_{S}[X_S\hhat{T}_S e^{\hhat{\mathcal{F}}(t)}\rho'_S(0)]
\;\;,
\end{equation}
where the initial condition is $\rho_S'(0)=Y_S\rho_S(0)$. We point out that, for a physical initial state $\rho_S(0)$ with even parity, $\rho_S'(0)$ has the same parity as $Y_S$, which might be odd.
However, since the results presented in the previous sections apply to initial states with arbitrary parity symmetry, Eq.~(\ref{eq:corr_system}) follows directly.

Remarkably, it is also possible to compute thermal correlations of the kind
\begin{equation}
\label{eq:corr_system_gen}
\begin{array}{lll}
C^{\text{th}}_{XY}(t)&=&\text{Tr}_{SE}[X_S(t_2) Y_S(t_1)\rho^{\text{th}}]\\
&=&\text{Tr}_{SE}[X_S(0) U(t_2-t_1) Y_S(0)\rho^{\text{th}}U^\dagger(t_2-t_1)]\;\;.
\end{array}
\end{equation}
where $t=t_2-t_1$ and where $\rho^{\text{th}}\propto \exp(-\beta H)$ is the combined system-environment thermal-equilibrium state. One possible way to proceed is to suppose this state to be separable (akin to the hypothesis of the quantum regression theorem \cite{PhysRevA.90.022110}) thereby reducing to solve an expression equivalent to Eq.~(\ref{eq:corr:special}). However, the thermal-equilibrium state usually includes entanglement between the system and the environment, i.e., it is not separable, i.e., Eq.~(\ref{eq:product}) does not hold. This prevents us from using the results given in section \ref{sec:InfluenceSuperoperator} directly. 

To make progress, we can use the following idea \cite{PhysRevLett.104.250401,doi:10.1143/JPSJ.81.063301,Tanimura_2014,Tanimura_2020,KondoPRL} instead. We suppose that at a time $-T<0$ the system+environment is in a separable non-equilibrium state $\rho(-T)=\rho^\text{eq}\rho_S(-T)$. We then assume the existence of a thermal equilibration time $T^{\text{th}}\ll T$, so that the equality $\rho^{\text{th}}=U(T)\rho(-T)U^\dagger(T)$ holds. Using this identity in Eq.~(\ref{eq:corr_system_gen}) we can write
\begin{equation}
\label{eq:corr_system_gen_2}
\begin{array}{lll}
C^{\text{th}}_{XY}(t)&=&\text{Tr}_{SE}[X_S(0) \rho^Y(t)]\;\;,
\end{array}
\end{equation}
where $t>0$ and 
\begin{equation}
\label{eq:def_rhoY}
\rho^Y(t)=U(t) Y_S(0) U(T)\rho(-T)U^\dagger(T)U^\dagger(t)\;\;.
\end{equation}
In order to compute this quantity, it is possible to generalize the reasoning developed in section \ref{sec:InfluenceSuperoperator} and section \ref{sec:HEOM}  to find (see Appendix \ref{app:computing_correlations}) that the formal time-derivative of the density matrix has the same form as Eq.~(\ref{eq:motion}), i.e.
\begin{equation}
\label{eq:HEOM_Y}
\displaystyle\dot{\rho}^Y_S(t)=\displaystyle\sum_j \hhat{A}^j(t) \hhat{T}_S  \hhat{\Theta}_j (t)\rho^Y_S(t)\;\;,
\end{equation}
but with a different initial condition given by
\begin{equation}
\label{eq:new_initial_condition}
\begin{array}{lll}
\rho_S^Y(0)
&=&Y_S\hhat{T}_S e^{\hhat{\mathcal{F}}_{T}(0)}\rho_S^Y(-T)\;\;.
\end{array}
\end{equation}
This result offers the following strategy to compute the correlations $C^{\text{th}}(t)$.
\begin{itemize}
\item[i.] Solve the HEOM from time $-T$ (with initial condition $\rho_S(-T)$) to time $0$  to obtain a collection of auxiliary density matrices $\rho^{(n)}_{j_n\cdots j_1}(0)$. For $T$ much longer than the thermal equilibration time, $\rho_S(0)=\rho^{(0)}(0)=\rho^\text{th}$ represents the thermal state of the system+environment.
\item[ii.] The HEOM are local in time, implying that the matrices $\rho^{(n)}_{j_n\cdots j_1}(0)$ must contain all the  information about the dynamics from time $-T$ to $0$ (needed to further propagate the state further in time). This information is equivalent to that contained in the formal expression $\hhat{T}_Se^{\hhat{\mathcal{F}}_{T}(0)}\rho_S^Y(-T)$. From another point of view, these matrices also represent the system-environment entanglement \cite{PhysRevLett.104.250401,doi:10.1143/JPSJ.81.063301,Tanimura_2014,Tanimura_2020,KondoPRL}.
\item[iii.] Using Eq.~(\ref{eq:rho_n}), the initial condition $\rho_S^Y(0)$ in Eq.~(\ref{eq:new_initial_condition}) can be implemented by multiplying each auxiliary density matrix by $Y_S$, i.e., $\rho^{(n)}_{j_n\cdots j_1}(0)\rightarrow Y_S\rho^{(n)}_{j_n\cdots j_1}(0)$.
\item[iv.] As implied by Eq.~(\ref{eq:HEOM_Y}), the density matrix $\rho_S^Y(t)$ at time $t$ can be computed by solving the same HEOM as before with initial condition given by the auxiliary density matrices defined in (iii).
\item[v.] By using the matrix $\rho_S^Y(t)$ computed in (iv), the thermal correlation in Eq.~(\ref{eq:corr_system_gen_2}) can be computed as $C_{XY}^\text{th}(t)=\text{Tr}_S[X_S(0)\rho_S^Y(t)]$, by definition of partial trace.
\end{itemize}

In conclusion, we showed that the possibility to apply the influence superoperator and the HEOM to initial states with arbitrary symmetry can be used to compute thermal correlation functions which characterize the equilibrium properties of the system+environment.
\section{Conclusions}
We presented \emph{a canonical derivation of an influence superoperator which encodes the full dynamical effects of a Fermionic environment linearly coupled to a Fermionic quantum system}. Such a superoperator can be used to generate the reduced system dynamics without restrictions in terms of the parity of the initial state. In a Markovian regime where the environment acts as an ideal quantum white noise, the formalism becomes equivalent to a generalized Lindblad Master equation. In general, the expression for the Fermionic influence superoperator represents a sufficient condition to deduce a version of the Hierarchical Equation of Motion which can be applied to states with arbitrary parity-symmetry, which is vital for the evaluating impurity correlation functions and spectra \cite{KondoPRL}.

\acknowledgements
We acknowledge fruitful discussions with Xiao Zheng, Ken Funo and Paul Menczel. 
M.C.~acknowledges support from NSAF No. U1930403. N.L.~acknowledges partial support from JST PRESTO through Grant No.~JPMJPR18GC. F.N.~is supported in part by: Nippon Telegraph and Telephone Corporation (NTT) Research, Japan Science and Technology Agency (JST) (via the Q-LEAP program, Moonshot R\&D Grant No.~JPMJMS2061, and the CREST Grant No.~JPMJCR1676), Japan Society for the Promotion of Science (JSPS) (via the KAKENHI Grant No.~JP20H00134 and the JSPS-RFBR Grant No.~JPJSBP120194828), Army Research Office (ARO) (Grant No.~W911NF-18-1-0358), Asian Office of Aerospace Research and Development (AOARD) (via Grant No.~FA2386-20-1-4069). F.N.~and N.L.~acknowledge the Foundational Questions Institute Fund (FQXi) via Grant No.~FQXi-IAF19-06. YNC acknowledges the support of the Ministry of Science and Technology, Taiwan (MOST Grants No. 110-2123-M-006-001), and the U.S. Army Research Office (ARO Grant No. W911NF-19-1-0081).

\newpage
\appendix

\section{Table of Symbols}
In the following table we review the meaning of the most relevant symbols used throughout the article.
\renewcommand{\arraystretch}{1.9}{
\begin{table}[h!]
\begin{tabularx}{.5\textwidth}{|c|X|}
\hline
\textbf{Symbol}          & \thead{\textbf{Description}} \\
\hline
$\mathcal{S}$& System/Environment physical domain: $\mathcal{S}=S/E$.\\
\hline
$O_{\mathcal{S}}$& Generic operator in the domain $\mathcal{S}$.\\
\hline
$\hat{O}_S$&System operator equivalent to $O_S$ but commuted to the right of all environmental operators.\\
\hline
$\hhat{O}_{\mathcal{S}}$&Generic  superoperator in the domain ${\mathcal{S}}$.\\
\hline
$P_{\mathcal{S}}$&Parity operator in the domain ${\mathcal{S}}$: $P_{\mathcal{S}}=\prod_{k\in \mathcal{S}} \exp[i\pi f^\dagger_k f_k]$, where $f_k$ destroys a Fermion in the domain $\mathcal{S}$.\\
\hline
$\hhat{P}_\mathcal{S}$&Parity superoperator: $\hhat{P}_\mathcal{S}[\cdot]=P_\mathcal{S}[\cdot]P_\mathcal{S}$\\
\hline
$\hhat{P}^{e/o}_{\mathcal{S}}$& Projector in the even/odd parity sector: $\hhat{P}^\text{e/o}_{\mathcal{S}}[\cdot]=P^\text{e}_{\mathcal{S}}\cdot P^\text{e/o}_{\mathcal{S}}+P^\text{o/e}_{\mathcal{S}}\cdot P^\text{o}_{\mathcal{S}}$.\\
\hline
$O^{\text{e/o}}_{\mathcal{S}}$& Even/odd part of the operator $O_{\mathcal{S}}$: $O^{\text{e/o}}_{\mathcal{S}}=\hhat{P}^{e/o}_{\mathcal{S}} [O_{\mathcal{S}}]$. \\
\hline
$c^{\sigma}_k$& Annihilation/creation ($\sigma=\pm 1$) operator for the $k$th Fermion in the environment.\\
\hline
$\bar{\sigma}$& Opposite of $\sigma$: $\bar{\sigma}=-\sigma$.\\
\hline
$B^\sigma$& Environmental coupling operator: $B^\sigma=\sum_{k\in E} g_k c^\sigma_k$.\\
\hline
$s$& System coupling operator.\\
\hline
$\rho^\text{eq}$&Equilibrium state for the environment: $\rho^\text{eq}_E=\prod_{k\in E} [e^{-\beta (\omega_k-\mu_E) c^\dagger_k c_k}/(1+\exp[-\beta(\omega_k-\mu)])]$.\\
\hline
$C^{\sigma}(t_2,t_1)$& Correlation function: $C^{\sigma}(t_2,t_1)=\text{Tr}_E\left[B^\sigma(t_2)B^{\bar{\sigma}}(t_1)\rho^\text{eq}_E\right]$.\\
\hline
$\bar{C}^{\sigma}(t_2,t_1)$& Complex conjugate of the correlation function.\\
\hline
$\hhat{T}_{\mathcal{S}}$& Time-ordering superoperator in the domain $\mathcal{S}$.\\
\hline
$J(\omega)$& Spectral density: $J(\omega)=\pi\sum_{k\in E}g_k^2\delta(\omega-\omega_k)$.\\
\hline
$\beta,\mu$& Inverse temperature and chemical potential.\\
\hline
$n^\text{eq}(\omega)$& Fermi equilibrium distribution: $(\exp[\beta(\omega-\mu)]+1)^{-1}$.\\
\hline
$\hhat{B}^{\lambda}_{q}[\cdot]$&Environmental superoperator $\hhat{B}^{\lambda}_{q}[\cdot]=\delta_{q,1}B^\lambda[\cdot]+\delta_{q,-1}\hhat{P}_E[\cdot B^\lambda]$\\
\hline
$\hhat{S}^{\lambda}_{q}[\cdot]$&System superoperator $\hhat{S}^{\lambda}_{q}[\cdot]=\delta_{q,1}s^\lambda[\cdot]-\delta_{q,-1}\hhat{P}_S[\cdot s^\lambda]$\\
\hline
\end{tabularx}
\caption{List of symbols}
\end{table}
}

\section{Fermionic influence superoperator}
Here, we present the detailed reasoning and calculations supporting each subsection of Section \ref{sec:InfluenceSuperoperator}.
\subsection{A parity-friendly formalism}
We start by presenting more details on the definition of ``hat'' operators (see \cite{Schwarz}) and on the Fermionic partial trace.
\subsubsection{``Hat'' operators}
\label{sec:Parity_1}
Given a Fermionic system,  its Hilbert space $\mathcal{H}$ is naturally endowed with a $\mathbb{Z}_2$-graded structure $\mathcal{H}=\mathcal{H}^\text{e}\oplus\mathcal{H}^\text{o}$ due to the action of the parity operator see, for example,  \cite{Fidkowski_1,Bultinck_1,Bultinck_2,Ryu_1}. Here, vectors in $\mathcal{H}^\text{e/o}$ are homogeneous, i.e., they have well defined ($0/1$) parity. This structure is also inherited by 
operators  $O:\mathcal{H}\rightarrow \mathcal{H}$ which can also be decomposed into their even and odd part, i.e., $O=O^\text{e}+O^\text{o}$.  

When we compose two or more Fermionic systems (having Hilbert spaces $\mathcal{H}_1$ and $\mathcal{H}_2$) the physical anticommutation rules require a compatibility between the tensor-product  and the graded structure. To see this, it is possible to consider a graded tensor product $\otimes_g$ which, within the operator algebra, is characterized by the following identity
\begin{equation}
(O^x_1\otimes_g O^y_2)(O^{\prime x'}_1\otimes_gO^{\prime y'}_2)\!=\!(-1)^{x'y}O^x_1 O^{\prime x}_1\otimes_g O^y_2O^{\prime y'}_2
\end{equation}
where $x,x',y,y'=\text{e/o}$, and which characterizes the physical Fermionic statistics under particle-exchange.  This equation is equivalent to the following, perhaps more evocative, definitions
\begin{equation}
\label{eq:ferm_Fid}
\begin{array}{lll}
O_1^xO_2^y&=&(O^x_1\otimes_g O^y_2)\\
O_2^yO_1^x&=&(-1)^{xy}(O^x_1\otimes_g O^y_2)\;\;,
\end{array}
\end{equation}
where $O^x\equiv O^x_1\otimes \mathbb{I}_2$ and $O^y\equiv \mathbb{I}_1\otimes O^y_2$. 
Using these equations, it is possible define creation/annihilation operators with proper Fermionic statistics so that the full Hilbert space can be constructed by acting on the composite vacuum $\ket{0}\equiv\ket{0}_1\otimes_g\ket{0}_2$.

In order to systematically deal with the signs appearing as a consequence of the graded structure, we follow \cite{Schwarz}. It is in fact possible to map the graded-tensor product between two Fermionic systems  (which, for us, are the Environment $E$ and the system $S$), into a non-graded tensor product $\otimes$ through the substitution
\begin{equation}
\label{eq:hat_graded_app}
O^x_1\otimes_g O^y_2\mapsto\delta_{y,0} O^x_1\otimes \hat{O}^y_2+\delta_{y,1}O^x_1P_1 \otimes \hat{O}^y_2\;\;,
\end{equation}
where $P_1$ is the parity operator in the space $\mathcal{H}_1$.  Identifying $1\mapsto E$, $2\mapsto S$ and for a generic operator $O_S$ with no given parity symmetry, Eq.~(\ref{eq:hat_graded_app}) leads to the more direct identification
\begin{equation}
\label{eq:hat_app}
O_S= \hat{O}^\text{e}_S+P_E \hat{O}_S^\text{o}\;\;,
\end{equation}
where, explicitly
\begin{equation}
P_E=\prod_{k\in E} \exp{[i\pi c^\dagger_{k}c_{k}]}
\end{equation} 
is the parity operator over the environment variables and where up-indexes $\text{e/o}$ labels the even and odd part, i.e., 
\begin{equation}
\begin{array}{lll}
{O}_S^\text{e}&=&P_S^\text{e}{O}_SP_S^\text{e}+P_S^\text{o}{O}_SP_S^\text{o}\\
{O}^\text{o}_S&=&P_S^\text{e}{O}_SP_S^\text{o}+P_S^\text{o}{O}_SP_S^\text{e}\;\;,
\end{array}
\end{equation}
where 
\begin{equation}
\begin{array}{lll}
P_S^\text{e}&=&(P_S+1)/2\\
P_S^\text{o}&=&(1-P_S)/2\;\;.
\end{array}
\end{equation}
Intuitively in its ``hat'' version, a system operator is to be placed on the right of any environmental operator. As a consequence, Eq.~(\ref{eq:ferm_Fid}) is replaced by the ``Bosonic''-like $O^x_E \hat{O}^y_S=\hat{O}^y_S O^x_E$. This notation is  extremely practical to use. In fact, after using it in the initial condition in Eq.~(\ref{eq:product}) and in the interaction Hamiltonian in Eq.~(\ref{eq:H_I}), it allows to treat the tensor structure between system and environment as if it was Bosonic, while still being assured that all Fermionic signs are correctly accounted for.
\subsubsection{Partial trace in Fermionic systems}
\label{sec:Parity_2}
Within the graded structure of the Environment+System Hilbert space, a basis of vectors can be written as
\begin{equation}
\ket{v_E,v_S}\equiv \left(\prod_{i\in v_E}c^\dagger_{i}\right)\left(\prod_{j\in v_S}c^\dagger_{j}\right)\ket{0}\;\;,
\end{equation}  
where $v_E$ ($v_S$) is the ordered set specifying which environmental (system) Fermions are present. We also explicitly define the duals as
\begin{eqnarray}{lll}
\bra{v_S,v_E}&\equiv&\displaystyle \ket{v_E,v_S}^\dagger\\
&=&\displaystyle
\bra{0}\left(\prod_{j\in \tilde{v}_S}c_{j}\right)\left(\prod_{i\in \tilde{v}_E}c_{i}\right)\;\;,
\end{eqnarray}  
where $\tilde{v}_{E/S}$ denotes the sets $v_{E/S}$ inverted in their ordering.
Here, $c^\dagger_{i/j}$ are creation operators for Fermions in the environment and system. Using these definitions, we can write the partial trace of an operator $O_{ES}$ over the environment as
\begin{equation}
\label{eq:p_trace}
\begin{array}{lll}
\text{Tr}_E O_{ES}&\equiv&\!\!\!\displaystyle\sum_{v_E,v_S,v'_S}\bra{v_S,v_E}O_{ES}\ket{v_E,v'_S}\ketbra{v_S}{v'_S}.
\end{array}
\end{equation}
We now use this definition to prove useful identities. First, we point out that, unfortunately, for Fermions, in general we are prevented from using the, otherwise, very convenient
\begin{equation}
\label{eq:question_mark}
\text{Tr}_E (O_E O_S)\overset{?}{=}\text{Tr}_E(O_E)O_S\;\;.
\end{equation}
 To see this explicitly, we can consider an environment (system) made out of a single Fermion $c$ ($d$). We can further consider $O_E\rightarrow\mathbb{I}_E$ and $O_S\rightarrow d^\dagger$. In this simple case, using  Eq.~(\ref{eq:p_trace}) we obtain
\begin{equation}
\begin{array}{lll}
\text{Tr}_E (\mathbb{I}_E d^\dagger)&=&\displaystyle\bra{1,0}d^\dagger\ket{0,0}\ketbra{1}{0}+\bra{1,1}d^\dagger\ket{1,0}\ketbra{1}{0}\\
&=&\displaystyle\bra{1,0}(d^\dagger+cd^\dagger c^\dagger)\ket{0,0}\cdot\ketbra{1}{0}\\
&=&0\\
&\neq&\text{Tr}_E(\mathbb{I}_E) d^\dagger\\
&=&2d^\dagger\;\;,
\end{array}
\end{equation}
which is enough to conclude that, in general,
\begin{equation}
\label{eq:tr_ineq}
\text{Tr}_E (O_E O_S)\neq\text{Tr}_E(O_E)O_S\;\;.
\end{equation}
At the same time, it is possible to prove that the analogous version with ``hat'' operators holds, i.e.,
\begin{equation}
\label{eq:tr_hat}
\text{Tr}_E (O_E \hat{O}_S)=\text{Tr}_E(O_E)\hat{O}_S\;\;.
\end{equation}
In fact, since the partial trace over $E$ must involve an even number of environmental operators in order to give a non-zero result, and using Eq.~(\ref{eq:p_trace}), we have
\begin{widetext}
 \begin{equation}
\begin{array}{lll}
\text{Tr}_E O_E\hat{O}_S&=&\text{Tr}_E O^\text{e}_E\hat{O}_S\\
&=&\displaystyle\sum_{v_E,v_S,v'_S}\bra{0}\left(\prod_{j\in \tilde{v}_E} c_{j}\right)\left(\prod_{i\in \tilde{v}_E}c_{i}\right)O^\text{e}_E \left(\prod_{i\in v_S}c^\dagger_{i}\right)\hat{O}_S\left(\prod_{j\in v'_S} c^\dagger_{j}\right)\ket{0}\cdot\ketbra{v_S}{v'_S}\\
&=&\displaystyle\sum_{v_E,v_S,v'_S}\bra{0}\left(\prod_{i\in \tilde{v}_E}c_{i}\right) O^\text{e}_E \left(\prod_{i\in v_E}c^\dagger_{i}\right)\left(\prod_{j\in \tilde{v}_S} c_{j}\right)\hat{O}_S\left(\prod_{j\in v'_S} c^\dagger_{j}\right)\ket{0}\cdot\ketbra{v_S}{v'_S}\;\;,
\end{array}
\end{equation}
\end{widetext}
where in the last step we observed that the number of environmental operators involved is even.
Each of the Fermions present in the matrix elements in the equation above have to be appear an even number of times in order for the result to be non-zero. As a consequence, inserting an identity in between the environment and the system operators is equivalent to introducing $\ketbra{0}{0}$. We then have
\begin{equation}
\label{eq:Tr}
\begin{array}{l}
\text{Tr}_E [O_E\hat{O}_S]=\displaystyle\sum_{v_E}\bra{0}\left(\prod_{i\in \tilde{v}_E}c_{i}\right)O^\text{e}_E \left(\prod_{i\in v_E}c^\dagger_{i}\right)\ket{0}\\
\displaystyle\phantom{=}\sum_{v_S,v'_S}\bra{0}\left(\prod_{j\in \tilde{v}_S} c_{j}\right)\hat{O}_S\left(\prod_{j\in v'_S} c^\dagger_{j}\right)\ket{0}\cdot\ketbra{v_S}{v'_S}\\
=\displaystyle\text{Tr}_E\left[O^\text{e}_E\right]\bra{0}\left(\prod_{j\in \tilde{v}_S} c_{j}\right)\hat{O}_S\left(\prod_{j\in v'_S} c^\dagger_{j}\right)\ket{0}\cdot\ketbra{v_S}{v'_S}\\
=\text{Tr}_E\left(O_E\right) \hat{O}_S\;\;,
\end{array}
\end{equation}
where in the last step we reintroduced the odd part to the operator $O_E$ (since it gives a zero contribution to the trace) thereby proving Eq.~(\ref{eq:tr_hat}).

While Eq.~(\ref{eq:tr_hat}) does generalize Eq.~(\ref{eq:question_mark}), only valid for Bosonic fields, it is not enough for our purposes and we need the further generalization
\begin{equation}
\label{eq:key_relation_app}
\begin{array}{l}
\text{Tr}_{ES}[A_S O_E \hat{O}_S]=\\
\text{Tr}_E[O_E] \text{Tr}_S[A_S\hat{O}^\text{e}_S]+\text{Tr}_E [P_E O_E] \text{Tr}_S[A_S\hat{O}^\text{o}_S],
\end{array}
\end{equation}
for all system operators $A_S$. This  can be proven directly as
\begin{equation}
\label{eq:proof_trace}
\begin{array}{l}
\text{Tr}_{ES}[A_S O_E \hat{O}_S]=\text{Tr}_{ES}[A_S O_E (\hat{O}^\text{e}_S+\hat{O}^\text{o}_S)]\\
=\text{Tr}_{ES}[A^\text{e}_S O_E \hat{O}^\text{e}_S]+\text{Tr}_{ES}[A^\text{o}_S O_E \hat{O}^\text{o}_S]\\
=\text{Tr}_{ES}[\hat{A}^\text{e}_S O_E \hat{O}^\text{e}_S]+\text{Tr}_{ES}[P_E\hat{A}^\text{o}_S O_E \hat{O}^\text{o}_S]\\
=\text{Tr}_{ES}[ O_E \hat{A}^\text{e}_S\hat{O}^\text{e}_S]+\text{Tr}_{ES}[P_E O_E \hat{A}^\text{o}_S\hat{O}^\text{o}_S]\\

=\text{Tr}_EO_E\text{Tr}_S \hat{A}_S^\text{e}\hat{O}^\text{e}_S+
\text{Tr}_EP_EO_E\text{Tr}_S \hat{A}_S^\text{o}\hat{O}^\text{o}_S\\

=\text{Tr}_EO_E\text{Tr}_S \hat{A}_S\hat{O}^\text{e}_S+
\text{Tr}_EP_EO_E\text{Tr}_S \hat{A}_S\hat{O}^\text{o}_S\\
=\text{Tr}_SA_S\{\text{Tr}_E[O_E] \hat{O}^\text{e}_S+\text{Tr}_E [P_E O_E] \hat{O}^\text{o}_S\}\;\;,
\end{array}
\end{equation}
where, explicitly, we observed that traces with an odd number of system-operators must be zero to justify the second and sixth equality. To justify the third and forth equalities we used the definition of hat-operator in Eq.~(\ref{eq:hat_app}) and its properties. We further used the identity in Eq.~(\ref{eq:tr_hat}) in the fifth equality and finished noticing that, once the environmental degrees of freedom have been traced out, hat-operators are equivalent to normal ones.
The identity in Eq.~(\ref{eq:key_relation_app}) has a key role in characterizing how to find an expression for the reduced density matrix which is capable of computing the correct expectation values. In section \ref{app:Dyson}, we explicitly see that the full density matrix can be written as a linear combination of terms taking the form $O_E \hat{O}_S$, see also the simplified version in Eq.~(\ref{eq:decomposition_0}), i.e., $\rho(t)=\sum_i\rho_E^i
\hat{\rho}_S^i$, which leads to
\begin{equation}
\label{eq:key_relation_app_2}
\begin{array}{l}
\text{Tr}_{ES}[A_S \rho(t)]=\displaystyle\sum_i \text{Tr}_{ES}[A_S\rho_E^i
\hat{\rho}_S^i]\\
=\displaystyle\sum_i\text{Tr}_SA_S\{\text{Tr}_E[\rho^i_E] \hat{\rho}^{i,\text{e}}_S+\text{Tr}_E [P_E \rho^i_E] \hat{\rho}^{i,\text{o}}_S\}.
\end{array}
\end{equation}
Using the fact that $A_S$ is a generic system operator, we can compare the previous equation to the defining property of the reduced density matrix in Eq.~(\ref{eq:density_matrix_implicit}), i.e., the ability to compute expectation values 
\begin{equation}
\text{Tr}_{ES}[A_S \rho(t)]\equiv\text{Tr}_SA_S\rho_S(t)\;\;,
\end{equation}
to derive the expression for the reduced density matrix
\begin{equation}
\label{eq:effective_1_app}
{\rho}_S(t)=\sum_i\text{Tr}_E[{\rho}^i_E] {\hat{{\rho}}}^{i,\text{e}}_S+\text{Tr}_E [P_E {\rho}^i_E] {\hat{{\rho}}}^{i,\text{o}}_S\;\;,
\end{equation}
i.e., Eq.~(\ref{eq:effective_1}) in the main text.
\subsection{Reduced Dyson series}
\label{app:Dyson}
The starting point of this section is the Dyson series for the environment+system in Eq.~(\ref{eq:series_0}), which reads
\begin{equation}
\label{eq:Reduced_appendix}
{\rho}(t)=\displaystyle\sum_{n=0}^\infty\frac{(-i)^n}{n!}\hhat{T}^\text{b}\int_0^t \left[\prod_{i=1}^n d t_i\hhat{H}^\times_{I}(t_i)\right]\rho(0),
\end{equation}
where the time-ordering superoperator is defined as
\begin{equation}
\label{eq:Tb_app}
\begin{array}{lll}
\hhat{T}^b[\hhat{H}_I(t_{P(n)})\cdots \hhat{H}_I(t_{P(1)})]=\hhat{H}_I(t_{n})\cdots \hhat{H}_I(t_{1}),
\end{array}
\end{equation}
where  $t_n\geq\cdots\geq t_1$ and $P$ is a permutation.
We begin by analyzing in more detail  the superoperator $\hhat{H}_I^X[\cdot]=[H_I,\cdot]$.
When it acts on an operator of the form $O_E\hat{O}_S$, we have, using Eq.~(\ref{eq:decomposition_Hamiltonian}) and omitting the time dependences,
\begin{equation}
\label{eq:HI}
\begin{array}{lll}
H^X_I[O_E\hat{O}_S]&=&{}[H_I,O_E\hat{O}_S]\\
&=&\displaystyle P_E B^\dagger\hat{s}O_E \hat{O}_S-P_E B\hat{s}^\dagger O_E\hat{O}_S\\
&&-O_E \hat{O}_S P_E B^\dagger\hat{s}+O_E \hat{O}_S P_E B\hat{s}^\dagger\\
&=&\displaystyle P_E B^\dagger O_E\hat{s}\hat{O}_S-P_E B O_E\hat{s}^\dagger_1\hat{O}_S\\
&&-\displaystyle O_E P_E B^\dagger\hat{O}_S\hat{s}+O_E P_E B\hat{O}_S\hat{s}^\dagger.
\end{array}
\end{equation}
For reasons that will become apparent later (section \ref{app:Wick}), we now introduce the full-parity superoperator $\hhat{P}=\hhat{P}_E\hhat{P}_S$ before the terms where operators act on the right of $O_E\hat{O}_S$, i.e., the last two terms in the expression above. Here, $\hhat{P}_E[\cdot]=P_E[\cdot] P_E$, where $P_E=\prod_{k\in E} \exp{[i\pi c^\dagger_k c_k]}$ and $\hhat{P}_S[\cdot]=P_S[\cdot] P_S$, where $P_S=\prod_{j\in S} \exp{[i\pi d^\dagger_j d_j]}$, with $c_k$ ($d_j$) the $k$th ($j$th) Fermion in the environment (system). The introduction of the operator $\hhat{P}$ is ``harmless'' (\cite{Saptsov}, pag.~5) when the overall parity of $O_E\hat{O}_S$ is even. However, since we are interested in analyzing a more general situation,  we also introduce it in the odd parity sector, which requires an extra minus sign. We then write, for $O_E\hat{O}_S$ even
\begin{equation}
\label{eq:HI_even}
\begin{array}{l}
H^X_I[O_E\hat{O}_S]=\displaystyle \left(P_E B^\dagger O_E\right)\hat{s}\hat{O}_S\\
\phantom{=}\displaystyle-\left(P_E B O_E\right)\hat{s}^\dagger_1\hat{O}_S-\left(\hhat{P}_EO_E P_E B^\dagger\right)\hhat{P}_S\hat{O}_S\hat{s}\\
\phantom{=}\displaystyle+\left(\hhat{P}_EO_E P_E B\right)\hhat{P}_S\hat{O}_S\hat{s}^\dagger\\
=\displaystyle\sum_{\lambda,q}\hhat{B}_q^{\prime \lambda}[O_E]\hhat{S}_{q}^{\bar{\lambda}}[\hat{O}_S]\;\;,
\end{array}
\end{equation}
and, for $O_E\hat{O}_S$ odd,
\begin{equation}
\label{eq:HI_odd}
\begin{array}{l}
H^X_I[O_E\hat{O}_S]=\displaystyle \left(P_E B^\dagger O_E\right)\hat{s}\hat{O}_S\\
\phantom{=}\displaystyle-\left(P_E B O_E\right)\hat{s}^\dagger_1\hat{O}_S
+\left(\hhat{P}_EO_E P_E B^\dagger\right)\hhat{P}_S\hat{O}_S\hat{s}\\
\phantom{=}\displaystyle-\left(\hhat{P}_EO_E P_E B\right)\hhat{P}_S\hat{O}_S\hat{s}^\dagger\\
=\displaystyle\sum_{\lambda,q}\hhat{B}_q^{\prime \lambda}[O_E]\hhat{S}_{q}^{\prime\bar{\lambda}}[\hat{O}_S]\;\;.
\end{array}
\end{equation}
Here, down-indexes take the values $\pm 1$ and specify if the operator acts on the left ($+1$) or right ($-1$) of its argument. Up-indexes take the values $\pm 1$ and distinguish the presence ($+1$) or absence ($-1$) of daggers in the definition. We also used the notation $\bar{\lambda}\equiv-\lambda$ . Explicitly, the quantities $\hhat{B}_q^\lambda$ and $\hhat{S}_q^\lambda$ are defined as
\begin{equation}
\label{eq:PEB}
\begin{array}{lll}
\hhat{B}_{1}^{\prime1}(t)[\cdot]&=&P_E B^\dagger(t) [\cdot]\\
\hhat{B}_{1}^{\prime-1}(t)[\cdot]&=&P_E B(t) [\cdot]\\
\hhat{B}_{-1}^{\prime1}(t)[\cdot]&=&\hhat{P}_E[[\cdot]P_E B^\dagger(t)] \\
\hhat{B}_{-1}^{\prime-1}(t)[\cdot]&=&\hhat{P}_E[[\cdot]P_E B(t)]\;\;,
\end{array}
\end{equation}
and
\begin{equation}
\label{eq:SS_op_1}
\begin{array}{lll}
\hhat{S}_{1}^{-1}(t)[\cdot]&=&\hat{s}(t) [\cdot]\\
\hhat{S}_{1}^{1}(t)[\cdot]&=&-\hat{s}^\dagger(t) [\cdot]\\
\hhat{S}_{-1}^{-1}(t)[\cdot]&=&-\hhat{P}_S[[\cdot]\hat{s}(t)] \\
\hhat{S}_{-1}^{1}(t)[\cdot]&=&\hhat{P}_S[[\cdot]\hat{s}^\dagger(t)]\;\;.
\end{array}
\end{equation}
The only difference between the even and odd case is the definition of the system superoperators, which take an extra minus sign when $\hhat{P}_S$ appear, i.e.,
\begin{equation}
\label{eq:SS_op_2}
\begin{array}{lll}
\hhat{S}_{1}^{\prime-1}(t)[\cdot]&=&\hat{s}(t) [\cdot]\\
\hhat{S}_{1}^{\prime1}(t)[\cdot]&=&-\hat{s}^\dagger(t) [\cdot]\\
\hhat{S}_{-1}^{\prime-1}(t)[\cdot]&=&\hhat{P}_S[[\cdot]\hat{s}(t)] \\
\hhat{S}_{-1}^{\prime1}(t)[\cdot]&=&-\hhat{P}_S[[\cdot]\hat{s}^\dagger(t)]\;\;.
\end{array}
\end{equation}
Now, using Eq.~(\ref{eq:HI_even}) and Eq.~(\ref{eq:HI_odd}) we derive the first order contribution to the Dyson equation in Eq.~(\ref{eq:series_0}) as
\begin{equation}
\label{eq:temp}
\begin{array}{lll}
{}[H_I(t_1),\rho(0)]\!\!&=&\!\!\!\displaystyle\sum_{\lambda,q} \left[\hhat{B}^{\prime \lambda}_q(t_1)[\rho^\text{eq}_E]\hhat{S}_q^{\bar{\lambda}}(t_1)[\hat{\rho}^\text{e}_S(0)]\right.\\
&+&\displaystyle\left.\!\!\!\! \hhat{B}^{\prime \lambda}_q(t_1)[\rho^\text{eq}_EP_E]\hhat{S}_q^{\prime\bar{\lambda}}(t_1)[\hat{\rho}^\text{o}_S(0)]\right]
\end{array}
\end{equation}
where we used the initial condition written in Eq.~(\ref{eq:initial_condition}).

Since the superoperators $\hhat{S}$ involve hat-operators $\hat{s}$, the result above is in the form $\sum_{j}O_E^j \hat{O}_S^j$. Furthermore, since the Hamiltonian $H_I$ is even in the fields, each term in Eq.~(\ref{eq:temp}) has the same overall parity as the part of the density matrix for the system (the initial condition) that they are acting upon (for example, the first term is even as it acts on the even part $\hat{\rho}^\text{e}_S(0)$).
As a consequence, we can use this symmetry, together with linearity, to explicitly write all perturbative terms in Eq.~(\ref{eq:series_0}). For example, the second order term $T[H_I(t_2),[H_I(t_1),\rho(0)]]$ becomes
\begin{equation}
\begin{array}{l}
\displaystyle\sum_{q,\lambda} \hhat{T}_E\hhat{B}^{\prime \lambda_2}_{q_2}(t_2)\hhat{B}^{\prime \lambda_1}_{q_1}(t_1)[\rho^\text{eq}_E]\hhat{T}_S\hhat{S}_{q_2}^{\bar{\lambda}_2}(t_2)\hhat{S}_{q_1}^{\bar{\lambda}_1}(t_1)[\hat{\rho}^\text{e}_S(0)]+\\
\displaystyle \hhat{T}_E\hhat{B}^{\prime \lambda_2}_{q_2}(t_2)\hhat{B}^{\prime \lambda_1}_{q_1}(t_1)[\rho^\text{eq}_EP_E]\hhat{T}_S\hhat{S}_{q_2}^{\prime \bar{\lambda}_2}(t_2)\hhat{S}_{q_1}^{\prime \bar{\lambda}_1}(t_1)[\hat{\rho}^\text{o}_S(0)],
\end{array}
\end{equation}
where used the short-hand $\sum_{q,\lambda}\equiv\sum_{q_1,\lambda_1}\sum_{q_2,\lambda_2}$, and where we  factorized the time-ordering operator for the full system $\hhat{T}^b=\hhat{T}_E\hhat{T}_S$ into time-ordering for the system $\hhat{T}_S$ and the environment $\hhat{T}_E$. Since these two newly defined superoperators act on a sequence of system and environmental field operators which have the same time-ordering, we can define them to be of Fermionic type, i.e.,
\begin{equation}
\hhat{T}_SO_S(t_{P(n)})\cdots O_S(t_{P(1)})=(-1)^{\#P}O_S(t_{n})\cdots O_S(t_{1}),
\end{equation}
where  $t_n\geq\cdots\geq t_1$, and where $P$ is a permutation with parity $\#P$. The same definition applies to $\hhat{T}_E$, upon changing $O_S\rightarrow O_E$. The importance of this choice becomes apparent in section \ref{sec:Influence}.

By iteratively using the arguments above, we can write the density matrix $\rho(t)$ for the full environment+system as
\begin{widetext}
\begin{equation}
\label{eq:useful_for_corr}
\begin{array}{lll}
\rho(t)&=&\displaystyle\sum_{n=0}^\infty\frac{(-i)^n}{n!}\int_0^t \left(\prod_{i=1}^nd t_i\right)\left\{\sum_{q_n,\lambda_n,\cdots,q_1,\lambda_1}\right.\left.\left[\hhat{T}_E\hhat{B}^{\prime \lambda_n}_{q_n}(t_n)\cdots\hhat{B}^{\prime \lambda_1}_{q_1}(t_1)[\rho^\text{eq}_E]\right]\hhat{T}_S\left[\hhat{S}_{q_n}^{\bar{\lambda}_n}(t_n)\cdots\hhat{S}_{q_1}^{\bar{\lambda}_1}(t_1)\right][\hat{\rho}^\text{e}_S(0)]\right.\\
&&+\left.\displaystyle\sum_{q_n,\lambda_n,\cdots,q_1,\lambda_1} \left[\hhat{T}_E \hhat{B}^{\prime \lambda_n}_{q_n}(t_n)\cdots\hhat{B}^{\prime \lambda_1}_{q_1}(t_1)[\rho^\text{eq}_EP_E]\right]\hhat{T}_S\left[\hhat{S}_{q_n}^{\prime \bar{\lambda}_n}(t_n)\cdots\hhat{S}_{q_1}^{\prime \bar{\lambda}_1}(t_1)\right][\hat{\rho}^\text{o}_S(0)]\right\}\;\;.
\end{array}
\end{equation}
\end{widetext}
Here, we explicitly remark the \emph{absence} of the operator $P_E$ in front of the environmental operators acting on $\rho_E^\text{eq}P_E$. However, such an operator will \emph{appear} in the corresponding correlation functions as we are about to show.

Crucially, the expression above shows that the density matrix has a decomposition in the form given by Eq.~(\ref{eq:decomposition_0}), i.e., as a sum of terms in which environmental operators multiply ``hat'' system operators. Explicitly,
\begin{equation}
\label{eq:identifications_1}
\rho(t)=\sum_i\rho_E^{i}\hat{\rho}_S^{i}\equiv\sum_{i_\text{e}}\rho_E^{ \prime i_\text{e}}\hat{\rho}_S^{\prime i_\text{e}}+\sum_{i_\text{o}}\rho_E^{\prime \prime i_\text{o}}\hat{\rho}_S^{\prime \prime i_\text{o}}\;\;.
\end{equation}
 The terms in Eq.~(\ref{eq:identifications_1}) are defined through the following identifications
\begin{equation}
\label{eq:identifications_2}
\begin{array}{lll}
\displaystyle\sum_{i_\text{e}/i_\text{o}}&\rightarrow&\displaystyle\sum_{n=0}^\infty\frac{(-i)^n}{n!}\int_0^t \left(\prod_{i=1}^nd t_i\right)\sum_{q_n,\lambda_n,\cdots,q_1,\lambda_1}\\
\rho_E^{\prime i_\text{e}}&\rightarrow&\hhat{T}_E\hhat{B}^{\prime \lambda_n}_{q_n}(t_n)\cdots\hhat{B}^{\prime \lambda_1}_{q_1}(t_1)[\rho^\text{eq}_E]\\
\hat{\rho}_S^{\prime i_\text{e}}&\rightarrow&\hhat{T}_S\hhat{S}_{q_n}^{\bar{\lambda}_n}(t_n)\cdots\hhat{S}_{q_1}^{\bar{\lambda}_1}(t_1)[\hat{\rho}^\text{e}_S(0)]\\
\rho_E^{\prime \prime i_\text{o}}&\rightarrow&\hhat{T}_E \hhat{B}^{\prime \lambda_n}_{q_n}(t_n)\cdots\hhat{B}^{\prime \lambda_1}_{q_1}(t_1)[\rho^\text{eq}_E P_E]\\
\hat{\rho}_S^{\prime \prime i_\text{o}}&\rightarrow&\hhat{T}_S\hhat{S}_{q_n}^{\prime \bar{\lambda}_n}(t_n)\cdots\hhat{S}_{q_1}^{\prime \bar{\lambda}_1}(t_1)[\hat{\rho}^\text{o}_S(0)]\;\;.
\end{array}
\end{equation}
We note that  the full density matrix $\rho(t)$ in Eq.~(\ref{eq:identifications_1}) could be, equivalently, written as
\begin{equation}
\label{eq:def_from_main}
\sum_{i}\hhat{T}_E\hhat{\rho}_E^{ \prime i}[\rho_E^\text{eq}]\hhat{T}_S\hhat{\rho}_S^{\prime i}[\rho_S^\text{e}(0)]+\hhat{T}_E\hhat{\rho}_E^{\prime \prime i}[\rho_E^\text{eq} P_E]\hhat{T}_S\hat{\rho}_S^{\prime \prime i}[\rho_S^\text{o}(0)]\;\;,
\end{equation}
where 
\begin{equation}
\begin{array}{lll}
\hhat{\rho}_E^{\prime i}&\rightarrow&\hhat{B}^{\prime \lambda_n}_{q_n}(t_n)\cdots\hhat{B}^{\prime \lambda_1}_{q_1}(t_1)\\
\hhat{\rho}_S^{\prime i}&\rightarrow&\hhat{S}_{q_n}^{\bar{\lambda}_n}(t_n)\cdots\hhat{S}_{q_1}^{\bar{\lambda}_1}(t_1)\\
\hhat{\rho}_E^{\prime \prime i}&\rightarrow& \hhat{B}^{\prime \lambda_n}_{q_n}(t_n)\cdots\hhat{B}^{\prime \lambda_1}_{q_1}(t_1)\\
\hhat{\rho}_S^{\prime \prime i}&\rightarrow&\hhat{S}_{q_n}^{\prime \bar{\lambda}_n}(t_n)\cdots\hhat{S}_{q_1}^{\prime \bar{\lambda}_1}(t_1)\;\;,
\end{array}
\end{equation}
which gives the explicit definitions to the quantities presented in Eq.~(\ref{eq:def_from_main}) in the main text.\\
From now on, for clarity of exposition, we omit the primes and double primes in $\rho^{i_\text{e/o}}_E$ and $\rho^{i_\text{e/o}}_S$  and, with a further abuse of notation, use the indexes $i_{\text{e/o}}$ as the way  to uniquely identify them.
As we remarked in section \ref{sec:Parity}, we can use the decomposition in Eq.~(\ref{eq:identifications_1}) into Eq.~(\ref{eq:effective_0})
\begin{equation}
\begin{array}{l}
\text{Tr}_{ES} A_S \rho(t)=\\
=\displaystyle\sum_i \text{Tr}_SA_S\{\text{Tr}_E[\rho_E^i] \hat{\rho}^{i,\text{e}}_S+\text{Tr}_E [P_E \rho_E^i] \hat{\rho}^{i,\text{o}}_S\}\\
=\displaystyle\sum_{i_e} \text{Tr}_SA_S\{\text{Tr}_E[\rho_E^{i_e}] \hat{\rho}^{i_e,\text{e}}_S+\text{Tr}_E [P_E \rho_E^{i_e}] \hat{\rho}^{i_e,\text{o}}_S\}\\
\displaystyle\phantom{=}+\sum_{i_o} \text{Tr}_SA_S\{\text{Tr}_E[\rho_E^{i_o}] \hat{\rho}^{i_o,\text{e}}_S+\text{Tr}_E [P_E \rho_E^{i_o}] \hat{\rho}^{i_o,\text{o}}_S\},
\end{array}
\end{equation}
which, by direct comparison with Eq.~(\ref{eq:density_matrix_implicit})), allows to write the reduced density matrix as Eq.~(\ref{eq:effective_1}) which reads
\begin{eqnarray}
\label{eq:effective_1_alternative}
{\rho}_S&=&\displaystyle\sum_i\text{Tr}_E[{\rho}^i_E] {\hat{{\rho}}}^{i,\text{e}}_S+\text{Tr}_E [P_E {\rho}^i_E] {\hat{{\rho}}}^{i,\text{o}}_S\nonumber\\
&=&\displaystyle\sum_{i_\text{e}}\text{Tr}_E[{\rho}^{i_\text{e}}_E] {\hat{{\rho}}}^{i_\text{e},\text{e}}_S+\text{Tr}_E [P_E {\rho}^{i_\text{e}}_E] {\hat{{\rho}}}^{i_\text{e},\text{o}}_S\\
&&+\displaystyle\sum_{i_\text{o}}\text{Tr}_E[{\rho}^{i_\text{o}}_E] {\hat{{\rho}}}^{i_\text{o},\text{e}}_S+\text{Tr}_E [P_E {\rho}^{i_\text{o}}_E] {\hat{{\rho}}}^{i_\text{o},\text{o}}_S\nonumber\;\;,
\end{eqnarray}
At this point, it is relevant to observe that Eq.~(\ref{eq:effective_1_alternative}) relies on an even/odd decomposition of the system operators $\hat{\rho}_S^{i_\text{o}}$ and $\hat{\rho}_S^{i_\text{e}}$ defined in Eq.~(\ref{eq:identifications_2}). In principle, despite the indexes-notation used, the parity of $\hat{\rho}_S^{i_\text{e/o}}$ depends on the order $n$ (so that they have the same parity as $\hat{\rho}^{\text{e/o}}_S(0)$ for $n$ even and opposite for $n$ odd). Explicitly, we have
\begin{alignat*}{4}
\hat{\rho}_S^{i_\text{e/o},\text{e/o}}&=&\hat{\rho}_S^{i_\text{e/o}}&~~\text{for} ~n~ \text{even}\\
&=&0&~~\text{for} ~n~ \text{odd}\\
\hat{\rho}_S^{i_\text{e/o},\text{o/e}}&=&0&~~\text{for} ~n~ \text{even}\\
&=&\hat{\rho}_S^{i_\text{e/o}}&~~\text{for} ~n~ \text{odd}\;\;.
\end{alignat*}
With this in mind, in Eq.~(\ref{eq:effective_1_alternative}) only the first and forth term survive for $n$ even and only the second and the third survive for $n$ odd, to get
\begin{widetext}
\setlength{\mathindent}{0pt}{\begin{eqnarray*}
\rho_S&=&\displaystyle\sum_{n=\text{even}}\frac{(-i)^n}{n!}\int_0^t \left(\prod_{i=1}^n d t_i\right)\sum_{q_n,\lambda_n\cdots q_1,\lambda_1}\left\{C^{\prime \lambda_n\cdots \lambda_1}_{q_n\cdots q_1}\hat{T}_S\left[\hhat{S}_{q_n}^{\bar{\lambda}_n}\cdots\hhat{S}_{q_1}^{\bar{\lambda}_1}\right]\hat{\rho}^\text{e}_S(0)
\displaystyle+D^{\prime \lambda_n\cdots \lambda_1}_{q_n\cdots q_1}\hat{T}_S\left[\hhat{S}_{q_n}^{\prime \bar{\lambda}_n}\cdots\hhat{S}_{q_1}^{\prime \bar{\lambda}_1}\right]\hat{\rho}^\text{o}_S(0)\right\}\\
&&+\displaystyle\sum_{n=\text{odd}}\frac{(-i)^n}{n!}\int_0^t \left(\prod_{i=1}^n d t_i\right)\sum_{q_n,\lambda_n\cdots q_1,\lambda_1}\left\{\tilde{C}^{\prime \lambda_n\cdots \lambda_1}_{q_n\cdots q_1}\hat{T}_S\left[\hhat{S}_{q_n}^{\bar{\lambda}_n}\cdots\hhat{S}_{q_1}^{\bar{\lambda}_1}\right]\hat{\rho}^\text{e}_S(0)+\tilde{D}^{\prime \lambda_n\cdots \lambda_1}_{q_n\cdots q_1}\hat{T}_S\left[\hhat{S}_{q_n}^{\prime \bar{\lambda}_n}\cdots\hhat{S}_{q_1}^{\prime \bar{\lambda}_1}\right]\hat{\rho}^\text{o}_S(0)\right\},
\end{eqnarray*}}
\end{widetext}
 where 
 \begin{equation}
 \label{eq:corr_app}
\begin{array}{lll}
 C^{\prime \lambda_n\cdots \lambda_1}_{q_n\cdots q_1}&=&\text{Tr}_E\hat{T}_E\left[\hhat{B}^{\prime \lambda_n}_{q_n}\cdots\hhat{B}^{\prime \lambda_1}_{q_1}\right][\rho^\text{eq}_E]\\
 D^{\prime \lambda_n\cdots \lambda_1}_{q_n\cdots q_1}&=& \text{Tr}_E\hat{T}_E\left[P_E\hhat{B}^{\prime \lambda_n}_{q_n}\cdots\hhat{B}^{\prime \lambda_1}_{q_1}\right][\rho^\text{eq}_EP_E]\\
\end{array}
 \end{equation}
 and 
 \begin{align*}
      \tilde{C}^{\prime \lambda_n\cdots \lambda_1}_{q_n\cdots q_1}&=&\text{Tr}_E\hat{T}_E\left[P_E\hhat{B}^{\prime \lambda_n}_{q_n}\cdots\hhat{B}^{\prime \lambda_1}_{q_1}\right][\rho^\text{eq}_E]\\
 \tilde{D}^{\prime \lambda_n\cdots \lambda_1}_{q_n\cdots q_1}&=& \text{Tr}_E\hat{T}_E\left[\hhat{B}^{\prime \lambda_n}_{q_n}\cdots\hhat{B}^{\prime \lambda_1}_{q_1}\right][\rho^\text{eq}_EP_E]\;\;.
 \end{align*}
For $n$ odd, the correlation functions are zero as they contain and odd number of creation/annihilation operators for Fermions and the equilibrium state is a thermal state (hence, even). As a consequence, we can write
\begin{widetext}
\begin{equation}
\label{eq:wide_for_corr}
\begin{array}{lll}
\rho_S&=&\displaystyle\sum_{n=0}^\infty\frac{(-i)^n}{n!}\int_0^t \left(\prod_{i=1}^n d t_i\right)\sum_{q_n,\lambda_n\cdots q_1,\lambda_1}\left\{C^{\prime \lambda_n\cdots \lambda_1}_{q_n\cdots q_1}\hat{T}_S\left[\hhat{S}_{q_n}^{\bar{\lambda}_n}\cdots\hhat{S}_{q_1}^{\bar{\lambda}_1}\right]\hat{\rho}^\text{e}_S(0)+D^{\prime \lambda_n\cdots \lambda_1}_{q_n\cdots q_1}\hat{T}_S\left[\hhat{S}_{q_n}^{\prime \bar{\lambda}_n}\cdots\hhat{S}_{q_1}^{\prime \bar{\lambda}_1}\right]\hat{\rho}^\text{o}_S(0)\right\}.
\end{array}
\end{equation}
\end{widetext}
It is interesting to realize how the two $P_E$ operators explicitly appearing (other $P_E$ operator are implicit in the definition of the fields $B$)  in the correlation functions $D^{\prime \lambda_n\cdots \lambda_1}_{q_n\cdots q_1}$ given in Eq.~(\ref{eq:corr_app}) have different origin. The $P_E$  multiplying $\rho_E^\text{eq}$ ultimately originates from the decomposition of the \emph{system} initial condition in Eq.~(\ref{eq:initial_condition}). The remaining $P_E$ originates from the properties of the partial trace for Fermions, i.e., from Eq.~(\ref{eq:effective_1}).

As a final step, we point out that the operators  $P_E$ implicitly present in the correlation functions in Eq.~(\ref{eq:corr_app}) through the definition of the superoperators $\hhat{B}$ in Eq.~(\ref{eq:PEB})  always appear on the left of the operators $B$ (here we do not consider the operators $P_E$ originating from $\hhat{P}_E$). As a consequence, using the cyclic property of the trace, environmental correlations will always involve terms in the form $\text{Tr}[P_E B^{\lambda_1}\cdots P_E B^{\lambda_n}\rho_E^\text{eq}]$ apart from the possible presence of an extra $P_E$ from $\hhat{P}_E$. Since each field $B$ has parity one, and since only correlation functions for even $n$ contribute, we can always anticommute the $P_E$ with the fields and removing them using $P_E^2=1$ (the presence of the extra $P_E$ from $\hhat{P}_E$ is irrelevant for this line of thought). This corresponds to effectively remove all the original $P_E$ in front of the fields $B$ in Eq.~(\ref{eq:PEB}), by adding an extra $(-i)^n$ factor. This leads to 
\begin{equation}
\label{eq:temp_12}
\begin{array}{l}
\rho_S=\displaystyle\sum_{n=0}^\infty\frac{(-1)^n}{n!}\int_0^t \left(\prod_{i=1}^n d t_i\right)\\
\displaystyle\left[\sum_{q_n,\lambda_n\cdots q_1,\lambda_1}C^{\lambda_n\cdots \lambda_1}_{q_n\cdots q_1}\hat{T}_S\left[\hhat{S}_{q_n}^{\bar{\lambda}_n}\cdots\hhat{S}_{q_1}^{\bar{\lambda}_1}\right]\hat{\rho}^\text{e}_S(0)\right.\\
\displaystyle+\left.\sum_{q_n,\lambda_n\cdots q_1,\lambda_1}D^{\lambda_n\cdots \lambda_1}_{q_n\cdots q_1}\hat{T}_S\left[\hhat{S}_{q_n}^{\prime \bar{\lambda}_n}\cdots\hhat{S}_{q_1}^{\prime \bar{\lambda}_1}\right]\hat{\rho}^\text{o}_S(0)\right],
\end{array}
\end{equation}
where
\begin{equation}
\label{eq:temp_12_cd}
 \begin{array}{lll}
 C^{\lambda_n\cdots \lambda_1}_{q_n\cdots q_1}&=&\text{Tr}_E\hat{T}_E\left[\hhat{B}^{ \lambda_n}_{q_n}\cdots\hhat{B}^{ \lambda_1}_{q_1}\right][\rho^\text{eq}_E]\\
 D^{\lambda_n\cdots \lambda_1}_{q_n\cdots q_1}&=& \text{Tr}_E\hat{T}_E\left[P_E\hhat{B}^{ \lambda_n}_{q_n}\cdots\hhat{B}^{ \lambda_1}_{q_1}\right][\rho^\text{eq}_EP_E]\;\;,
\end{array}
\end{equation}
with
\begin{equation}
\label{eq:PEB_simpler}
\begin{array}{lll}
\hhat{B}_{1}^{1}(t)[\cdot]&=& B^\dagger(t) [\cdot]\\
\hhat{B}_{1}^{-1}(t)[\cdot]&=& B(t) [\cdot]\\
\hhat{B}_{-1}^{1}(t)[\cdot]&=&\hhat{P}_E[[\cdot] B^\dagger(t)] \\
\hhat{B}_{-1}^{-1}(t)[\cdot]&=&\hhat{P}_E[[\cdot] B(t)]\;\;.
\end{array}
\end{equation}
 It is actually possible to simplify this expression even further. To achieve this, we analyze the correlations $ D^{\lambda_n\cdots \lambda_1}_{q_n\cdots q_1}$. Our goal is to remove the two $P_E$ explicitly appearing in Eq.~(\ref{eq:temp_12_cd}). We begin by observing that, if the fields $\hhat{B}$ were normal operators (i.e., not superoperators), we could simply use the cyclic property of the trace and conclude that the presence of the $P_E$ is irrelevant. However, this reasoning does not hold with superoperators because the  operators they introduce might act either on the left or on the right of the density matrix, changing the relative position of the two $P_E$.\\
Nevertheless, we can imagine to move the $P_E$ (which multiplies $\rho^\text{eq}_E$) on the left, until it gets next to the remaining $P_E$. As we do this, we get an extra minus sign each time one of the down-indexes of the $\hhat{B}$ is $+1$, i.e., it acts on the left of the density matrix (hence it is ``in between'' the first and the second $P_E$). However,  the down-indexes $q_1,\dots,q_n$ also label the system superoperators $\hhat{S}$. As a consequence, the two $P_E$ can be effectively dropped by adding a minus sign each time a $-1$ appears in the down indexes of the superoperators $\hhat{S}$. This is, for us, extremely convenient as such a minus sign is exactly what differentiates the operators $\hhat{S}^{\prime}$ from $\hhat{S}$, see Eq.~(\ref{eq:SS_op_1}) and Eq.~(\ref{eq:SS_op_2}). This last consideration allows \emph{to write the reduced density matrix in a form which does not need to explicitly distinguish which parity sector we are acting upon}, i.e., 
\begin{equation}
\label{eq:Dyson_app}
\begin{array}{l}
\rho_S=\displaystyle\sum_{n=0}^\infty\frac{(-1)^n}{n!}\int_0^t \left(\prod_{i=1}^n d t_i\right)\\
\phantom{=}\displaystyle\left[\sum_{q_n,\lambda_n\cdots q_1,\lambda_1}C^{\lambda_n\cdots \lambda_1}_{q_n\cdots q_1}\hat{T}_S\left[\hhat{S}_{q_n}^{\bar{\lambda}_n}\cdots\hhat{S}_{q_1}^{\bar{\lambda}_1}\right]\hat{\rho}^\text{e}_S(0)\right.\\
\phantom{=}\displaystyle+\left.\sum_{q_n,\lambda_n\cdots q_1,\lambda_1}C^{\lambda_n\cdots \lambda_1}_{q_n\cdots q_1}\hat{T}_S\left[\hhat{S}_{q_n}^{ \bar{\lambda}_n}\cdots\hhat{S}_{q_1}^{ \bar{\lambda}_1}\right]\hat{\rho}^\text{o}_S(0)\right]\\

=\displaystyle\sum_{n=0}^\infty\frac{(-1)^n}{n!}\int_0^t \left(\prod_{i=1}^n d t_i\right)\\
\phantom{=}\displaystyle\sum_{q_n,\lambda_n\cdots q_1,\lambda_1}C^{\lambda_n\cdots \lambda_1}_{q_n\cdots q_1}\hat{T}_S\left[\hhat{S}_{q_n}^{\bar{\lambda}_n}\cdots\hhat{S}_{q_1}^{\bar{\lambda}_1}\right]\hat{\rho}_S(0)\;\;,
\end{array}
\end{equation}
which is Eq.~(\ref{eq:reduced_density_0}) in the main text.
\subsection{Wick's theorem}
\label{app:Wick}
In this section we review the proof of the Wick's theorem for superoperators in \cite{Saptsov} and analyze the time-ordered case.
\subsubsection{Wick's theorem for superoperators}
\label{sec:Wick:super:elegant}
To keep this article self-contained, to adapt the notation, and to highlight its elegance, in this section we briefly review the proof of the Wick's theorem for Fermionic superoperators developed by Saptsov and Wegewijs in \cite{Saptsov}.\\

The main object of study are correlations of the form
\begin{equation}
S_n=\text{Tr}(\hhat{c}_{q_n}^{p_n}\cdots \hhat{c}_{q_1}^{p_1}\rho_E^\text{eq})\;\;,
\end{equation}
where $\rho_E^\text{eq}=\exp[{-\beta\sum_k (\omega_k-\mu) c^\dagger_k c_k}]/Z^\text{eq}_E$, with $ Z^\text{eq}_{E}=\prod_k (1+\exp[-\beta(\omega_k-\mu)])$, and
where $p=(\lambda,k)$ is a multi-index so that $\lambda=\pm 1$ defines the presence ($\lambda=+1$) or absence ($\lambda=-1$) of a dagger and $k$ is an external index labeling the Fermions of the bath. The index $q=\pm 1$ specifies whether the operator acts on the left ($q=1$) or right $q=-1$. Explicitly, we have
\begin{equation}
\begin{array}{lll}
\hhat{c}^p_q[\cdot]&=&\hhat{c}^{\lambda,k}_q[\cdot]\\
&=&\delta_{q,+1} c^{p}[\cdot]+\delta_{q,-1}[\cdot]c^{p}\;\;,
\end{array}
\end{equation}
where 
\begin{equation}
c^p=c^{(\lambda,k)}=(\delta_{\lambda,1}c^\dagger_{k}+\delta_{\lambda,-1}c_{k})\;\;.
\end{equation}
Using this notation, the usual Fermionic anticommutation rules read
\begin{equation}
\{c^{p},c^{p'}\}=\delta_{p,\bar{p}'}\;\;,
\end{equation}
where $p=(\lambda,k)$ and $\bar{p}\equiv (\bar{\lambda},k)$ with $\bar{\lambda}=-\lambda$.\\

The main issue to prove a Wick's theorem for superoperators is that no definite commutation or anticommutation rules hold for superoperators. We can see this explicitly as
\begin{equation}
\begin{array}{l}
\{\hhat{c}_{q_1}^{p_1},\hhat{c}_{q_2}^{p_2}\}(\cdot)=\\
=\delta_{q_1,+1}\delta_{q_2,+1}\delta_{p_1,\bar{p}_2}(\cdot)+2\delta_{q_1,+1}\delta_{q_2,-1}c^{p_1}(\cdot)c^{p_2}\\
\phantom{=}+2\delta_{q_1,-1}\delta_{q_2,+1}c^{p_2}(\cdot)c^{p_1}+\delta_{q_1,-1}\delta_{q_2,-1}\delta_{p_1,\bar{p}_2} (\cdot)\\
= \delta_{p_1,\bar{p}_2} \left(\delta_{q_1,+1}\delta_{q_2,+1}+\delta_{q_1,-1}\delta_{q_2,-1}\right)\\
\phantom{=}+2\delta_{q_1,+1}\delta_{q_2,-1}c^{p_1}(\cdot)c^{p_2}+2\delta_{q_1,-1}\delta_{q_2,+1}c^{p_2}(\cdot)c^{p_1}\;\;.
\end{array}
\end{equation}
The factor $2$ in the second line appears as a consequence of the fact that $\hhat{c}^{p}_{1}$ and $\hhat{c}^{p'}_{-1}$ commute, i.e., $[\hhat{c}^{p}_{1},\hhat{c}^{p'}_{-1}]=0$. The elegant consideration presented in \cite{Saptsov} is to consider the modified fields
\begin{equation}
\begin{array}{lll}
\hhat{\mathcal{J}}^p_q=\delta_{q,+1} \hhat{c}^p_q+\delta_{q,-1} \hhat{P}_E\hhat{c}^p_q\;\;,
\end{array}
\end{equation}
where $\hhat{P}_E[\cdot]=P_E\cdot P_E$. This ``harmless'' (see \cite{Saptsov}, pag. 5) definition has profound effects as, now,
\begin{equation}
\begin{array}{l}
\{\hhat{\mathcal{J}}_{q_1}^{p_1},\hhat{\mathcal{J}}_{q_2}^{p_2}\}(\cdot)=\delta_{q_1,+1}\delta_{q_2,+1}\delta_{p_1,\bar{p}_2} \{c^{p_1},c^{\bar{p}_1}\}(\cdot)\\
\phantom{=}+\delta_{q_1,+1}\delta_{q_2,-1}(c^{p_1}P_E(\cdot)c^{p_2}P_E+P_E c^{p_1}(\cdot)c^{p_2}P_E)\\
\phantom{=}+\delta_{q_1,-1}\delta_{q_2,+1}(P_E c^{p_2}(\cdot)c^{p_1}P_E + c^{p_2}P_E(\cdot)c^{p_1}P_E)\\
\phantom{=}+\delta_{q_1,-1}\delta_{q_2,-1}\delta_{p_1,\bar{p}_2}P^2_E (\cdot)\left(c^{p_2}P_E c^{\bar{p}_1}P_E+c^{p_1}P_E c^{\bar{p}_2}P_E\right)\\

=\delta_{q_1,+1}\delta_{q_2,+1}\delta_{p_1,\bar{p}_2}(\cdot)+q_1 \delta_{q_1,-1}\delta_{q_2,-1}\delta_{p_1,\bar{p}_2}(\cdot)\\
= q_1\delta_{p_1,\bar{p}_2}\delta_{q_1,q_2}\;\;,
\end{array}
\end{equation}
which starts to resemble the Fermionic anticommutation rules. To complete the mapping, it is possible \cite{Saptsov} to introduce
\begin{equation}
\begin{array}{lll}
\hhat{J}^p_{q}&=&\displaystyle\frac{1}{\sqrt{2}}\left(\delta_{q,+1}\sum_{q'}q'\hhat{\mathcal{J}}^p_{q'}+\delta_{q,-1}\sum_{q'}\hhat{\mathcal{J}}^p_{q'}\right)\\
&=&\displaystyle\frac{\left[\delta_{q,+1}\left(\hhat{c}^p_1-\hhat{P}_E\hhat{c}^p_{-1}\right)+\delta_{q,-1}\left(\hhat{c}^p_1+\hhat{P}_E\hhat{c}^p_{-1}\right)\right]}{\sqrt{2}}.
\end{array}
\end{equation}
For future reference, these expressions can be inverted to obtain the Fermionic operators as
\begin{equation}
\begin{array}{lll}
\hhat{c}^p_1&=&\displaystyle\frac{1}{\sqrt{2}}\left(\hhat{J}^p_{-1}+\hhat{J}^p_{+1}\right)\\
\hhat{P}_E\hhat{c}^p_{-1}&=&\displaystyle\frac{1}{\sqrt{2}}\left(\hhat{J}^p_{-1}-\hhat{J}^p_{+1}\right)\;\;.
\end{array}\label{eq:transformation}
\end{equation}
We now have, defining $\bar{q}=-q$,
\begin{equation}
\label{eq:sp_comm}
\begin{array}{lll}
\{\hhat{J}_{q_1}^{p_1},\hhat{J}_{q_2}^{p_2}\}&=&\frac{1}{2}\delta_{q_1,+1}\delta_{q_2,+1}\{\sum_{q}q\hhat{\mathcal{J}}^{p_1}_q,\sum_{q}q\hhat{\mathcal{J}}^{p_2}_q\}\\
&&+\frac{1}{2}\delta_{q_1,+1}\delta_{q_2,-1}\{\sum_{q}q\hhat{\mathcal{J}}^{p_1}_q,\sum_{q}\hhat{\mathcal{J}}^{p_2}_q\}\\
&&+\frac{1}{2}\delta_{q_1,-1}\delta_{q_2,+1}\{\sum_{q}\hhat{\mathcal{J}}^{p_1}_q,\sum_{q}q\hhat{\mathcal{J}}^{p_2}_q\}\\
&&+\frac{1}{2}\delta_{q_1,-1}\delta_{q_2,-1}\{\sum_{q}\hhat{\mathcal{J}}^{p_1}_q,\sum_{q}\hhat{\mathcal{J}}^{p_2}_q\}\\

&=&\frac{1}{2}\delta_{q_1,+1}\delta_{q_2,+1}\sum_{q}\sum_{q'}qq'\{\hhat{\mathcal{J}}^{p_1}_q,\hhat{\mathcal{J}}^{p_2}_{q'}\}\\
&&+\frac{1}{2}\delta_{q_1,+1}\delta_{q_2,-1}\sum_{q}\sum_{q'}q\{\hhat{\mathcal{J}}^{p_1}_q,\hhat{\mathcal{J}}^{p_2}_{q'}\}\\
&&+\frac{1}{2}\delta_{q_1,-1}\delta_{q_2,+1}\sum_{q}\sum_{q'}q'\{\hhat{\mathcal{J}}^{p_1}_q,\hhat{\mathcal{J}}^{p_2}_{q'}\}\\
&&+\frac{1}{2}\delta_{q_1,-1}\delta_{q_2,-1}\sum_{q}\sum_{q'}\{\hhat{\mathcal{J}}^{p_1}_q,\hhat{\mathcal{J}}^{p_2}_{q'}\}\\

&=&\frac{1}{2}\delta_{q_1,+1}\delta_{q_2,+1}\delta_{p_1,\bar{p}_2}\sum_{q}qqq\\
&&+\frac{1}{2}\delta_{q_1,+1}\delta_{q_2,-1}\delta_{p_1,\bar{p}_2}\sum_{q}qq\\
&&+\frac{1}{2}\delta_{q_1,-1}\delta_{q_2,+1}\delta_{p_1,\bar{p}_2}\sum_{q}qq\\
&&+\frac{1}{2}\delta_{q_1,-1}\delta_{q_2,-1}\delta_{p_1,\bar{p}_2}\sum_{q}q\\

&=&\delta_{q_1,+1}\delta_{q_2,-1}\delta_{p_1,\bar{p}_2}+\delta_{q_1,-1}\delta_{q_2,+1}\delta_{p_1,\bar{p}_2}\\

&=&\delta_{q_1,\bar{q}_2}\delta_{p_1,\bar{p}_2}\;\;,
\end{array}
\end{equation}
which elegantly resembles the Fermionic anticommutation rules. But this is not all, as other important relations hold. One is the fluctuation-dissipation-like relation
\begin{equation}
\begin{array}{lll}
\hhat{J}^{p}_{\pm1}(\rho_E^\text{eq})&=&(\hhat{c}^p_1\mp\hhat{P}_E \hhat{c}^p_{-1})(\rho_E^\text{eq})\\
&=&{c}^\lambda_k\rho_E^\text{eq}\mp{P}_E \rho_E^\text{eq}{c}^\lambda_k P_E\\
&=&(e^{\lambda\beta(\omega_{k}-\mu)}\pm1)\rho_E^\text{eq} {c}^\lambda_k\;\;,
\end{array}
\end{equation}
where we used Eq.~(\ref{eq:comm}) and the fact that $\rho^\text{eq}_E$ is even in the number of Fermionic operators. We then have
\begin{equation}
\label{eq:sp_fd}
\begin{array}{lll}
\hhat{J}^{p}_{-1}(\rho^\text{eq}_E)&=&\frac{(e^{\lambda\beta(\omega_{k}-\mu)}-1)}{(e^{\lambda\beta(\omega_{k}-\mu)}+1)}\hhat{J}^{p}_{+1}(\rho^\text{eq}_E)\\
&=&\tanh(\lambda\beta(\omega_{k}-\mu)/2)\hhat{J}^{p}_{+1}(\rho^\text{eq}_E)\;\;.
\end{array}
\end{equation}
Another important relation is the ``closure''
\begin{equation}
\label{eq:sp_closure}
\text{Tr}\left[\hhat{J}^q_{+1}\cdot\right]=0\;\;,
\end{equation}
which is proved by  using the cyclic property of the trace as
\begin{equation}
\begin{array}{lll}
\text{Tr}\left[\hhat{J}^p_{+1}\cdot\right]&=&\text{Tr}\left[(\hhat{c}^p_1-\hhat{P}_E \hhat{c}^p_{-1})\cdot\right]\\
&=&\text{Tr}\left[{c}^{\lambda}_{k} \cdot-{P}_E \cdot{c}^{\lambda}_{k} P_E\right]\\
&=&\text{Tr}\left[{c}^{\lambda}_{k} \cdot- {c}^{\lambda}_{k}\cdot\right]\\
&=&0\;\;.
\end{array}
\end{equation}
Everything is now ready to prove Wick's theorem. We consider
\begin{equation}
S_n=\text{Tr}\left[J^{p_1}_{q_1}\cdots J^{p_n}_{q_n}\rho^\text{eq}_E\right]\;\;,
\end{equation}
which is non-zero only for even $n$. In this case, if $q_n=+1$, we can anticommute it on the left and then use the closure property to derive
\begin{equation}
\begin{array}{lll}
S_n&=&\displaystyle\sum_{i=1}^{n-1}(-1)^{\#P_{i,n}}\text{Tr}\left[\{J^{p_n}_{q_n},J^{p_i}_{q_i}\}J^{p_1}_{q_1}\underbrace{\cdots}_{i,n} J^{p_n}_{q_n}\rho_\beta\right]\\
&=&\displaystyle\sum_{i=1}^{n-1}(-1)^{\#P_{i,n}}\text{Tr}[\{J^{p_i}_{q_i},J^{p_n}_{q_n}\}\rho^\text{eq}_E]S^{i,n}_{n-2}\\
&=&\displaystyle\sum_{i=1}^{n-1}(-1)^{\#P_{i,n}}\langle J^{p_i}_{q_i}J^{p_n}_{q_n}\rangle_E S^{i,n}_{n-2}\;\;,
\end{array}
\end{equation}
where the underbrace indicates the indexes labeling the missing operators and where we used Eq.~(\ref{eq:sp_comm}) in the second step and we used the closure property Eq.~(\ref{eq:sp_closure}) in the last step.
We also defined $S^{i,n}_{n-2}=\text{Tr}\left[J^{p_1}_{q_1}\underbrace{\cdots}_{i,n} J^{p_n}_{q_n}\rho^\text{eq}_E\right]$ and $\langle\cdot\rangle_{E}=\text{Tr}[\cdot\rho_E^\text{eq}]$. Here, $\#P_{i,n}$ is the number of transpositions needed to bring $J^{p_n}_{q_n}$ and $J^{p_i}_{q_i}$ adjacent \cite{Saptsov}.

If $q_n=-1$, we cannot apply the closure relation directly. However, we can first use the fluctuation-dissipation relation Eq.~(\ref{eq:sp_fd}) to obtain
\begin{equation}
\begin{array}{lll}
S_n&=&\displaystyle t_n\sum_{i=1}^{n-1}(-1)^{\#P_{i,n}}\text{Tr}[\{J^{p_n}_{\bar{q}_n},J^{p_i}_{q_i}\}\rho_\beta]S^{i,n}_{n-2}\\
&=&\displaystyle t_n\sum_{i=1}^{n-1}(-1)^{\#P_{i,n}}\text{Tr}[J^{p_i}_{q_i}J^{p_n}_{\bar{q}_n}\rho_E]S^{i,n}_{n-2}\\
&=&\displaystyle\sum_{i=1}^{n-1}(-1)^{\#P_{i,n}}\text{Tr}[J^{p_i}_{q_i}J^{p_n}_{{q}_n}\rho_E]S^{i,n}_{n-2}\\
&=&\displaystyle\sum_{i=1}^{n-1}(-1)^{\#P_{i,n}}\langle J^{p_i}_{q_i}J^{p_n}_{{q}_n}\rangle_E S^{i,n}_{n-2}\;\;,
\end{array}
\end{equation}
where $q_n=(\lambda_n,k_n)$ is a multi-index and defined $t_n\equiv\tanh({\lambda_n}\beta(\omega_{k_n}-\mu)/2)$. In order to derive the second line we used the closure property Eq.~(\ref{eq:sp_closure}) and to obtain the third line we used the fluctuation-dissipation relation Eq.~(\ref{eq:sp_fd}) again.\\
Proceeding this way iteratively, we prove
\begin{equation}
\label{eq:WWick}
\text{Tr}\left[J^{p_1}_{q_1}\cdots J^{p_n}_{q_n}\rho^\text{eq}_E\right]=\sum_{c\in C_n}(-1)^{\#c}\prod_{(i,j)\in c}\langle J^{p_i}_{q_i}J^{p_j}_{q_j}\rangle_E\;\;.
\end{equation}
Here, each full-contraction $c\in C_n$ is one of the possible sets of ordered pairs (or just, contractions) $(i_c,j_c)$, $i_c<j_c$ over the set $\mathbb{N}_n=\{1,\cdots,n\}$. We further denoted by $\#c$ the parity of the contraction $c$, i.e., the parity of the permutation needed to order the set $\mathbb{N}_n$, such that all pairs in $c$ are adjacent.

To conclude, we observe that, in order to use this form of Wick's theorem, the superoperators have to be written in terms  of $\hhat{c}^p_1$ and $\hhat{P}_E \hhat{c}^p_{-1}$ defined in Eq.~(\ref{eq:transformation}). Using $B(t)=\sum_k g_k c_k e^{-i\omega_k t}$, we can write the superoperators $\hhat{B}$ defined in Eq.~(\ref{eq:PEB_main}) as
\begin{equation}
\begin{array}{lll}
\hhat{B}^\lambda_1(t)&=&\sum_k g_k \hhat{c}^p_1 e^{-i\omega_k t}\\
\hhat{B}^\lambda_{-1}(t)&=&\sum_k g_k \hhat{P}_E\hhat{c}^p_{-1} e^{-i\omega_k t}
\end{array}
\end{equation} 
where $p=(\lambda,k)$. This shows that, as long as the correlations are written in terms of the superoperators $\hhat{B}$ above, we can indeed use the Wick's theorem in Eq.~(\ref{eq:WWick}), justifying the reasoning done in section \ref{sec:Wick}.
\subsubsection{Time-ordering in Wick's theorem}
The form of the Wick's theorem in Eq.~(\ref{eq:WWick}) implies that if $P^a$ is a single swap between two adjacent superoperators (let us say between $J^{p_a}_{q_a}$ and $J^{p_{a+1}}_{q_{a+1}}$), the parity $\#c$ of each full-contraction $c$ will provide an extra minus-sign unless $(a,a+1)\in c$. In fact, the parity of the permutation needed to order the set (after applying $P^a$) such that all pairs are adjacent is $-(\#c)$ when $(a,a+1)\notin c$. 

When $(a,a+1)\in c$, there is no extra-sign as  $a$ and $a+1$, even if swapped, are already adjacent. This slight imperfection with respect to total antisymmetry implies that special care needs to be taken with respect to the order in which the original sequence appears inside the correlation. However, total antisymmetry can be restored by simply considering Fermionic time-ordering of the original sequence. In this case,  supposing $t_n\geq\cdots\geq t_1$, we have
\begin{equation}
\text{Tr}\left[\hhat{T}[J^{p_{P(1)}}_{q_{P(1)}}\cdots J^{p_{P(n)}}_{q_{P(n)}}]\rho^\text{eq}_E\right]=(-1)^{\#P}
W[J^{p_n}_{q_n}\cdots J^{p_1}_{q_1}],
\end{equation}
where $P$ is a generic permutation  and where
\begin{equation}
W[^{p_n}_{q_n}\cdots J^{p_1}_{q_1}]=\sum_{c\in \bar{C}_n}(-1)^{\#c}\prod_{i,j\in c}\langle J^{p_i}_{q_i}J^{p_j}_{q_j}\rangle_E\;\;.
\end{equation}
Here, $\bar{C}_n$ is the set of contractions over the set $\bar{\mathbb{N}}_n=\{n,\cdots,1\}$.
Importantly, since $i,j$ are ordered as in the sequence given to $W$, we can always include an additional time-ordering in the definition to obtain
\begin{equation}
\label{eq:WT}
\text{Tr}\left[\hhat{T} J^{p_{P(1)}}_{q_{P(1)}}\cdots J^{p_{P(n)}}_{q_{P(n)}}]\rho_E^\text{eq}\right]=(-1)^{\#P}W_T[J^{p_n}_{q_n}\cdots J^{p_1}_{q_1}]\;\;,
\end{equation}
 where
\begin{equation}
W_T[J^{p_1}_{q_1}\cdots J^{p_n}_{q_n}]=\sum_{c\in \bar{C}_n}(-1)^{\#c}\prod_{i,j\in c}\langle \hhat{T}[J^{p_i}_{q_i}J^{p_j}_{q_j}]\rangle_E\;\;,
\end{equation}
which fulfills
\begin{equation}
\label{eq:WT_minus}
W_T[P^a[J^{p_n}_{q_n}\cdots J^{p_1}_{q_1}]]=-W_T[J^{p_n}_{q_n}\cdots J^{p_1}_{q_1}]\;\;.
\end{equation}
In fact,
\begin{equation}
\begin{array}{l}
W_T[P^{a}[J^{p_n}_{q_n}\cdots J^{p_1}_{q_1}]]=\\
\displaystyle=-\sum_{c\in C_n}(-1)^{\#c}\prod_{(i,j)\in c,(a,b)\not\in c}\langle \hhat{T}[J^{p_i}_{q_i}J^{p_j}_{q_j}]\rangle_E\\
\displaystyle\phantom{=}+\sum_{c\in C_n}(-1)^{\#c}\prod_{(i,j)\in c,(a,b)\in c}\langle \hhat{T}[P^a[J^{p_i}_{q_i}J^{p_j}_{q_j}]]\rangle_E\\

\displaystyle=-\sum_{c\in C_n}(-1)^{\#c}\prod_{(i,j)\in c,(a,b)\not\in c}\langle \hhat{T}[J^{p_i}_{q_i}J^{p_j}_{q_j}]\rangle_E\\
\displaystyle\phantom{=}-\sum_{c\in C_n}(-1)^{\#c}\prod_{(i,j)\in c,(a,b)\in c}\langle \hhat{T}[J^{p_i}_{q_i}J^{p_j}_{q_j}]\rangle_E\\

\displaystyle=-\sum_{c\in C_n}(-1)^{\#c}\prod_{(i,j)\in c}\langle \hhat{T}[J^{p_i}_{q_i}J^{p_j}_{q_j}]\rangle_E\;\;.
\end{array}
\end{equation}
Since we can always decompose the generic permutation $P$ appearing in Eq.~(\ref{eq:WT}) in terms of transpositions $P^a$, by 
repetitive application of Eq.~(\ref{eq:WT_minus})  we obtain
\begin{equation}
\label{eq:Wick_raw}
\text{Tr}\left[\hhat{T}[J^{p_{P(1)}}_{q_{P(1)}}\cdots J^{p_{P(n)}}_{q_{P(n)}}]\rho^\text{eq}_E\right]=
W_T[J^{p_{P(1)}}_{q_{P(1)}}\cdots J^{p_{P(n)}}_{q_{P(n)}}]\;\;.
\end{equation}
This is an rather convenient result as we can apply the Wick's operator $W$ directly to the original sequence, independently from its order.

\subsubsection{Commutation relations with equilibrium distribution}
In this subsection we prove the relation
\begin{equation}
\label{eq:comm}
c^\lambda_k\rho_E^\text{eq}=e^{\lambda\beta(\omega_k-\mu)}\rho_E^\text{eq} c^\lambda_k\;\;,
\end{equation}
where $\lambda=\pm1$ and where $\rho_E^\text{eq}=\exp[-\beta\sum_k(\omega_k-\mu)c^\dagger_k c_k]/Z^\text{eq}_E$ with $ Z^\text{eq}_{E}=\prod_k (1+\exp[-\beta(\omega_k-\mu)])$.
To start we have
\begin{equation}
\begin{array}{lll}
c_k\rho_E^\text{eq}&=&\displaystyle c_k e^{-\beta(\omega_k-\mu) c^\dagger_k c_k}e^{-\beta\sum_{j\neq k}(\omega_j-\mu) c^\dagger_j c_j}/ Z^\text{eq}_{E}\\
&=&c_k[1+(e^{-\beta(\omega_k-\mu)}-1)c^\dagger_k c_k]\\
&&\times e^{-\beta\sum_{j\neq k}(\omega_j-\mu) c^\dagger_j c_j}/ Z^\text{eq}_{E}\\
&=&e^{-\beta(\omega_k-\mu)}c_k e^{-\beta\sum_{j\neq k}(\omega_j-\mu) c^\dagger_j c_j}/ Z^\text{eq}_{E}\;\;,
\end{array}
\end{equation}
 We also have 
\begin{equation}
\begin{array}{lll}
\rho_E^\text{eq} c_k&=&e^{-\beta(\omega_k-\mu) c^\dagger_k c_k} c_k e^{-\beta\sum_{j\neq k}(\omega_j-\mu) c^\dagger_j c_j}/ Z^\text{eq}_{E}\\
&=&[1+(e^{-\beta(\omega_k-\mu)}-1)c^\dagger_k c_k] \\
&&\times c_k e^{-\beta\sum_{j\neq k}(\omega_j-\mu) c^\dagger_j c_j}/ Z^\text{eq}_{E}\\
&=&c_k e^{-\beta\sum_{j\neq k}(\omega_j-\mu) c^\dagger_j c_j}/ Z^\text{eq}_{E}\;\;,
\end{array}
\end{equation}
so that, by comparison, we obtain
\begin{equation}
\label{eq:comm_a}
c_k\rho_E^\text{eq}=e^{-\beta(\omega_k-\mu)}\rho_E^\text{eq} c_k\;\;.
\end{equation}
Similarly,
\begin{equation}
\begin{array}{lll}
c^\dagger_k\rho_E^\text{eq}&=&c^\dagger_k e^{-\beta(\omega_k-\mu) c^\dagger_k c_k}e^{-\beta\sum_{j\neq k}(\omega_j-\mu) c^\dagger_j c_j}/ Z^\text{eq}_{E}\\
&=&c^\dagger_k[1+(e^{-\beta(\omega_k-\mu)}-1)c^\dagger_k c_k]\\
&&\times e^{-\beta\sum_{j\neq k}(\omega_j-\mu) c^\dagger_j c_j}/ Z^\text{eq}_{E}\\
&=&c^\dagger_k e^{-\beta\sum_{j\neq k}(\omega_j-\mu) c^\dagger_j c_j}/ Z^\text{eq}_{E}\;\;,
\end{array}
\end{equation}
and
\begin{equation}
\begin{array}{lll}
\rho_E^\text{eq} c^\dagger_k&=&e^{-\beta(\omega_k-\mu) c^\dagger_k c_k} c^\dagger_k e^{-\beta\sum_{j\neq k}(\omega_j-\mu) c^\dagger_j c_j}/ Z^\text{eq}_{E}\\
&=&[1+(e^{-\beta(\omega_k-\mu)}-1)c^\dagger_k c_k]\\
&&\times c^\dagger_k e^{-\beta\sum_{j\neq k}(\omega_j-\mu) c^\dagger_j c_j}/ Z^\text{eq}_{E}\\
&=&e^{-\beta(\omega_k-\mu)} c^\dagger_k e^{-\beta\sum_{j\neq k}(\omega_j-\mu) c^\dagger_j c_j}/ Z^\text{eq}_{E}\;\;,
\end{array}
\end{equation}
so that
\begin{equation}
\label{eq:comm_b}
c^\dagger_k\rho_E^\text{eq}=e^{\beta(\omega_k-\mu)}\rho_E^\text{eq} c^\dagger_k\;\;.
\end{equation}
Together, Eq.~(\ref{eq:comm_a}) and Eq.~(\ref{eq:comm_b}) prove Eq.~(\ref{eq:comm}).
\subsection{Influence Superoperator}
In this section we explicitly derive the expression for the influence superoperator $\hhat{\mathcal{F}}(t)$ in Eq.~(\ref{eq:main_result}) and for the superoperator $\hhat{W}$ in Eq.~(\ref{eq:KernelW}).  We also provide a relation between the factorial and the double factorial which is used to re-sum the reduced Dyson series.\\

\subsubsection{Expression for the influence superoperator}
\label{sec:func_sup_app}
Given the arguments in the main text, from Eq.~(\ref{eq:temp_reduced_main}) we find that the reduced density matrix depends on the quantity
\begin{equation}
\hhat{T}_S\int_0^t dt_2 dt_1\hhat{W}(t_2,t_1)
\end{equation}
To proceed, note the following symmetry 
\begin{equation}
\label{eq:temp_symm_app}
\hhat{T}_S\hhat{W}(t_2,t_1)=\hhat{T}_S\hhat{W}(t_1,t_2)\;\;.
\end{equation}
In fact, 
\begin{equation}
\begin{array}{l}
\hhat{T}_SW(t_1,t_2)=\\
\displaystyle\sum_{q_1,q_2,\lambda_1,\lambda_2} C^{\lambda_2,\lambda_1}_{q_2,q_1}(t_1,t_2)\hhat{T}_S\hhat{S}_{q_2}^{\bar{\lambda}_2}(t_1)\hhat{S}_{q_1}^{\bar{\lambda}_1}(t_2)\\
=-\displaystyle\sum_{q_1,q_2,\lambda_1,\lambda_2} C^{\lambda_1,\lambda_2}_{q_1,q_2}(t_2,t_1)\hhat{T}_S\hhat{S}_{q_2}^{\bar{\lambda}_2}(t_1)\hhat{S}_{q_1}^{\bar{\lambda}_1}(t_2)\\
=\displaystyle\sum_{q_1,q_2,\lambda_1,\lambda_2} C^{\lambda_1,\lambda_2}_{q_1,q_2}(t_2,t_1)\hhat{T}_S\hhat{S}_{q_1}^{\bar{\lambda}_1}(t_2)\hhat{S}_{q_2}^{\bar{\lambda}_2}(t_1)\\
=\hhat{T}_SW(t_2,t_1)\;\;,
\end{array}
\end{equation}
where we used the fact that
\begin{equation}
\begin{array}{lll}
C^{\lambda_2,\lambda_1}_{q_2,q_1}(t_1,t_2)&=&\text{Tr}_E\left[\hhat{T}_E\hhat{B}^{ \lambda_2}_{q_2}(t_1)\hhat{B}^{ \lambda_1}_{q_1}(t_{2})[\rho_E(0)]\right]\\
&=&-\text{Tr}_E\left[\hhat{T}_E\hhat{B}^{ \lambda_1}_{q_1}(t_{2})\hhat{B}^{ \lambda_2}_{q_2}(t_1)[\rho_E(0)]\right]\\
&=&-C^{\lambda_1,\lambda_2}_{q_1,q_2}(t_2,t_1)\;\;.
\end{array}
\end{equation}
In turn, this means that 
\begin{equation}
\begin{array}{l}
\displaystyle\hhat{T}_S\int_0^t dt_2 dt_1\hhat{W}(t_2,t_1)\\
=\displaystyle\hhat{T}_S\int_0^t dt_2 \int_0^{t} dt_1[\theta(t_2-t_1)+\theta(t_1-t_2)]\hhat{W}(t_2,t_1)\\

=\displaystyle\left(\int_0^t dt_2 \int_0^{t_2} dt_1+\int_0^t dt_1 \int_{0}^{t_1} dt_2\right)\hhat{T}_S\hhat{W}(t_2,t_1)\\
=2\displaystyle\hhat{T}_S\int_0^t dt_2 \int_0^{t_2} dt_1\hhat{W}(t_2,t_1)\;\;,
\end{array}
\end{equation}
where, in the last step, we used Eq.~(\ref{eq:temp_symm_app}). The expression above allows to write Eq.~(\ref{eq:F_first}).
\subsubsection{Expression for the superoperator $\hhat{W}$}
\label{sec:W_app}
We start from the expression of the superoperator $\hhat{W}$ defined in Eq.~(\ref{eq:F}) which reads
\begin{equation}
\label{eq:KernelW_app}
W(t_2,t_1)=\sum_{q_1,q_2,\lambda_1,\lambda_2} C^{\lambda_2,\lambda_1}_{q_2,q_1}(t_2,t_1)\hhat{S}_{q_2}^{\bar{\lambda}_2}(t_2)\hhat{S}_{q_1}^{\bar{\lambda}_1}(t_1)\;\;,
\end{equation}
where the two-point correlations can be written as, see Eq.~(\ref{eq:two_point_correlations}),
\begin{equation}
C^{\lambda_2,\lambda_1}_{q_2,q_1}(t_2,t_1)=\text{Tr}_E\left[\hhat{T}_E\hhat{B}^{ \lambda_2}_{q_2}(t_2)\hhat{B}^{ \lambda_1}_{q_1}(t_{1})[\rho_E(0)]\right]\;\;,
\end{equation}
along with Eq.~(\ref{eq:PEB_main}) which describes the superoperators $\hhat{B}$ and $\hhat{S}$
\begin{equation}
\label{eq:PEB_app}
\begin{array}{lll}
\hhat{B}_{1}^{1}[\cdot]&=&B^\dagger [\cdot],\\
\hhat{B}_{1}^{-1}[\cdot]&=& B [\cdot],\\
\hhat{B}_{-1}^{1}[\cdot]&=&\hhat{P}_E[\cdot B^\dagger],\\
\hhat{B}_{-1}^{-1}[\cdot]&=&\hhat{P}_E[\cdot B],
\end{array}~
\begin{array}{lll}
\hhat{S}_{1}^{-1}[\cdot]&=&\hat{s} [\cdot]\\
\hhat{S}_{1}^{1}[\cdot]&=&-\hat{s}^\dagger [\cdot]\\
\hhat{S}_{-1}^{-1}[\cdot]&=&-\hhat{P}_S[\cdot\hat{s}] \\
\hhat{S}_{-1}^{1}[\cdot]&=&\hhat{P}_S[\cdot\hat{s}^\dagger], 
\end{array}
\end{equation}
Explicitly, the superoperators $\hhat{B}$ read
\begin{eqnarray}
\hhat{B}_{1}^{1}(t)[\cdot]&=&\sum_k g_k c^\dagger_k(t) [\cdot]\\
\hhat{B}_{1}^{-1}(t)[\cdot]&=&  \sum_k g_k c_k(t)  [\cdot]\\
\hhat{B}_{-1}^{1}(t)[\cdot]&=&\sum_k g_k\hhat{P}_E[[\cdot]c^\dagger_k(t)] \\
\hhat{B}_{-1}^{-1}(t)[\cdot]&=&\sum_k g_k c_k \hhat{P}_E[[\cdot] c_k(t)]\;\;,
\end{eqnarray}
Note that among the $16$ terms in Eq.~(\ref{eq:KernelW_app}), only $8$ are non-trivial. This is due to the fact that, since $\rho_E(0)$ is even, the extra-constraint $\delta_{\lambda_2,\bar{\lambda}_1}$ appears (indexes $\lambda=\pm 1$ correspond to creation/annihilation operators). The non-zero contributions are, for $t_2\geq t_1$,
\begin{widetext}
\begin{equation}
\label{eq:non-zero}
\resizebox{\textwidth}{!}{$\begin{array}{lll}
\text{Tr}_E \left(\hhat{B}^{  1}_{1}(t_2)\hhat{B}^{  -1}_{1}(t_1)\rho_E(0)\right) \hhat{S}^{  -1}_{1}(t_2)\hhat{S}^{  1}_{1}(t_1)[\cdot]&=&\text{Tr}_E \left(B^\dagger(t_2)B(t_1)[\rho_E(0)]\right)\hhat{S}^{  -1}_{1}(t_2)\hhat{S}^{  1}_{1}(t_1)[\cdot]\\
\text{Tr}_E \left(\hhat{B}^{  1}_{-1}(t_2)\hhat{B}^{  -1}_{1}(t_1)\rho_E(0)\right) \hhat{S}^{  -1}_{-1}(t_2)\hhat{S}^{  1}_{1}(t_1))[\cdot]&=&\text{Tr}_E \left(P_E B(t_1)[\rho_E(0)]B^\dagger(t_2)P_E\right)\hhat{S}^{  -1}_{-1}(t_2)\hhat{S}^{  1}_{1}(t_1))[\cdot]\\
\text{Tr}_E \left(\hhat{B}^{  -1}_{1}(t_2)\hhat{B}^{  1}_{1}(t_1)\rho_E(0)\right) \hhat{S}^{  1}_{1}(t_2)\hhat{S}^{  -1}_{1}(t_1)[\cdot]&=&\text{Tr}_E \left(B(t_2)B^\dagger(t_1)[\rho_E(0)]\right)\hhat{S}^{  1}_{1}(t_2)\hhat{S}^{  -1}_{1}(t_1)[\cdot]\\
\text{Tr}_E \left(\hhat{B}^{  -1}_{-1}(t_2)\hhat{B}^{  1}_{1}(t_1)\rho_E(0)\right) \hhat{S}^{  1}_{-1}(t_2)\hhat{S}^{  -1}_{1}(t_1)[\cdot]&=&\text{Tr}_E \left(P_E B^\dagger(t_1)[\rho_E(0)]B(t_2)P_E\right)\hhat{S}^{  1}_{-1}(t_2)\hhat{S}^{  -1}_{1}(t_1)[\cdot]\\
\\
\text{Tr}_E \left(\hhat{B}^{  1}_{1}(t_2)\hhat{B}^{  -1}_{-1}(t_1)\rho_E(0)\right) \hhat{S}^{  -1}_{1}(t_2)\hhat{S}^{  1}_{-1}(t_1)[\cdot]&=&\text{Tr}_E \left( B^\dagger(t_2)P_E[\rho_E(0)]B(t_1)P_E\right) \hhat{S}^{  -1}_{1}(t_2)\hhat{S}^{  1}_{-1}(t_1)[\cdot]\\
\text{Tr}_E \left(\hhat{B}^{  1}_{-1}(t_2)\hhat{B}^{  -1}_{-1}(t_1)\rho_E(0)\right) \hhat{S}^{  -1}_{-1}(t_2)\hhat{S}^{  1}_{-1}(t_1)[\cdot]&=&\text{Tr}_E \left( P_E P_E [\rho_E(0)] B(t_1)P_E B^\dagger (t_2)P_E\right)\hhat{S}^{  -1}_{-1}(t_2)\hhat{S}^{  1}_{-1}(t_1)[\cdot]\\
\text{Tr}_E \left(\hhat{B}^{  -1}_{1}(t_2)\hhat{B}^{  1}_{-1}(t_1)\rho_E(0)\right) \hhat{S}^{  1}_{1}(t_2)\hhat{S}^{  -1}_{-1}(t_1)[\cdot]&=&\text{Tr}_E \left( B(t_2)P_E [\rho_E(0)]  B^\dagger (t_1)P_E\right)\hhat{S}^{  1}_{1}(t_2)\hhat{S}^{  -1}_{-1}(t_1)[\cdot]\\
\text{Tr}_E \left(\hhat{B}^{  -1}_{-1}(t_2)\hhat{B}^{  1}_{-1}(t_1)\rho_E(0)\right) \hhat{S}^{  1}_{-1}(t_2)\hhat{S}^{  -1}_{-1}(t_1)[\cdot]&=&\text{Tr}_E \left( P_E P_E [\rho_E(0)] B^\dagger(t_1)P_E  B(t_2)P_E\right)\hhat{S}^{  1}_{-1}(t_2)\hhat{S}^{  -1}_{-1}(t_1)[\cdot]\;\;.
\end{array}$}
\end{equation}
\end{widetext}
Above, we kept the superoperators acting on the system as the time-ordering acts at the superoperator level. We can then write
\begin{equation}
\label{eq:W_even_sup}
\begin{array}{l}
W(t_2,t_1)=\\
C^{\sigma=1}(t_2,t_1)\left[\hhat{S}^{  -1}_{1}(t_2)\hhat{S}^{  1}_{1}(t_1)+\hhat{S}^{  -1}_{-1}(t_2)\hhat{S}^{  1}_{1}(t_1)\right]\\
+C^{\sigma=-1}(t_2,t_1)\left[\hhat{S}^{  1}_{1}(t_2)\hhat{S}^{  -1}_{1}(t_1)+\hhat{S}^{  1}_{-1}(t_2)\hhat{S}^{  -1}_{1}(t_1)\right]\\
+C^{\sigma=-1}(t_1,t_2)\left[-\hhat{S}^{  -1}_{1}(t_2)\hhat{S}^{  1}_{-1}(t_1)-\hhat{S}^{  -1}_{-1}(t_2)\hhat{S}^{  1}_{-1}(t_1)\right]\\
+C^{\sigma=1}(t_1,t_2)\left[-\hhat{S}^{  1}_{1}(t_2)\hhat{S}^{  -1}_{-1}(t_1)-\hhat{S}^{  1}_{-1}(t_2)\hhat{S}^{  -1}_{-1}(t_1)\right]\;\;,
\end{array}
\end{equation}
where we defined
\begin{equation}
\label{eq:corr_sigma_app}
\begin{array}{lll}
C^{\sigma=1}(t_2,t_1)&=&\text{Tr}_E\left[B^\dagger(t_2)B(t_1)\rho_E(0)\right]\\
C^{\sigma=-1}(t_2,t_1)&=&\text{Tr}_E\left[B(t_2)B^\dagger(t_1)\rho_E(0)\right]\;\;.
\end{array}
\end{equation}
We can group the terms to obtain Eq.~(\ref{eq:KernelW}) in the main text as
\begin{widetext}
\begin{equation}
\label{eq:W_app}
\begin{array}{l}
W(t_2,t_1)=\\
=\hhat{S}^{  -1}_{1}(t_2)\left[C^{\sigma=1}(t_2,t_1)\hhat{S}^{  1}_{1}(t_1)-C^{\sigma=-1}(t_1,t_2)\hhat{S}^{  1}_{-1}(t_1)\right]+\hhat{S}^{  -1}_{-1}(t_2)\left[C^{\sigma=1}(t_2,t_1)\hhat{S}^{  1}_{1}(t_1)-C^{\sigma=-1}(t_1,t_2)\hhat{S}^{  1}_{-1}(t_1)\right]\\
\phantom{=}+\hhat{S}^{  1}_{1}(t_2)\left[-C^{\sigma=1}(t_1,t_2)\hhat{S}^{  -1}_{-1}(t_1)+C^{\sigma=-1}(t_2,t_1)\hhat{S}^{  -1}_{1}(t_1)\right]+\hhat{S}^{  1}_{-1}(t_2)\left[-C^{\sigma=1}(t_1,t_2)\hhat{S}^{  -1}_{-1}(t_1)+C^{\sigma=-1}(t_2,t_1)\hhat{S}^{  -1}_{1}(t_1)\right]\\

=\left[\hhat{S}^{  -1}_{1}(t_2)+\hhat{S}^{  -1}_{-1}(t_2)\right]\left[C^{\sigma=1}(t_2,t_1)\hhat{S}^{  1}_{1}(t_1)-C^{\sigma=-1}(t_1,t_2)\hhat{S}^{  1}_{-1}(t_1)\right]\\
\phantom{=}-\left[\hhat{S}^{  1}_{-1}(t_2)+\hhat{S}^{  1}_{1}(t_2)\right]\left[C^{\sigma=1}(t_1,t_2)\hhat{S}^{  -1}_{-1}(t_1)-C^{\sigma=-1}(t_2,t_1)\hhat{S}^{  -1}_{1}(t_1)\right]\\

=\displaystyle\sum_{\sigma=\pm}A^\sigma(t_2) B^\sigma(t_2,t_1)\;\;.
\end{array}
\end{equation}
\end{widetext}
where 
\begin{widetext}
\begin{equation}
\begin{array}{lll}
A^\sigma(t)&=&\sigma\left(\hhat{S}^{  -\sigma}_{\sigma}(t)+\hhat{S}^{  -\sigma}_{-\sigma}(t)\right)=\displaystyle\hat{s}^{\bar{\sigma}}(t)[\cdot]-\hhat{P}_S[[\cdot]\hat{s}^{\bar{\sigma}}(t)]\\
B^{\sigma=1}(t_2,t_1)&=&C^{\sigma=1}(t_2,t_1)\hhat{S}^{  1}_{1}(t_1)-C^{\sigma=-1}(t_1,t_2)\hhat{S}^{  1}_{-1}(t_1)=-C^{\sigma=1}(t_2,t_1)\hat{s}^\dagger(t_1)[\cdot]-C^{\sigma=-1}(t_1,t_2) P_S[[\cdot]\hat{s}^\dagger(t_1)]\\
B^{\sigma=-1}(t_2,t_1)&=&C^{\sigma=1}(t_1,t_2)\hhat{S}^{  -1}_{-1}(t_1)-C^{\sigma=-1}(t_2,t_1)\hhat{S}^{  -1}_{1}(t_1)=-C^{\sigma=1}(t_1,t_2)P_S[[\cdot]\hat{s}(t_1)]-C^{\sigma=-1}(t_2,t_1)\hat{s}(t_1)[\cdot]\;\;,
\end{array}
\end{equation}
\end{widetext}
It is further possible to derive the more compact notation
\begin{equation}
\begin{array}{l}
B^{\sigma}(t_2,t_1)\\
=-\left(C^{\sigma}(t_2,t_1)\hat{s}^\sigma(t_1)[\cdot]+C^{\bar{\sigma}}(t_1,t_2) P_S[[\cdot]\hat{s}^\sigma(t_1)]\right)\\
=-\left(C^{\sigma}(t_2,t_1)\hat{s}^\sigma(t_1)[\cdot]+\bar{C}^{\bar{\sigma}}(t_2,t_1) P_S[[\cdot]\hat{s}^\sigma(t_1)]\right)\;\;,
\end{array}
\end{equation}
where we used Eq.~(\ref{eq:corr_id_1}).
\subsubsection{Proof of Eq.~(\ref{eq:W_widetext})}
\label{app:W_explicit}
The starting point of this section is Eq.~(\ref{eq:F}) which describes the influence superoperator as
\begin{equation}
\begin{array}{lll}
\displaystyle\hhat{\mathcal{F}}(t)&=&\displaystyle\int_0^t dt_{2}\int_0^{t_2} dt_{1} \hhat{W}(t_2,t_1)
\end{array}
\end{equation}
where
\begin{equation}
\begin{array}{lll}
 \hhat{W}(t_2,t_1)[\cdot]&=&\displaystyle\sum_{\sigma=\pm}\hhat{A}^\sigma(t_2)\hhat{B}^\sigma(t_2,t_1)[\cdot]\\
 
 &=&\displaystyle -{C}^{(1)}s_2 s_1^\dagger[\cdot]+{C}^{(1)}\hhat{P}_S[s_1^\dagger\cdot s_2]\\
 
&&\displaystyle -\bar{C}^{(-1)}s_2\hhat{P}_S[ \cdot s_1^\dagger]+\bar{C}^{(-1)}\hhat{P}_S[\hhat{P}_S[\cdot s_1^\dagger] s_2]\\

 &&\displaystyle -{C}^{(-1)}s^\dagger_2 s_1[\cdot]+{C}^{(-1)}\hhat{P}_S[s_1\cdot s^\dagger_2]\\
 
&&\displaystyle -\bar{C}^{(1)}s^\dagger_2\hhat{P}_S[ \cdot s_1]+\bar{C}^{(1)}\hhat{P}_S[\hhat{P}_S[\cdot s_1] s^\dagger_2]\;\;,

\end{array}
\end{equation}
where we used the short-hands $C^{(1)}=C^{\sigma=1}(t_2,t_1)$, $C^{(-1)}=C^{\sigma=-1}(t_2,t_1)$, $s^\sigma_1=s^\sigma(t_1)$, and $s^\sigma_2=s^\sigma(t_2)$. We now define $ \hhat{W}_\pm(t_2,t_1)$ as the composition of $\hhat{W}$ with the projectors $\hhat{P}^\text{e/o}_S$ onto the even/odd sector, i.e.,

\begin{equation}
    \begin{array}{lll}
    \hhat{W}_\pm(t_2,t_1)[\cdot]&=&\hhat{W}(t_2,t_1)[\hhat{P}^{\text{e/o}}_S[\cdot]]\\
    
     &=&\displaystyle -{C}^{(1)}s_2 s_1^\dagger\cdot\pm{C}^{(1)}s_1^\dagger\cdot s_2\\
 
&&\displaystyle \pm\bar{C}^{(-1)}s_2 \cdot s_1^\dagger-\bar{C}^{(-1)}\cdot s_1^\dagger s_2\\

 &&\displaystyle -{C}^{(-1)}s^\dagger_2 s_1\cdot\pm{C}^{(-1)}s_1\cdot s^\dagger_2\\
 
&&\displaystyle \pm\bar{C}^{(1)}s^\dagger_2 \cdot s_1-\bar{C}^{(1)}\cdot s_1 s^\dagger_2\\

&=&-C^{(1)}[s_2,s_1^\dagger\cdot]_{\mp}-\bar{C}^{(-1)}[\cdot s^\dagger_1,s_2]_\mp\\
&&-C^{(-1)}[s^\dagger_2,s_1\cdot]_\mp-\bar{C}^{(1)}[\cdot s_1,s_2^\dagger]_\mp
    \end{array}
\end{equation}
Using $C^\sigma(t_2,t_1)=\bar{C}^\sigma(t_1,t_2)$, we can write
\begin{equation}
    \begin{array}{lll}
    \hhat{W}_\pm(t_2,t_1)[\cdot]&=&\displaystyle

-\sum_{\sigma=\pm}C^{\sigma}(t_2,t_1)[s^{\bar{\sigma}}(t_2),s^\sigma(t_1)\cdot]_{\mp}\\

&&\displaystyle
-\sum_{\sigma=\pm}C^{\sigma}(t_1,t_2)[s^{\bar{\sigma}}(t_1),s^\sigma(t_2)\cdot]_{\mp},
    \end{array}
\end{equation}
which proves Eq.~(\ref{eq:W_widetext}) in the main text.
\subsubsection{A relation between factorial and double factorial}
\label{app:factorial}
The double factorial of an integer $n$ is defined as
\begin{equation}
\begin{array}{llll}
n!!&=&n (n-2)\cdots 2&\text{for~}n~\text{even}\\
n!!&=&n (n-2)\cdots 1&\text{for~}n~\text{odd}\;\;.
\end{array}
\end{equation}
We can see that if we multiply the double factorials of two consecutive numbers, we can ``fill the gaps'' with respect to the definition of factorial. Explicitly, 
\begin{equation}
\label{eq:double_factorial_0}
n!! (n-1)!!=n!\;\;.
\end{equation}
Another interesting connection between double and single factorial is
\begin{equation}
\label{eq:double_factorial_1}
(2n)!!=2n (2n-2)\cdots 2=2^n n!\;\;.
\end{equation}
Using Eq.~(\ref{eq:double_factorial_0}) and Eq.~(\ref{eq:double_factorial_1}) we have
\begin{equation}
(2n-1)!!=\displaystyle\frac{(2n)!}{(2n)!!}=\displaystyle\frac{(2n)!}{2^n n!}\;\;.
\end{equation}

\section{Applications}
Here, we provide details on the derivations of the results presented in Section \ref{sec:Applications}.
\subsection{Markovian regime}
\label{app:Markovian}
In section \ref{sec:Markovian} we analyzed the idealized  conditions under which the correlations characterizing the environment take the form in Eq.~(\ref{eq:corr_delta}), i.e.
\begin{equation}
\label{eq:corr_delta_app}
C^\sigma(t_2,t_1)= \Gamma^\sigma\delta(t_2-t_1)\;\;,
\end{equation}
where
\begin{equation}
\label{eq:a_sigma}
\Gamma^\sigma=\Gamma (1-\sigma +2\sigma n_0)\;\;.
\end{equation}
This Markovian regime leads to drastic simplifications in Eq.~(\ref{eq:main_result}). In fact, all superoperators present in $\hhat{W}$ are evaluated at the same point in time making the time-ordering procedure much easier to handle. Specifically, using Eq.~(\ref{eq:corr_delta_app}), into Eq.~(\ref{eq:F}), we have
\begin{widetext}
\begin{equation}
\label{eq:F_app}
\begin{array}{lll}
\hhat{\mathcal{F}}(t) [\cdot]&=&-\displaystyle\frac{1}{2}\int_0^t dt'\sum_\sigma\left(s^{\bar{\sigma}}(t')[\cdot]-\hhat{P}_S[[\cdot] s^{\bar{\sigma}}(t')]\right)\left(\Gamma^\sigma s^\sigma(t')[\cdot]+\bar{\Gamma}^{\bar{\sigma}}\hhat{P}_S[[\cdot] s^\sigma(t')]\right)\;\;,
\end{array}
\end{equation}
\end{widetext}
where we used $\int_0^{t_2} d t_1 \delta(t_2-t_1)=1/2$,(see Eq. 5.3.12 in \cite{Gardiner}). Using Eq.~(\ref{eq:der}), this also means that, in the Shroedinger picture,
\begin{equation}
\dot{\rho}^\text{Shr}_S(t)=-i[H_S,\rho^\text{Shr}_S(t)]+L[\rho^\text{Shr}_S(t)]\;\;,
\end{equation}
where  
\begin{equation}
L[\cdot]=U(t)\frac{d\hhat{\mathcal{F}}(t) }{dt}U^\dagger(t)[\cdot]\;\;,
\end{equation}
with $U(t)=\exp(-iH_S t)$. For clarity of notation, from now on we will omit the label ``$\text{Shr}$''. Using the definition of operators in the interaction frame, $s^\sigma(t)=U^\dagger(t)s^\sigma U(t)$ and taking the derivative of Eq.~(\ref{eq:F_app}), we find
\begin{equation}
\label{eq:temp_L}
\begin{array}{lll}
L[\cdot]&=&-\displaystyle(s\cdot-\hhat{P}_S[\cdot s])(\Gamma^{\sigma=1} s^\dagger\cdot+\bar{\Gamma}^{\sigma=-1}\hhat{P}_S[\cdot s^\dagger])/2\\
&&-\displaystyle(s^\dagger\cdot-\hhat{P}_S[\cdot s^\dagger])(\Gamma^{\sigma=-1} s\cdot+\bar{\Gamma}^{\sigma=1}\hhat{P}_S[\cdot s])/2.
\end{array}
\end{equation}
Note that  $L[\cdot]$ preserves the parity of its argument, i.e., it maps even (odd) operators into even (odd) operators. Using the decomposition
\begin{equation}
\rho_S(t)=\rho_S^\text{e}(t)+\rho_S^\text{o}(t)\;\;,
\end{equation}
we can write the action of the superoperators $\hhat{P}_S$ to write
\begin{equation}
\label{eq:lindblad_decomposition_app}
\dot{\rho}_S(t)=-i[H_S,\rho_S(t)]+L^\text{e}[\rho^\text{e}_S(t)]+L^\text{o}[\rho^\text{o}_S(t)]\;\;,
\end{equation}
where 
\begin{equation}
\rho_S^\text{e/o}=\hhat{P}^\text{e/o}[\rho_S]\;\;,
\end{equation}
in terms of
\begin{equation}
\begin{array}{lll}
\hhat{P}^\text{e}&=&P^\text{e}\cdot P^\text{e}+P^\text{o}\cdot P^\text{o}\\
\hhat{P}^\text{o}&=&P^\text{e}\cdot P^\text{o}+P^\text{o}\cdot P^\text{e}\;\;,
\end{array}
\end{equation}
with 
\begin{equation}
\begin{array}{lll}
P^\text{e}&=&(P_S+1)/2\\
P^\text{o}&=&(1-P_S)/2\;\;.
\end{array}
\end{equation}
The even/odd dissipators in Eq.~(\ref{eq:lindblad_decomposition_app}) are defined as,
\begin{equation}
\begin{array}{lll}
L^\text{e}[\cdot]&=&-\displaystyle(\Gamma^{\sigma=1}ss^\dagger\cdot+\bar{\Gamma}^{\sigma=-1}\cdot s^\dagger s\\
&&-\bar{\Gamma}^{\sigma=-1}s\cdot s^\dagger-\Gamma^{\sigma=1}s^\dagger\cdot s)/2\\
&&-\displaystyle(\Gamma^{\sigma=-1}s^\dagger s\cdot+\bar{\Gamma}^{\sigma=1}\cdot s s^\dagger\\
&&-\bar{\Gamma}^{\sigma=1}s^\dagger\cdot s-\Gamma^{\sigma=-1}s\cdot s^\dagger)/2\\
&=&\displaystyle\left(\Gamma^{\sigma=1}[2s^\dagger\cdot s-ss^\dagger\cdot-\cdot ss^\dagger]\right.\\
&&+\left.\Gamma^{\sigma=-1}[2s\cdot s^\dagger-s^\dagger s\cdot-\cdot s^\dagger s]\right)/2\;\;.
\end{array}
\end{equation}
When the argument is odd, terms involving one and only one $\hhat{P}_S$ change sign with respect to the even case. This leads to
\begin{equation}
\begin{array}{lll}
L^\text{o}[\cdot]
&=&\displaystyle\left(\Gamma^{\sigma=1}[-2s^\dagger\cdot s-ss^\dagger\cdot-\cdot ss^\dagger]\right.\\
&&+\left.\Gamma^{\sigma=-1}[-2s\cdot s^\dagger+s^\dagger s\cdot+\cdot s^\dagger s]\right)/2\;\;.
\end{array}
\end{equation}
Using Eq.~(\ref{eq:a_sigma}) into Eq.~(\ref{eq:lindblad_decomposition_app}) we obtain the following explicit Lindblad equation in the Shroedinger picture
\begin{equation}
\begin{array}{lll}
\dot{\rho}_S(t)
&=&-i[H_S,\rho_S(t)]\\
&&+\Gamma(1-n_0)D_s[\rho^\text{e}_S(t)]+\Gamma n_0 D_{s^\dagger}[\rho^\text{e}_S(t)]\\
&&+\Gamma(1-n_0)D'_s[\rho^\text{o}_S(t)]+\Gamma n_0 D'_{s^\dagger}[\rho^\text{o}_S(t)]\;\;,
\end{array}
\end{equation}
where $D_s[\cdot]=2 s[\cdot]s^\dagger-s^\dagger(t)s[\cdot]-[\cdot]s^\dagger s$,  $D_{s^\dagger}=2s^\dagger(t)[\cdot]s(t)-s s^\dagger[\cdot]-[\cdot]s s^\dagger$, $D'_s[\cdot]=-2 s[\cdot]s^\dagger-s^\dagger(t)s[\cdot]-[\cdot]s^\dagger s$,  $D'_{s^\dagger}=-2s^\dagger(t)[\cdot]s(t)-s s^\dagger[\cdot]-[\cdot]s s^\dagger$. In a more compact form, this equation becomes Eq.~(\ref{eq:Lindblad}).

\subsection{Hierarchical equations of motion}
Here, we provide the details of the derivation of the Hierarchical Equations of motion. 
\subsubsection{An expression for the influence superoperator}
\label{sec:expression_influence_HEOM_app}
Here, we explicitly derive Eq.~(\ref{eq:F_with_exp}), i.e., the expression for the influence superoperator when the correlations in Eq.~(\ref{eq:correlations_main}) are given by the ansatz in Eq.~(\ref{eq:ansats_correlations}). In fact, using such an ansatz, the superoperator $\hhat{W}$ in Eq.~(\ref{eq:KernelW}) reads
\begin{equation}
\begin{array}{lll}
\hhat{W}(t_2,t_1)[\cdot]&=&\displaystyle\sum_{\sigma}\hhat{A}^\sigma(t_2) \hhat{B}^\sigma(t_2,t_1)[\cdot]\\
&=&\displaystyle -\sum_{\sigma}\hhat{A}^\sigma(t_2)\left\{C^\sigma(t_2,t_1)\hat{s}^\sigma(t_1)[\cdot]\right.\\
&&\left.+\bar{C}^{\bar{\sigma}}(t_2,t_1)\hhat{P}_S[[\cdot] \hat{s}^\sigma(t_1)]\right\}\\

&=&\displaystyle -\sum_{n,\sigma}\hhat{A}^\sigma(t_2)\left\{a_n^\sigma e^{-b^\sigma_n(t_2,t_1)}\hat{s}^\sigma(t_1)[\cdot]\right.\\
&&\left.+\bar{a}_n^{\bar{\sigma}} e^{-\bar{b}^{\bar{\sigma}}_n(t_2,t_1)}\hhat{P}_S[[\cdot] \hat{s}^\sigma(t_1)]\right\}\\

&=&\displaystyle -\sum_{n,\sigma}\hhat{A}^\sigma(t_2)e^{-b^\sigma_n(t_2,t_1)}\\
&&\times\left\{a_n^\sigma \hat{s}^\sigma(t_1)[\cdot]+\bar{a}_n^{\bar{\sigma}} \hhat{P}_S[[\cdot] \hat{s}^\sigma(t_1)]\right\}\;\;,
\end{array}
\end{equation}
where in the last step we used the very convenient Eq.~(\ref{eq:Cbarbar_app}).
Using the definition in Eq.~(\ref{eq:AB}), i.e., 
\begin{equation}
\hhat{\mathcal{B}}^{\sigma}_n(t)[\cdot]=\displaystyle -\left(a_n^{\sigma}\hat{s}^\sigma(t)[\cdot]+\bar{a}_n^{\bar{\sigma}}P_S[[\cdot]\hat{s}^\sigma(t)]\right)\;\;,
\end{equation}
we can write
\begin{equation}
\begin{array}{lll}
\hhat{W}(t_2,t_1)[\cdot]&=&\displaystyle\sum_{n,\sigma}\hhat{A}^\sigma(t_2)e^{-b^\sigma_n(t_2,t_1)}\hhat{\mathcal{B}}^\sigma_n(t_1)\;\;,
\end{array}
\end{equation}
which, using Eq.~(\ref{eq:F}), immediately leads to Eq.~(\ref{eq:F_with_exp}) in the main text, i.e., 
\begin{equation}
\begin{array}{lll}
\hhat{\mathcal{F}}(t)&=& \displaystyle\int_0^t d t_2\int_0^{t_2} dt_1\sum_{n,\sigma}  A^\sigma(t_2) e^{-b_n^\sigma(t_2-t_1)}\hhat{\mathcal{B}}^\sigma_n(t_1)\;\;.
\end{array}
\end{equation}
\subsubsection{HEOM}
Here, we present all details to derive a generalized version of the HEOM valid in both even- and odd-parity sector which contains the usual expression for the HEOM in the even-parity sector.
The starting point is the expression for the $n$th auxiliary density matrix defined in Eq.~(\ref{eq:rho_n}) which, omitting the time-dependence for the density matrices, reads as
\begin{equation}
\label{eq:rho_n_def_app}
\rho^{(n)}_{j_n\cdots j_1}(t)=\alpha^n\hhat{T}_S \hhat{\Theta}_{j_n}(t)\cdots \hhat{\Theta}_{j_1}(t)\rho^{(0)}(t)\;\;,
\end{equation}
with $\rho^{(0)}(t)\equiv\rho_S(t)$. The superoperators $\hhat{\Theta}$ are defined as
\begin{equation}
\hhat{\Theta}_j(t)\equiv\hhat{\Theta}^\sigma_m(t)=\displaystyle\int_0^{t} d\tau e^{-b_j(t-\tau)}\hhat{\mathcal{B}}_j(\tau)\;\;,
\end{equation}
where we defined the multi-index $j=(m,\sigma)$ and, consistently $b_j\equiv b_m^\sigma$ and $\hhat{\mathcal{B}}_j\equiv\hhat{\mathcal{B}}^\sigma_m$. The derivative of $\hhat{\Theta}$ is given by
\begin{equation}
\label{eq:der_1_app}
\displaystyle\frac{d}{dt}{\hhat{\Theta}}_j(t)=\displaystyle -b_m^\sigma\hhat{\Theta}_j(t)+\hhat{\mathcal{B}}_j(\tau)\;\;,
\end{equation}
In order to compute the derivative of the previous auxiliary density matrixes, we further need the derivative of $\rho^{(0)}(t)={\rho}_S(t)$. Using, Eq.~(\ref{eq:der_int}) 
\begin{equation}
\label{eq:temp_ddtrho_s}
\begin{array}{lll}
\displaystyle\frac{d}{dt}{\rho}_S(t)&=&\displaystyle \hhat{T}_S\left(\frac{d}{dt}\hhat{\mathcal{F}}(t) \right){\rho}_S(t)\;\;.
\end{array}
\end{equation}
From Eq.~(\ref{eq:F_with_exp}), the time-derivative of $\hhat{\mathcal{F}}(t) $ is simply given by
\begin{equation}
\begin{array}{lll}
\displaystyle\frac{d}{dt}{\hhat{\mathcal{F}}}(t)&=&\displaystyle\sum_{n,\sigma}  \hhat{A}^\sigma(t)\int_0^{t} d\tau e^{-b_m^\sigma(t-\tau)}\hhat{\mathcal{B}}^\sigma_m(\tau)\\
&=&\displaystyle\sum_{m,\sigma}  \hhat{A}^\sigma(t)\hhat{\Theta}^\sigma_n(t)\\
&\equiv&\displaystyle\sum_{j}  \hhat{A}^j(t)\hhat{\Theta}_j(t)\;\;,
\end{array}
\end{equation}
where $\hhat{A}^j(t)\equiv\hhat{A}^\sigma(t)$ which, redundantly, makes $\hhat{A}$ also a (trivial) function of $n$. Inserting the equation above in Eq.~(\ref{eq:temp_ddtrho_s}), we find
\begin{align*}
\displaystyle\frac{d}{dt}{\rho}_S(t)&=&\displaystyle \hhat{T}_S\sum_{m,\sigma}  \hhat{A}^\sigma(t)\hhat{\Theta}^\sigma_m(t){\rho}_S(t)\\
&=&\displaystyle \sum_\sigma\hhat{A}^\sigma(t)\hhat{T}_S\sum_{m}  \hhat{\Theta}^\sigma_m(t){\rho}_S(t)\\
&\equiv&\displaystyle \sum_j\hhat{A}^j(t)\hhat{T}_S \hhat{\Theta}_j(t){\rho}_S(t)\;\;.\numberthis \label{eq:der_2_app}
\end{align*}
Using Eq.~(\ref{eq:der_1_app}) and Eq.~(\ref{eq:der_2_app}) we can write the derivative of the auxiliary density matrices in Eq.~(\ref{eq:rho_n_def_app}) as
\begin{widetext}
\begin{equation}
\label{eq:rho_n_app}
\begin{array}{lll}
\displaystyle\dot{\rho}^{(n)}_{j_n\cdots j_1}(t)&=&\displaystyle\alpha^n\hhat{T}_S \sum_{k=1}^n\hhat{\Theta}_{j_n}(t)\cdots\left[-b_{j_k}\hhat{\Theta}_{j_k}(t)+\hhat{\mathcal{B}}_{j_k}(t)\right]\cdots \hhat{\Theta}_{j_1}(t)\rho^{(0)}(t)\\
&&+\displaystyle\alpha^n\hhat{T}_S \hhat{\Theta}_{j_n}(t)\cdots \hhat{\Theta}_{j_1}(t)\sum_{j_{n+1}}\hhat{A}^{j_{n+1}}(t)\hhat{\Theta}_{j_{n+1}}(t)\rho^{(0)}(t)\\
&=&\displaystyle\sum_{k=1}^n(-b_{j_k})\rho^{(n)}_{j_n\cdots j_1}+\alpha\sum_{k=1}^n (-1)^{n-k}\hhat{\mathcal{B}}_{j_k}(t)\rho^{(n-1)}_{j_n\cdots j_{k+1}j_{k-1}\cdots j_1}+\displaystyle\alpha^{-1}\sum_{j_{n+1}}\hhat{A}^{j_{n+1}}(t)\rho^{(n+1)}_{j_{n+1}\cdots j_1}\;\;,
\end{array}
\end{equation}
\end{widetext}
where the superoperators $\hhat{A}$ and $\hhat{\mathcal{B}}$ are given by Eq.~(\ref{eq:A_and_B}) and Eq.~(\ref{eq:AB}), i.e., 
\begin{equation}
\label{eq:def_app}
\begin{array}{lll}
\hhat{A}^j(t)\equiv\hhat{A}^\sigma(t)&=&\displaystyle\hat{s}^{\bar{\sigma}}(t)[\cdot]-\hhat{P}_S[[\cdot]\hat{s}^{\bar{\sigma}}(t)]\\
\hhat{\mathcal{B}}_j(t)\equiv\hhat{\mathcal{B}}^{\sigma}_m(t)&=&\displaystyle -\left(a_m^{\sigma}\hat{s}^\sigma(t)[\cdot]+\bar{a}_m^{\bar{\sigma}}P_S[[\cdot]\hat{s}^\sigma(t)]\right).
\end{array}
\end{equation}
In the last step of Eq.~(\ref{eq:rho_n_app}), we accounted for the minus signs originating when moving the superoperators $\hhat{\mathcal{B}}_{j_k}(t)$  on the very left (a sign appears each time $\hhat{B}$ moves across a $\hhat{\Theta}$). On the contrary, signs appearing when moving the superoperators $\hhat{A}^{j_{n+1}}(t)$ on the very left are always compensated by the ones appearing when moving $\hhat{\Theta}_{j_{n+1}}(t)$ which also needs to be brought on the left in order to be able to use Eq.~(\ref{eq:rho_n_def_app}).

We can now go back to the Shroedinger picture by multiplying each iteration of the HEOM by $U\cdot U^\dagger$ where $U=\exp(-iH_St)$ is the free evolution of the system. Using $U d/dt(\hat{O}) U^\dagger =d/dt(U \hat{O} U^\dagger)-\mathcal{L}\hat{O}$, where $\hhat{\mathcal{L}}=-i[H_S,\cdot]$ we derive the  generalized Hierarchical Equations of Motion
\begin{equation}
\label{eq:HEOMalpha}
\begin{array}{lll}
\displaystyle\dot{\rho}^{\text{Schr},(n)}_{j_n\cdots j_1}
&=&\displaystyle(\hhat{\mathcal{L}}-\sum_{k=1}^nb_{j_k})\rho^{\text{Schr},(n)}_{j_n\cdots j_1}\\
&&\displaystyle +\alpha\sum_{k=1}^n (-1)^{n-k}\hhat{\mathcal{B}}^{j_k}\rho^{\text{Schr},(n-1)}_{j_n\cdots j_{k+1}j_{k-1}\cdots j_1}\\
&&+\displaystyle\alpha^{-1}\sum_{j_{n+1}}\hhat{A}^{j_{n+1}}\rho^{\text{Schr},(n+1)}_{j_{n+1}\cdots j_1}\;\;.
\end{array}
\end{equation}
Here, the adjective ``generalized'' is motivated by the fact that the previous expression can be applied to both even- and odd-parity sectors.
If we now assume $\rho^{(0)}(t)$ to be have a definite parity symmetry, then the parity superoperators inside the definitions in Eq.~(\ref{eq:def_app}) translate into signs dependent on the iteration index $n$. For example, assuming $\rho^{(0)}(t)$ to be physical, hence even, when $A^{\sigma}$ acts on $\rho^{(0)}$, the parity operator adds a minus sign (note that $\hhat{P}_S$ acts on the density matrix multiplied by the odd operator $\hat{s}$), while  when it acts on $\rho^{(n)}$, the parity operator is trivial. We can then write, omitting the label $\text{Schr}$,
\begin{eqnarray*}{lll}
\displaystyle\dot{\rho}^{(n)}_{j_n\cdots j_1}
&=&\displaystyle(\hhat{\mathcal{L}}-\sum_{k=1}^nb_{j_k})\rho^{(n)}_{j_n\cdots j_1}\\
&&\displaystyle -\alpha\sum_{k=1}^n (-1)^{n-k}\hhat{\mathcal{C}}_n^{j_k}\rho^{(n-1)}_{j_n\cdots j_{k+1}j_{k-1}\cdots j_1}\\
&&+\displaystyle\alpha^{-1}\sum_{j_{n+1}}\hhat{\mathcal{A}}_n^{\sigma_{n+1}}\rho^{(n+1)}_{j_{n+1}\cdots j_1}\;\;,
\end{eqnarray*}
where 
\begin{equation}
\label{eq:HEOM_minus_1}
\begin{array}{lll}
\hhat{\mathcal{A}}_n^j[\cdot]&\equiv&\displaystyle\hat{s}^{\bar{\sigma}}[\cdot]+(-1)^n[\cdot]\hat{s}^{\bar{\sigma}}\\
\hhat{\mathcal{C}}_n^j[\cdot]&\equiv&\displaystyle a_n^{\sigma}\hat{s}^\sigma[\cdot]-(-1)^n\bar{a}_n^{\bar{\sigma}}[\cdot]\hat{s}^\sigma\;\;,
\end{array}
\end{equation}
where the notation is slightly redundant as it implies a trivial dependence of $\hhat{\mathcal{A}}$ on the index $m$, originating from the expansion of the correlation in Eq.~(\ref{eq:ansats_correlations}).
For the specific choice $\alpha=i$, we obtain 
\begin{equation}
\label{eq:rho_n_simplified_app}
\begin{array}{lll}
\displaystyle\dot{\rho}^{(n)}_{j_n\cdots j_1}
&=&\displaystyle(\hhat{\mathcal{L}}-\sum_{k=1}^nb_{j_k})\rho^n_{j_n\cdots j_1}- i\sum_{j_{n+1}}\hhat{\mathcal{A}}_n^{j_{n+1}}\rho^{(n+1)}_{j_{n+1}\cdots j_1}\\
&&\displaystyle-i\sum_{k=1}^n (-1)^{n-k}\hhat{\mathcal{C}}_n^{j_k}\rho^{(n-1)}_{j_n\cdots j_{k+1}j_{k-1}\cdots j_1}\;\;,
\end{array}
\end{equation}
which represents one of the \emph{standard expressions for the HEOM}, see, for example, Eq.~(38) in \cite{Lambert_Bofin}. In the Appendix \ref{sec:order2} we give an explicit derivation of these equations up to order 2.\\
\subsubsection{Explicit calculation up to order 2}
\label{sec:order2}
Here, we  more explicitly compute the HEOM up to order 2.
We start by taking the derivative of the quantity in Eq.~(\ref{eq:HEOM_00}), which is done using Eq.~(\ref{eq:HEOM_ansats}) and Eq.~(\ref{eq:motion}) to obtain (omitting the time-dependence for the density matrices)
\begin{widetext}
\begin{equation}
\label{eq:HEOM_1}
\begin{array}{lll}
\displaystyle\dot{\rho}_{m_1}^{\sigma_1}&=&\displaystyle\frac{d}{dt}\left[\alpha \hhat{T}_S  \Theta^{\sigma_1}_{m_1} (t)\rho_S\right]\\
&=&\displaystyle\alpha \hhat{T}_S\left[-b_{m_1}^{\sigma_1}\Theta_{m_1}^{\sigma_1}(t)+B_{m_1}^{\sigma_1}(t)\right]\rho_S+\alpha \hhat{T}_S \Theta_{m_1}^{\sigma_1}(t) \alpha^{-1}\sum_{m_2,\sigma_2}A^{\sigma_2}(t)\rho_{m_2}^{\sigma_2}\\
&=&\displaystyle -b_{m_1}^{\sigma_1}\rho_{m_1}^{\sigma_1}+\alpha B_{m_1}^{\sigma_1}(t)\rho_S+ \hhat{T}_S \Theta_{m_1}^{\sigma_1}(t) \sum_{m_2,\sigma_2}A^{\sigma_2}(t)\alpha \hhat{T}_S \Theta_{m_2}^{\sigma_2}\rho_S\\
&=&\displaystyle -b_{m_1}^{\sigma_1}\rho_{m_1}^{\sigma_1}+\alpha B_{m_1}^{\sigma_1}(t)\rho_S + \alpha^{-1} \sum_{m_2,\sigma_2}A^{\sigma_2}(t)\rho_{m_2,m_2}^{\sigma_2,\sigma_1}\;\;,
\end{array}
\end{equation}
\end{widetext}
where we defined
\begin{equation}
\label{eq:HEOM_11}
\begin{array}{lll}
\displaystyle\rho_{m_2,m_2}^{\sigma_2,\sigma_1}&=&\displaystyle\alpha^2 \hhat{T}_S \Theta_{m_2}^{\sigma_2}\Theta_{m_2}^{\sigma_2}(t) \rho_S\;\;.
\end{array}
\end{equation}
We moved the operator $\Theta_{m_2}^{\sigma_2}$ across $A^{\sigma_2}$ and $\Theta_{m_1}^{\sigma_1}$ resulting in a $+$ sign. We then further moved $A^{\sigma_2}$ across the two $\Theta$ operators, again resulting in a $+$ sign. Similarly, we can proceed onto the next order to obtain
\begin{widetext}
\begin{equation}
\begin{array}{lll}
\displaystyle\dot{\rho}_{m_2,m_1}^{\sigma_2,\sigma_1}&=&\displaystyle \alpha^2\frac{d}{dt}\left[\hhat{T}_S  \Theta^{\sigma_2}_{m_2} (t)\Theta^{\sigma_1}_{m_1} (t)\rho_S\right]\\
&=&\displaystyle\alpha^2 \hhat{T}_S\left[-b_{m_2}^{\sigma_2}\Theta_{m_2}^{\sigma_2}(t)+B_{m_2}^{\sigma_2}(t)\right]\Theta^{\sigma_1}_{m_1} (t)\rho_S+\alpha^2 \hhat{T}_S\Theta_{m_2}^{\sigma_2}(t)\left[-b_{m_1}^{\sigma_1}\Theta_{m_1}^{\sigma_1}(t)+B_{m_1}^{\sigma_1}(t)\right]\rho_S\\
&&+\displaystyle\alpha^2 \hhat{T}_S \Theta_{m_2}^{\sigma_2}(t)  \Theta_{m_1}^{\sigma_1}(t) \alpha^{-1}\sum_{n_3,\sigma_3}A^{\sigma_3}(t)\rho_{n_3}^{\sigma_3}\\
&=&\displaystyle -b_{m_2}^{\sigma_2}\rho_{m_2,m_1}^{\sigma_2,\sigma_1}+\alpha B_{m_2}^{\sigma_2}(t)\rho_{m_1}^{\sigma_1} -b_{m_1}^{\sigma_1}\rho_{m_2,m_1}^{\sigma_2,\sigma_1}-\alpha B_{m_1}^{\sigma_1}(t)\rho_{m_2}^{\sigma_2}\\
&&+\displaystyle\alpha^2 \hhat{T}_S \Theta_{m_2}^{\sigma_2}(t)  \Theta_{m_1}^{\sigma_1}(t) \alpha^{-1}\sum_{n_3,\sigma_3}A^{\sigma_3}(t)\alpha \hhat{T}_S\Theta_{m_2}^{\sigma_3}(t)\rho_S\\

&=&\displaystyle -b_{m_2}^{\sigma_2}\rho_{m_2,m_1}^{\sigma_2,\sigma_1}+\alpha B_{m_2}^{\sigma_2}(t)\rho_{m_1}^{\sigma_1} -b_{m_1}^{\sigma_1}\rho_{m_2,m_1}^{\sigma_2,\sigma_1}-\alpha B_{m_1}^{\sigma_1}(t)\rho_{m_2}^{\sigma_2}+\displaystyle\alpha^2\sum_{n_3,\sigma_3}A^{\sigma_3}(t) \hhat{T}_S \Theta_{m_2}^{\sigma_3}(t)\Theta_{m_2}^{\sigma_2}(t)  \Theta_{m_1}^{\sigma_1}(t)\rho_S\\

&=&\displaystyle -b_{m_2}^{\sigma_2}\rho_{m_2,m_1}^{\sigma_2,\sigma_1}+\alpha B_{m_2}^{\sigma_2}(t)\rho_{m_1}^{\sigma_1}-b_{m_1}^{\sigma_1}\rho_{m_2,m_1}^{\sigma_2,\sigma_1}-\alpha B_{m_1}^{\sigma_1}(t)\rho_{m_2}^{\sigma_2}+\displaystyle\alpha^{-1}\sum_{n_3,\sigma_3}A^{\sigma_3}(t) \rho_{n_3,m_2,m_1}^{\sigma_3,\sigma_2,\sigma_1}\\

&=&\displaystyle (-b_{m_1}^{\sigma_1}-b_{m_2}^{\sigma_2})\rho_{m_2,m_1}^{\sigma_2,\sigma_1}+\alpha\sum_{j=1}^2(-1)^{2-j}B^{\sigma_j}_{n_j}(t)\rho_{n_{3-j}}^{\sigma_{3-j}} +\alpha^{-1}\sum_{m_2,\sigma_3}A^{\sigma_3}(t)\rho_{m_3,m_2,m_1}^{\sigma_3,\sigma_2,\sigma_1}\;\;,
\end{array}
\end{equation}
\end{widetext}
where
\begin{equation}
\rho_{n_3,m_2,m_1}^{\sigma_3,\sigma_2,\sigma_1}(t)=\alpha^3\hhat{T}_S \Theta_{m_2}^{\sigma_3}(t)\Theta_{m_2}^{\sigma_2}(t)  \Theta_{m_1}^{\sigma_1}(t)\rho\;\;.
\end{equation}
Minus signs appear when the operators $\Theta$ cross each other or other operators. 
\subsection{Computing system correlation functions}
\label{app:computing_correlations}
Here, we show how the correlations at thermal equilibrium in Eq.~(\ref{eq:corr_system_gen}), i.e., 
\begin{equation}
\label{C_XY_app}
\begin{array}{lll}
C^{\text{th}}_{XY}(t)&=&\text{Tr}_{SE}[X_S(0) U(t_2-t_1) Y_S(0)\rho^{\text{th}}U^\dagger(t_2-t_1)]
\end{array}
\end{equation}
can be computed using the HEOM in Eq.~(\ref{eq:generalized_HEOM}). Our starting point is Eq.~(\ref{eq:corr_system_gen_2}) which writes the correlation as
\begin{equation}
\begin{array}{lll}
C^{\text{th}}_{XY}(t)&=&\text{Tr}_{SE}[X_S(0) \rho^Y(t)]\;\;,
\end{array}
\end{equation}
where
\begin{equation}
\rho^Y(t)=U(t) Y_S(0) U(T)\rho(-T)U^\dagger(T)U^\dagger(t)\;\;,
\end{equation}
in terms of a separable state $\rho(-T)$ and a time $T$ such that $\rho^\text{th}=U(T)\rho(-T)U^\dagger(T)$.

To start, using the same definitions which lead to Eq.~(\ref{eq:series_0}), we can write
\begin{equation}
\label{eq:Y}
\rho^Y(t)=\displaystyle\sum_{n=0}^\infty\frac{(-i)^n}{n!}\hhat{T}^\text{b}\hhat{Y}_S(0)\int_{-T}^t \left[\prod_{i=1}^n d t_i\hhat{H}^\times_{I}(t_i)\right]\rho(-T)\;\;,
\end{equation}
where we defined $\hhat{Y}_S(0)[\cdot]=Y_S(0)[\cdot]$ as the superoperator version of $Y_S(0)$ and used the (Bosonic) time-ordering to reposition it outside the integral. By using the decomposition in Eq.~(\ref{eq:simple_but_powerful}), we can write  $\hhat{Y}_S(0)[\cdot]=\hhat{Y}^\text{e}_S(0)+\hhat{Y}^\text{o}_S(0)$ where $\hhat{Y}^\text{e}_S(0)[\cdot]=\hat{Y}^\text{e}_S(0)[\cdot]$ and $\hhat{Y}^\text{o}_S(0)[\cdot]=\hhat{P}'_E(0) \hat{Y}^\text{o}_S(0)[\cdot]$ with $\hhat{P}'_E[\cdot]=P_E[\cdot]$. 

Due to the presence of $\hhat{Y}_S(0)$, in order to make progress in evaluating Eq.~(\ref{eq:Y}), we need to adapt the reasoning done to deduce Eq.~(\ref{eq:useful_for_corr}) from Eq.~(\ref{eq:Reduced_appendix}). We can write
\begin{widetext}
\begin{equation}
\begin{array}{lll}
\rho^Y(t)&=&\displaystyle\sum_{n=0}^\infty\frac{(-i)^n}{n!}\int_0^t \left(\prod_{i=1}^nd t_i\right)\sum_{q_n,\lambda_n,\cdots,q_1,\lambda_1}\\
&&\left\{\left[\hhat{T}_E\hhat{B}^{\prime \lambda_n}_{q_n}(t_n)\cdots\hhat{B}^{\prime \lambda_1}_{q_1}(t_1)[\rho^\text{eq}_E]\right]\hhat{T}_S\left[\hhat{Y}^\text{e}_S(0)\hhat{S}_{q_n}^{\bar{\lambda}_n}(t_n)\cdots\hhat{S}_{q_1}^{\bar{\lambda}_1}(t_1)\right][\hat{\rho}^\text{e}_S(-T)]\right.\\

&&+\displaystyle\left[\hhat{T}_E\hhat{P}'_E(0)\hhat{B}^{\prime \lambda_n}_{q_n}(t_n)\cdots\hhat{B}^{\prime \lambda_1}_{q_1}(t_1)[\rho^\text{eq}_E]\right]\hhat{T}_S\left[\hhat{Y}^\text{o}_S(0)\hhat{S}_{q_n}^{\prime\bar{\lambda}_n}(t_n)\cdots\hhat{S}_{q_1}^{\prime\bar{\lambda}_1}(t_1)\right][\hat{\rho}^\text{e}_S(-T)]\\

&&+\left.\displaystyle \left[\hhat{T}_E \hhat{B}^{\prime \lambda_n}_{q_n}(t_n)\cdots\hhat{B}^{\prime \lambda_1}_{q_1}(t_1)[\rho^\text{eq}_EP_E]\right]\hhat{T}_S\left[\hhat{Y}^\text{e}_S(0)\hhat{S}_{q_n}^{\prime\bar{\lambda}_n}(t_n)\cdots\hhat{S}_{q_1}^{\prime\bar{\lambda}_1}(t_1)\right][\hat{\rho}^\text{o}_S(-T)]\right.\\

&&+\left.\displaystyle\left[\hhat{T}_E \hhat{P}'_E(0)\hhat{B}^{\prime \lambda_n}_{q_n}(t_n)\cdots\hhat{B}^{\prime \lambda_1}_{q_1}(t_1)[\rho^\text{eq}_EP_E]\right]\hhat{T}_S\left[\hhat{Y}^\text{o}_S(0)\hhat{S}_{q_n}^{ \bar{\lambda}_n}(t_n)\cdots\hhat{S}_{q_1}^{ \bar{\lambda}_1}(t_1)\right][\hat{\rho}^\text{o}_S(-T)]\right\}.
\end{array}
\end{equation}
\end{widetext}
Here, it is important to keep the time dependence for all superoperators (including $\hhat{P}'_E(0)$) to allow for the action of time-ordering. As we defined in section \ref{sec:Dyson}, the time-orderings $\hhat{T}_S$ and $\hhat{T}_B$ are Fermionic when acting on the fields $\hhat{B}'$ and $\hhat{S}$. This definition is possible because the ordering of the fields $\hhat{B}$ inside $\hhat{T}_B$ mirrors that of the fields $\hhat{S}$ inside $\hhat{T}_S$. On the other hand, the current situation involving the operator $\hhat{Y}_S$ is not as symmetrical. For this reason, we consider the field $\hhat{Y}_S$ and $\hhat{P}'_E$ to be commuted under the action of $\hhat{T}_S$ and $\hhat{T}_E$ other than anticommuted.

We are now ready to take  the partial trace which leads to
\begin{widetext}
\setlength{\mathindent}{0pt}{\begin{equation}\label{eq:sbirula}\begin{array}{lll}
\rho^Y_S(t)&=&\displaystyle\sum_{n=\text{even}}\frac{(-i)^n}{n!}\int_0^t \left(\prod_{i=1}^n d t_i\right)\sum_{q_n,\lambda_n\cdots q_1,\lambda_1}\\
&&\left\{C^{\prime \lambda_n\cdots \lambda_1}_{q_n\cdots q_1}\hat{T}_S\left[\hhat{Y}^\text{e}_S(0)\hhat{S}_{q_n}^{\bar{\lambda}_n}\cdots\hhat{S}_{q_1}^{\bar{\lambda}_1}\right]\hat{\rho}^\text{e}_S(0)
\displaystyle+C^{\prime\prime \lambda_n\cdots \lambda_1}_{q_n\cdots q_1}\hat{T}_S\left[\hhat{Y}^\text{o}_S(0)\hhat{S}_{q_n}^{\prime\bar{\lambda}_n}\cdots\hhat{S}_{q_1}^{\prime\bar{\lambda}_1}\right]\hat{\rho}^\text{e}_S(0)\right.\\
&&+\left. D^{\prime \lambda_n\cdots \lambda_1}_{q_n\cdots q_1}\hat{T}_S\left[\hhat{Y}^\text{e}_S(0)\hhat{S}_{q_n}^{\prime \bar{\lambda}_n}\cdots\hhat{S}_{q_1}^{\prime \bar{\lambda}_1}\right]\hat{\rho}^\text{o}_S(0)
\displaystyle+D^{\prime\prime \lambda_n\cdots \lambda_1}_{q_n\cdots q_1}\hat{T}_S\left[\hhat{Y}^\text{o}_S(0)\hhat{S}_{q_n}^{ \bar{\lambda}_n}\cdots\hhat{S}_{q_1}^{ \bar{\lambda}_1}\right]\hat{\rho}^\text{o}_S(0)\right\}\;\;,
\end{array}
\end{equation}}
\end{widetext}
 where, since the operator $\hhat{Y}^\text{o}_S(0)$ changes the parity of the state, the correlations take the form
 \newpage
 \begin{equation}
 \label{eq:corr_app}
\begin{array}{lll}
 C^{\prime \lambda_n\cdots \lambda_1}_{q_n\cdots q_1}&=&\text{Tr}_E\hat{T}_E\left[\hhat{B}^{\prime \lambda_n}_{q_n}\cdots\hhat{B}^{\prime \lambda_1}_{q_1}\right][\rho^\text{eq}_E]\\
  C^{\prime\prime \lambda_n\cdots \lambda_1}_{q_n\cdots q_1}&=&\text{Tr}_E\hat{T}_E\left[P_E\hhat{P}'_E(0)\hhat{B}^{\prime \lambda_n}_{q_n}\cdots\hhat{B}^{\prime \lambda_1}_{q_1}\right][\rho^\text{eq}_E]\\
 D^{\prime \lambda_n\cdots \lambda_1}_{q_n\cdots q_1}&=& \text{Tr}_E\hat{T}_E\left[P_E\hhat{B}^{\prime \lambda_n}_{q_n}\cdots\hhat{B}^{\prime \lambda_1}_{q_1}\right][\rho^\text{eq}_EP_E]\\
  D^{\prime\prime \lambda_n\cdots \lambda_1}_{q_n\cdots q_1}&=& \text{Tr}_E\hat{T}_E\left[\hhat{P}'_E(0)\hhat{B}^{\prime \lambda_n}_{q_n}\cdots\hhat{B}^{\prime \lambda_1}_{q_1}\right][\rho^\text{eq}_E P_E]\;\;.
\end{array}
 \end{equation}
Now, we notice that the difference between the fields $\hhat{S}$ and $\hhat{S}'$ is, ultimately, just a sign when the down-indexes are negative, see Eq.~(\ref{eq:SS_op_1}) and Eq.~(\ref{eq:SS_op_2}). The same sign can be implemented in the bath correlations by adding two extra $P_E$, i.e., we can consider Eq.~(\ref{eq:sbirula}) with the substitutions $\hhat{S}'^\lambda_q\rightarrow\hhat{S}^\lambda_q$ and

 \begin{equation}
 \label{eq:corr_app_2}
\begin{array}{lll}
 {C}^{\prime\prime \lambda_n\cdots \lambda_1}_{q_n\cdots q_1} &\rightarrow& \text{Tr}_E\hat{T}_E\left[P^2_E\hhat{P}'_E(0)\hhat{B}^{\prime \lambda_n}_{q_n}\cdots\hhat{B}^{\prime \lambda_1}_{q_1}\right][\rho^\text{eq}_EP_E ]\\
 &&= {D}^{\prime\prime \lambda_n\cdots \lambda_1}_{q_n\cdots q_1}\\
  {D}^{\prime \lambda_n\cdots \lambda_1}_{q_n\cdots q_1} &\rightarrow& \text{Tr}_E\hat{T}_E\left[P^2_E\hhat{B}^{\prime \lambda_n}_{q_n}\cdots\hhat{B}^{\prime \lambda_1}_{q_1}\right][\rho^\text{eq}_E P^2_E]\\
&&= {C}^{\prime \lambda_n\cdots \lambda_1}_{q_n\cdots q_1}\;\;,
\end{array}
 \end{equation}
 which leads to
 \begin{widetext}
\setlength{\mathindent}{0pt}{\begin{equation}\label{eq:sbirula}\begin{array}{lll}
\rho^Y_S(t)&=&\displaystyle\sum_{n=\text{even}}\frac{(-i)^n}{n!}\int_0^t \left(\prod_{i=1}^n d t_i\right)\sum_{q_n,\lambda_n\cdots q_1,\lambda_1}\\
&&\left\{C^{\prime \lambda_n\cdots \lambda_1}_{q_n\cdots q_1}\hat{T}_S\left[\hhat{Y}^\text{e}_S(0)\hhat{S}_{q_n}^{\bar{\lambda}_n}\cdots\hhat{S}_{q_1}^{\bar{\lambda}_1}\right](\hat{\rho}^\text{e}_S(0)+\hat{\rho}^\text{o}_S(0))
\displaystyle+D^{\prime\prime \lambda_n\cdots \lambda_1}_{q_n\cdots q_1}\hat{T}_S\left[\hhat{Y}^\text{o}_S(0)\hhat{S}_{q_n}^{\bar{\lambda}_n}\cdots\hhat{S}_{q_1}^{\bar{\lambda}_1}\right](\hat{\rho}^\text{e}_S(0)+\hat{\rho}^\text{o}_S(0))\right\}\;\;.
\end{array}
\end{equation}}
\end{widetext}
It is not possible to further reduce $D''$ because of the presence of $\hhat{P}'_E(0)$ whose action adds a sign corresponding to  the number of times the fields $\hhat{B}^{\lambda_j}_{q_j}$ appear with $q_j=1$ and $t_j<0$. However, the same sign can be introduced on the system variables to write 
 \begin{widetext}
\setlength{\mathindent}{0pt}{\begin{equation}\label{eq:sbirula}\begin{array}{lll}
\rho^Y_S(t)&&=\displaystyle\sum_{n=\text{even}}\frac{(-i)^n}{n!}\int_0^t \left(\prod_{i=1}^n d t_i\right)\sum_{q_n,\lambda_n\cdots q_1,\lambda_1}\\
&&\left\{C^{\prime \lambda_n\cdots \lambda_1}_{q_n\cdots q_1}\hat{T}_S\left[\hhat{Y}^\text{e}_S(0)\hhat{S}_{q_n}^{\bar{\lambda}_n}\cdots\hhat{S}_{q_1}^{\bar{\lambda}_1}\right](\hat{\rho}^\text{e}_S(0)+\hat{\rho}^\text{o}_S(0))
\displaystyle+C^{\prime \lambda_n\cdots \lambda_1}_{q_n\cdots q_1}\hat{T}_S\left[\hhat{Y}^\text{o}_S(0)\hhat{P}_S(0)\hhat{S}_{q_n}^{\bar{\lambda}_n}\cdots\hhat{S}_{q_1}^{\bar{\lambda}_1}\right]P_S(\hat{\rho}^\text{e}_S(0)+\hat{\rho}^\text{o}_S(0))\right\}.
\end{array}
\end{equation}}
\end{widetext}
It is now possible to keep $Y_S(0)$ ``factorized'' on the left and follow all the reasoning which allowed us to deduce Eq.~(\ref{eq:main_result}) from Eq.~(\ref{eq:wide_for_corr}) to get
\begin{equation}
\begin{array}{lll}
\rho_S^Y(t)&=&\hhat{T}_S \hhat{Y}^\text{e}_S(0)e^{\hhat{\mathcal{F}}_{T}(t)}\rho_S^Y(-T)\\
&&+\hhat{T}_S\hhat{Y}^\text{o}_S(0)\hhat{P}_S(0)) e^{\hhat{\mathcal{F}}_{T}(t)}P_S\rho_S^Y(-T)\\

&=&\hhat{T}_S e^{\hhat{\mathcal{F}}_{T}(t)}[\hhat{Y}^\text{e}_S(0)+\hhat{Y}^\text{o}_S(0)\hhat{P}_S(0)P_S]\rho_S^Y(-T)\;\;,
\end{array}
\end{equation}
which is valid for $t\geq0$ and where 
\begin{equation}
\displaystyle\hhat{\mathcal{F}}_{T}(t)=\int_{-T}^t dt_{2}\int_{-T}^{t_2} dt_{1} \hhat{W}(t_2,t_1)\;\;.
\end{equation}
Interestingly, despite the presence of the operator $\hhat{Y}$, the formal time-derivative of the density matrix $\rho_S^Y(t)$ has the same form as Eq.~(\ref{eq:motion}), i.e.
\begin{equation}
\label{eq:rhoY_der}
\displaystyle\dot{\rho}^Y_S(t)=\displaystyle\sum_j \hhat{A}^j(t) \hhat{T}_S  \hhat{\Theta}_j (t)\rho^Y_S(t)\;\;.
\end{equation}
However, the presence of $\hhat{Y}$ gives rise to a different boundary condition which reads
\begin{equation}
\label{eq:new_initial_condition_app}
\begin{array}{lll}
\rho_S^Y(0)

&=&\hhat{T}_S e^{\hhat{\mathcal{F}}_{T}(0)}[\hhat{Y}^\text{e}_S(0)+\hhat{Y}^\text{o}_S(0)\hhat{P}_S(0)P_S]\rho_S^Y(-T)\\
&=&Y_S\hhat{T}_S e^{\hhat{\mathcal{F}}_{T}(0)}\rho_S^Y(-T)\;\;,
\end{array}
\end{equation}
\onecolumngrid
\begin{center}
\begin{figure}[t!]
    \includegraphics[scale=1.]{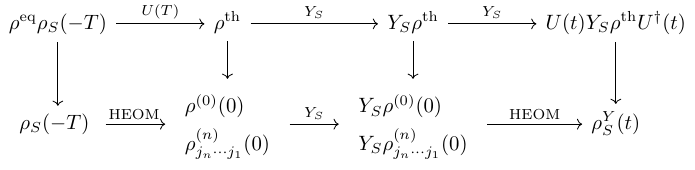}
    \caption{\raggedright Diagram showing how to generate the reduced $\rho_S^Y(t)$ needed to compute the correlation function $C^\text{th}_{XY}(t)$. The first row shows the time evolution in the system+environment while the second row the reduced system evolution. Down-arrows refer to the computation of the reduced-density matrices following the definition in Eq.~(\ref{eq:density_matrix_implicit}).}
    \label{fig_2}
\end{figure}
\end{center}
\twocolumngrid
where we used that all superoperators in $\hhat{F}(0)$ are evaluated at times $t<0$ and their number is even. 

The differential equation in Eq.~(\ref{eq:rhoY_der}) together with the initial condition in Eq.~(\ref{eq:new_initial_condition_app}) offer a direct way to compute the correlations in Eq.~(\ref{C_XY_app}). To achieve this, it is sufficient to show that the diagram in Fig.~(\ref{fig_2}) commutes.
We prove this justifying all the down arrows in \figref{fig_2}.
\begin{itemize}
\item[1.] $\rho_S(-T)$ is the reduced density matrix of $\text{Tr}_E[\rho^\text{eq}\rho_S(-T)]$. This is an immediate consequence of the identity in Eq.~(\ref{eq:effective_0}) and the fact that $\rho^\text{eq}$ has even parity, leading to $\text{Tr}_E[\rho^\text{eq}\rho_S(-T)]=\rho_S(-T)$.
\item[2.] $\rho^{(0)}(0)$ is the reduced density matrix of $\text{Tr}_E[\rho^\text{th}]$. This is a direct consequence of the meaning of the HEOM as given in Eq.~(\ref{eq:generalized_HEOM}).
\item[3.] $Y_S\rho^{(0)}(0)$ is the reduced density matrix of $\text{Tr}_E[Y_S\rho^\text{th}]$. This is a consequence of the definition of partial trace in Eq.~(\ref{eq:density_matrix_implicit}).
In fact, for all system operators $A_S$, we have $\text{Tr}_{ES}[A_S (Y_S \rho^\text{th})]=\text{Tr}_{ES}[A_S Y_S (\rho^\text{th})]=\text{Tr}_S[A_S Y_S\rho^{(0)}]$ where in the last equality we used the second down-arrow from the left. Since the superoperator associated with $Y_S$ is evaluated at time $0$, Eq.~(\ref{eq:rho_n}) implies that also the auxiliary density matrices $\rho^{(n)}_{j_n,\cdots,j_1}$ need to be multiplied by $Y_S$.
\item[4.] $\rho^Y_S(t)$ is the reduced density matrix of $U(t)Y_S\rho^\text{th}U^\dagger(t)$. This is a consequence of Eq.~(\ref{eq:rhoY_der}) which implies that the reduced density matrix $\rho^Y_S(t)$ can be computed using the usual HEOM equation, given in Eq.~(\ref{eq:generalized_HEOM}), with initial condition in Eq.~(\ref{eq:new_initial_condition_app}, i.e., $\rho^Y_S(0)=Y_S\hhat{T}_S e^{\hhat{\mathcal{F}}_{T}(0)}\rho_S^Y(-T)$. Using the results above, this initial condition does correspond to the auxiliary density matrices in the third place of the second row of the diagram. 
\end{itemize}
\subsubsection{Derivative of the reduced density matrix}
Here, we derive an expression for the time derivative of the reduced density matrix. We do this explicitly because the  Fermionic time-ordering always requires some extra-attention. 
The time derivative of the reduced density matrix in Eq.~(\ref{eq:main_result}) can be written as
\begin{equation}
\begin{array}{lll}
\displaystyle\frac{d}{dt}{\rho}_S(t)&=&\displaystyle\frac{d}{dt} \hhat{T}_S e^{\hhat{\mathcal{F}}(t) }\rho_S(0)\\
&=&\displaystyle \hhat{T}_S\sum_{n=0}^\infty\frac{d}{dt}  \frac{\hhat{\mathcal{F}(t)}^n}{n!}\rho_S(0)\;\;.
\end{array}
\end{equation}
Now, the derivative of a single $\hhat{\mathcal{F}}(t) $ is
\begin{equation}
\begin{array}{lll}
\displaystyle\frac{d}{dt}\hhat{\mathcal{F}}(t) &=&\displaystyle\frac{d}{dt}\int_0^t dt_2\int_0^{t_2}dt_1 \hhat{W}(t_2,t_1)\\
&=&\displaystyle\int_0^{t}dt_1 \hhat{W}(t,t_1)\\
&=&\displaystyle\sum_{i_2,k_2}\hhat{S}_{i_2}^{\bar{k}_2}(t)\int_0^{t}dt_1\sum_{i_1,k_1} C^{k_2,k_1}_{i_2,i_1}(t,t_1)\hhat{S}_{i_1}^{\bar{k}_1}(t_1),
\end{array}
\end{equation}
which, importantly, contains two system's superoperators. The T-product of the $n$th power of $\hhat{\mathcal{F}}(t) $ can be written as
\begin{equation}
\label{eq:derivative_rho_S}
\begin{array}{l}
\displaystyle \hhat{T}_S\frac{d}{dt}\hhat{\mathcal{F}}^n(t)=\displaystyle \hhat{T}_S\frac{d}{dt}[\hhat{\mathcal{F}}(t)\cdots\hhat{\mathcal{F}}(t)]\\
=\displaystyle \hhat{T}_S\frac{d}{dt}[\hhat{\mathcal{F}}(t)]\cdots\hhat{\mathcal{F}}(t)+\cdots+\hhat{T}_S\hhat{\mathcal{F}}(t)\cdots\frac{d}{dt}[\hhat{\mathcal{F}}(t)]\\
=\displaystyle n \hhat{T}_S\frac{d}{dt}[\hhat{\mathcal{F}}(t)]\underbrace{\hhat{\mathcal{F}}(t)\cdots\hhat{\mathcal{F}}(t)}_{n-1}\;\;,
\end{array}
\end{equation}
where, since the derivative of a single $\hhat{\mathcal{F}}(t) $ contains two system's superoperators, we can always move it in front without ``penalty'' signs from the Fermionic time ordering. Therefore, 
\begin{equation}
\label{eq:der_int}
\begin{array}{lll}
\displaystyle\frac{d}{dt}{\rho}_S(t)&=&\displaystyle \hhat{T}_S\left(\frac{d}{dt}\hhat{\mathcal{F}}(t) \right){\rho}_S(t)\;\;.
\end{array}
\end{equation}
To finish, we change to the Shroedinger frame defined as
\begin{equation}
\rho^\text{Shr}_S(t)=U(t){\rho}_S(t)U^\dagger(t)\;\;,
\end{equation}
where $U=e^{-iH_St}$, and where $H_S$ is the system's Hamiltonian. The time derivative in this frame reads
\begin{equation}
\label{eq:der}
\begin{array}{lll}
\displaystyle\frac{d}{dt}\rho^\text{Shr}_S(t)&=&\displaystyle \frac{d}{dt} U(t){\rho}_S(t)U^\dagger(t)\\
&=&\displaystyle -i[H,\rho_S(t)]+U(t)\frac{d}{dt}[{\rho}_S(t)]U^\dagger(t)\\
&=&\displaystyle -i[H,\rho_S(t)]\\
&&\displaystyle+U(t)[\hhat{T}_S\left(\frac{d}{dt}\hhat{\mathcal{F}}(t) \right){\rho}_S(t)]U^\dagger(t)\;\;.
\end{array}
\end{equation}
\section{Identities for the correlation functions}
\label{app:corr}
In this section, we derive constraints on the correlations $C^\sigma(t_2,t_1)$ defined in Eq.~(\ref{eq:corr_sigma_app}). To do this, we define the spectral density 
\begin{equation}
J(\omega)=\pi\sum_k g^2_k\delta(\omega-\omega_k)\;\;,
\end{equation}
which quantifies the strength of the interaction between the environment and the system. We then have
\begin{equation}
\label{eq:temp_C21}
\begin{array}{lll}
C^{\sigma=1}(t_2,t_1)&=&\displaystyle\text{Tr}_E[B^\dagger(t_2)B(t_1)\rho_E^\text{eq}]\\
&=&\displaystyle\sum_k g_k^2\text{Tr}_E[c^\dagger_k(t_2)c_k(t_1)\rho_E^\text{eq}]\\
&=&\displaystyle\sum_k g_k^2  e^{i\omega_k (t_2-t_1)}n^\text{eq}_k\\
&=&\displaystyle\frac{1}{\pi}\int d\omega J(\omega) e^{i\omega(t_2-t_1)} n^\text{eq}(\omega)\;\;,
\end{array}
\end{equation}
where $n^\text{eq}_k=\text{Tr}_E[c_k^\dagger c_k \rho_E^\text{eq}]$. The equilibrium thermal state for the environment is the Boltzmann distribution $\rho_E^\text{eq}=\exp{[-\beta\sum_k(\omega_k-\mu) c_k^\dagger c_k]}/Z^\text{eq}_E=\prod_k e^{-\beta(\omega_k-\mu) c_k^\dagger c_k}/(1+\exp[-\beta(\omega_k-\mu)])$, where $Z^\text{eq}_E=\text{Tr}_E \exp{[-\beta\sum_k(\omega_k-\mu) c_k^\dagger c_k]}=\prod_k (1+\exp[-\beta(\omega_k-\mu)])$. These definitions allow to write the Fermi-Dirac distribution $n^\text{eq}_k=\exp{[-\beta(\omega_k-\mu)]}/(1+\exp[-\beta(\omega_k-\mu)])=1/(\exp[\beta(\omega_k-\mu)]+1)$ which, in the continuum version, reads
\begin{equation}
\label{eq:FermiDirac}
n^\text{eq}(\omega)=\frac{1}{\exp[\beta(\omega-\mu)]+1}\;\;.
\end{equation}
We can also consider
\begin{equation}
\label{eq:temp_C21_prime_0}
\begin{array}{l}
C^{\sigma=-1}(t_2,t_1)=\text{Tr}_E[B(t_2)B^\dagger(t_1)\rho]\\
=\displaystyle\sum_k g_k^2\text{Tr}_E[c_k(t_2)c^\dagger_k(t_1)\rho]\\
=\displaystyle\sum_k g_k^2 e^{-i\omega_k(t_2-t_1)}\text{Tr}_E[c_kc^\dagger_k\rho]\\
=\displaystyle\sum_k g_k^2 e^{-i\omega_k(t_2-t_1)}(1-n^\text{eq}_k)\\
=\displaystyle\frac{1}{\pi}\int J(\omega) e^{-i\omega(t_2-t_1)}[1-n^\text{eq}(\omega)]\;\;.
\end{array}
\end{equation}
Both Eq.~(\ref{eq:temp_C21}) and Eq.~(\ref{eq:temp_C21_prime_0}) can be written together as
\begin{equation}
\label{eq:temp_corr_app}
\begin{array}{lll}
C^{\sigma}(t_2,t_1)=\displaystyle\int\frac{d\omega}{\pi} J(\omega) e^{i\sigma\omega(t_2-t_1)} \frac{1-\sigma+2\sigma n^\text{eq}(\omega)}{2},
\end{array}
\end{equation}
which is Eq.~(\ref{eq:temp_corr}) in the main article. 
Alternatively, we can also write
\begin{equation}
\label{eq:temp_C21_prime}
\begin{array}{l}
C^{\sigma=-1}(t_2,t_1)=\text{Tr}_E[B(t_2)B^\dagger(t_1)\rho]\\
=\displaystyle\sum_k g_k^2\text{Tr}_E[c_k(t_2)c^\dagger_k(t_1)\rho]\\
=\displaystyle\sum_k g_k^2 e^{-i\omega_k(t_2-t_1)}\text{Tr}_E[c_kc^\dagger_k\rho]\\
=\displaystyle\sum_k g_k^2 e^{\beta(\omega_k-\mu)}e^{-i\omega_k(t_2-t_1)}\text{Tr}_E[c^\dagger_k c_k\rho]\\
=\displaystyle e^{-\beta
\mu}\sum_k  g_k^2 e^{-i\omega_k[t_2-(t_1-i\beta)]} n^\text{eq}_k\\
=\displaystyle \frac{e^{-\beta
\mu}}{\pi}\int d\omega  J(\omega)e^{-i\omega[t_2-(t_1-i\beta)]} n^\text{eq}(\omega)\;\;,
\end{array}
\end{equation}
where we used Eq.~(\ref{eq:comm}), i.e.,
\begin{equation}
c^\dagger_k\rho_\beta=e^{\beta(\omega_k-\mu)}\rho_\beta c^\dagger_k\;\;.
\end{equation}
Inspection of Eq.~(\ref{eq:temp_C21}) and Eq.~(\ref{eq:temp_C21_prime}) directly leads to the following correspondence between time-reversal and conjugation, i.e., 
\begin{equation}
\label{eq:corr_id_1}
\begin{array}{lll}
\bar{C}^{\sigma}(t_2,t_1)&=&C^{\sigma}(t_1,t_2)\;\;,
\end{array}
\end{equation}
where the bar denotes complex conjugation and for $\sigma=\pm1$. At the same time, by comparing Eq.~(\ref{eq:temp_C21}) and Eq.~(\ref{eq:temp_C21_prime}) we arrive to the relation
\begin{equation}
\label{eq:Cprime}
C^{\sigma=-1}(t_1,t_2)=e^{-\beta\mu}C^{\sigma=1}(t_2-i\beta,t_1)\;\;.
\end{equation}
Using the ansatz in Eq.~(\ref{eq:ansats_correlations}), i.e.,
\begin{equation}
C^\sigma(t_2,t_1)=\sum_m a^\sigma_m e^{-b^\sigma_m (t_2-t_1)}\;\;,
\end{equation}
together with Eq.~(\ref{eq:Cprime}) and Eq.~(\ref{eq:corr_id_1}) we find
\begin{equation}
\begin{array}{lll}
C^{\sigma=-1}(t_1,t_2)&=&\bar{C}^{\sigma=-1}(t_2,t_1)\\
&=&\displaystyle\sum_m \bar{a}^{\sigma=-1}_m e^{-\bar{b}^{\sigma=-1}_m (t_2-t_1)}\\
C^{\sigma=-1}(t_1,t_2)&=&\displaystyle e^{-\beta\mu}\sum_m a_m^{\sigma=1}e^{-b^{\sigma=1}_m(t_2-t_1)}e^{i\beta b_m^{\sigma=1}}\;\;,
\end{array}
\end{equation}
which implies
\begin{equation}
\label{eq:ab_app}
\begin{array}{lll}
\bar{a}_m^{\sigma=-1}&=&e^{-\beta (\mu-ib_m^{\sigma=1})}a_m^{\sigma=1}\\
\bar{b}_m^{\sigma=-1}&=&b_m^{\sigma=1}\;\;.
\end{array}
\end{equation}
This allows us to explicitly write
\begin{eqnarray*}
C^{\sigma=1}(t_2,t_1)&=&\displaystyle\sum_m a^{\sigma=1}_m e^{-b^{\sigma=1}_m (t_2-t_1)}\\
C^{\sigma=-1}(t_1,t_2)&=&\displaystyle\sum_m \bar{a}^{\sigma=-1}_m e^{-\bar{b}^{\sigma=-1}_m (t_2-t_1)}\\
&=&\displaystyle\sum_m \bar{a}^{\sigma=-1}_m e^{-b^{\sigma=1}_m (t_2-t_1)}\;\;,
\end{eqnarray*}
which shows their similarity in the time-dependence in the exponent. Similarly,
\begin{equation}
\begin{array}{lll}
C^{\sigma=1}(t_1,t_2)&=&\displaystyle\sum_m \bar{a}^{\sigma=1}_m e^{-\bar{b}^{\sigma=1}_m (t_2-t_1)}\\
&=&\displaystyle\sum_m \bar{a}^{\sigma=1}_m e^{-{b}^{\sigma=-1}_m (t_2-t_1)}\\
C^{\sigma=-1}(t_2,t_1)&=&\displaystyle\sum_m {a}^{\sigma=-1}_m e^{-{b}^{\sigma=-1}_m (t_2-t_1)}\;\;.
\end{array}
\end{equation}
Similarly, since Eq.~(\ref{eq:ab_app}) implies $\bar{b}^{\bar{\sigma}}_m=b^\sigma_m$, we have
\begin{equation}
\label{eq:Cbarbar_app}
\begin{array}{lll}
\bar{C}^{\bar{\sigma}}(t_2,t_1)&=&\displaystyle\sum_m \bar{a}^{\bar{\sigma}}_m e^{-\bar{b}^{\bar{\sigma}}_m (t_2-t_1)}\\
&=&\displaystyle\sum_m \bar{a}^{\bar{\sigma}}_m e^{-{b}^{{\sigma}}_m (t_2-t_1)}\;\;.
\end{array}
\end{equation}
\nocite{apsrev41Control}
\bibliographystyle{apsrev4-2}

\bibliography{bib}
\end{document}